\documentclass[aps,prb,twocolumn,amsmath,amssymb,nofootinbib,eqsecnum,superscriptaddress,floatfix]{revtex4}

\usepackage{amsmath}
\usepackage{amssymb}
\usepackage{amsthm}
\usepackage[dvips]{color} 
\usepackage{graphicx}
\usepackage{dcolumn} 
\usepackage{bm} 
\usepackage[hypertex]{hyperref}
\usepackage{longtable}
\usepackage{ulem}   
\newcommand{\bs}[1]{{\boldsymbol{#1}}} 

\begin{document}

\title{ 
Superconductivity on the surface of topological insulators and
in two-dimensional noncentrosymmetric materials
      }

\author{Luiz Santos} 
\affiliation{
Department of Physics, 
Harvard University, 
17 Oxford Street, 
Cambridge, Massachusetts 02138.
USA
            } 

\author{Titus Neupert} 
\affiliation{
Condensed Matter Theory Laboratory,
RIKEN, Wako, Saitama 351-0198, Japan
            } 
\author{Claudio Chamon} 
\affiliation{
Physics Department, 
Boston University, 
Boston, Massachusetts 02215, USA
            } 
\author{Christopher Mudry} 
\affiliation{
Condensed Matter Theory Group, 
Paul Scherrer Institute, CH-5232 Villigen PSI,
Switzerland
            } 

\date{\today}

\begin{abstract}
We study the superconducting instabilities of a single species of
two-dimensional Rashba-Dirac fermions, as it pertains to the surface of a
three-dimensional time-reversal symmetric topological band insulator.
We also discuss the similarities as well as the differences between
this problem and that of superconductivity in two-dimensional
time-reversal symmetric noncentrosymmetric materials with spin-orbit 
interactions. The superconducting order parameter has both 
$s$-wave and $p$-wave components, 
even when the superconducting pair potential only
transfers either pure singlet or pure triplet pairs of electrons
in and out of the
condensate, a corollary to the nonconservation of spin due to the
spin-orbit coupling. We identify one single superconducting regime in
the case of superconductivity in the topological surface states 
(Rashba-Dirac limit), 
irrespective of the relative strength between singlet and
triplet pair potentials. In contrast, in the Fermi limit relevant to
the noncentrosymmetric materials we find two regimes depending
on the value of the chemical potential and the relative strength
between singlet and triplet potentials. We construct explicitly the
Majorana bound states in these regimes. In the single regime for the
case of the Rashba-Dirac limit, there exists one
and only one Majorana fermion bound to the core of an isolated
vortex. In the Fermi limit, there are always an even number 
(0 or 2 depending on the regime) of Majorana fermions bound
to the core of an isolated vortex.
In all cases, the vorticity
required to bind Majorana fermions is 
quantized in units of the flux quantum, in contrast to
the half flux in the case of two-dimensional 
$p^{\ }_{x}\pm\text{i}p^{\ }_{y}$
superconductors that break time-reversal symmetry.

\end{abstract}

\maketitle

\section{Introduction}
\label{sec: intro}

Bi$_2$Se$_3$ 
is an inversion-symmetric layered band insulator 
with a bulk gap estimated to be 
$0.35$ eV.%
~\cite{Mooser56,Black57,Xia09}
Density-functional theory predicts 
that Bi$_2$Se$_3$ supports a
single Rashba-Dirac cone of gapless surface states,
a prediction that has been verified using
angle-resolved photoemission spectroscopy.%
\cite{Xia09,Zhang09}
This remarkable attribute of Bi$_2$Se$_3$,
which has otherwise only been observed in 
the insulating alloys Bi$_{1-x}$Sb$_{x}$ so far,%
~\cite{Hsieh08,Hsieh09}
is the defining property of a 
three-dimensional 
(3D) time-reversal symmetric (TRS) topological band insulator.%
~\cite{Fu07,Moore07,Roy06}
In a recent work, Hor \emph{et al}.~\cite{Hor09}have reported the observation
of strongly type II superconductivity in Cu$_x$Bi$_2$Se$_3$ 
below 3.8 K when Cu is intercalated between the Bi$_2$Se$_3$ layers.%
They have also proposed to use Cu$_x$Bi$_2$Se$_3$
as a mean to induce superconducting correlations for the 
TRS topological surface states by the proximity effect.

The surface states in a 
$3D$ TRS topological band insulator
are reminiscent of the Bloch states of graphene in that, in both cases, 
their density of states vanishes linearly at the so-called 
Rashba-Dirac point.%
~\cite{Castro09}
However, they differ in a fundamental way from those in
graphene. For example, the surface of Bi$_2$Se$_3$
supports one Rashba-Dirac cone as opposed to two in graphene.
This difference is a manifestation of the fact that
inversion symmetry is maximally broken on the surface
of Bi$_2$Se$_3$ in that the kinetic energy is 
dominantly of the Rashba type, whereas the spin-orbit coupling
is for all intent and purposes negligible for graphene.
Consequently, the surface states of a
3D TRS topological band insulator
are not localized by weak TRS disorder,%
~\cite{Ostrovsky07}${}^{-}$\cite{Nomura07}
whereas Anderson localization rules in graphene.%
~\cite{Ostrovsky06}

Another difference with graphene, as we shall show in this paper as a
warm up, is that all states in the Rashba-Dirac sea contribute to the Pauli
magnetic susceptibility, which is anisotropic in that the in-plane and
out-of-plane components differ by a factor of 2. For comparison, the
Pauli magnetic susceptibility is isotropic in spin space and proportional
to the density of states at the Fermi surface in any electron gas
(including graphene) with small breaking of the spin-rotation symmetry
(SRS). This anisotropy and the fact that the Pauli susceptibility does
not only depend on the density of states at the Fermi level could
potentially be used as a simple diagnostic of a limit in which the
Rashba coupling is the largest energy scale.

The main emphasis of this paper will be on the superconducting
instabilities of the surface states in a 3D TRS topological band
insulator and on those in close relatives, i.e., two-dimensional (2D) TRS
noncentrosymmetric superconductors in a regime that has been little
studied so far. The theoretical studies of
noncentrosymmetric superconductors with TRS
usually assume the hierarchy of energy scales 
\begin{equation}
t\gg\alpha\gg\Delta
\label{eq: hierarchy A}
\end{equation}
where $t$ is the inversion-symmetric band width,
$\alpha>0$ is the spin-orbit coupling that preserves TRS
but breaks SRS, and
$\Delta>0$ is the single-particle superconducting gap.%
~\cite{Edelstein89}$^{-}$\cite{Kaur05}
The regime 
\begin{equation}
\alpha\gg\Delta\gg t
\label{eq: hierarchy B}
\end{equation}
is the one that applies to intrinsic superconducting instabilities 
of the surface states in a 3D TRS topological band insulator.
We will address the question of whether interesting phenomena
associated to superconductivity
occur upon exchanging the hierarchies
(\ref{eq: hierarchy A}) and (\ref{eq: hierarchy B}).

In the same way that (2D) TRS
band insulators have been classified according
to their topological character,%
\cite{Kane05a}$^{-}$\cite{Konig07}
Bogoliubov-de-Gennes (BdG) superconductors 
have also been given topological attributes
whenever they support gapless boundary states in
confined geometries.%
~\cite{Schnyder08}$^{-}$\cite{Kitaev09}
A necessary (but not sufficient)
condition for a 2D TRS superconductor
to be topologically non-trivial is that it is noncentrosymmetric,
or, equivalently, that it breaks SRS.
According to Refs.~\onlinecite{Schnyder08} and \onlinecite{Qi09a},
a sufficient condition is that,
for any weak and local TRS static disorder,
a 2D TRS superconductor
in an infinitely long strip geometry 
supports an odd number of Kramers' doublets of gapless edge states 
of which at least one wave function is extended along the edge
(see also Refs.~\onlinecite{Roy08}-\onlinecite{Yip09}
for varying alternative criteria).

Applying this definition of a topological superconducting phase to the
superconducting instabilities of surface states in a 3D TRS
topological band insulator immediately leads to a paradox: What is the
meaning of the boundary of a boundary?  
A more meaningful question to
ask might be: What are the spectral properties of TRS-breaking
vortices if the
surface states in a 3D TRS topological band insulator support a type
II superconducting order? Do they bind mid gap states generically,
zero modes in particular, or not? These are questions that we address
in this paper.

Defects in a type II superconductor are vortices. 
On the one hand, Caroli \emph{et al}.~\cite{Caroli64}have shown
that vortices in an $s$-wave TRS and SRS superconductor
bind nonvanishing-energy bound states 
with a mean level spacing of the order
of the superconducting gap squared divided by the
Fermi energy.%
On the other hand, Jackiw and Rossi in Ref.%
~\onlinecite{Jackiw81}
found a single bound state that is exponentially localized
around the core of a unit-flux vortex in a 2D $s$-wave
relativistic superconductor with a vanishing density of
states (Rashba-Dirac point). The energy of this bound state 
is precisely pinned to the
Fermi energy (see also Ref.%
~\onlinecite{Weinberg81}
for the corresponding index theorem and Refs.%
~\onlinecite{Kopnin91} 
and 
\onlinecite{Volovik93}
for examples of nonrelativistic zero modes bound to vortices).
A midgap state bound to the core of a vortex
does not carry an electric charge, 
for it is an eigenstate of the generator of the 
particle-hole symmetry (PHS) obeyed by any BdG
Hamiltonian. It is thus charge neutral and as such
is a physical realization of a Majorana fermion. 
Majorana fermions were also found to be
exponentially localized to the core
of a vortex in a $p^{\ }_{x}\pm \text{i}p^{\ }_{y}$
type II superconductor by
Read and Green and by Ivanov in Refs.~\onlinecite{Read00} 
and \onlinecite{Ivanov01}, respectively.
More importantly, they showed that these Majorana fermions
obey non-Abelian braiding statistics.
Theoretical proposals to nucleate Majorana fermions 
have been made relying on 2D TRS noncentrosymmetric superconductors,
\cite{Lu08,Sato09,Tanaka09a} 
or on proximity effects at the 2D 
interface between band insulators, superconductors, and ferromagnets.%
~\cite{Fu08,Sau10,Lee10}

We will show in this paper that, 
when the dispersion is Rashba-Dirac like,
there is a single zero mode bound to the core of an isolated vortex 
with unit circulation, and thus a single Majorana bound state.
The mechanism, in the case of singlet pairing, is precisely
that of Jackiw and Rossi.~\cite{Jackiw81} This zero mode remains
for arbitrary ratios of triplet and singlet pairing, 
with the pairing potentials $\Delta^{\ }_{\mathrm{t}}$ 
and $\Delta^{\ }_{\mathrm{s}}$, respectively, 
and also as the
chemical potential $\mu$ is varied. The stability of a singly degenerate
zero mode is guaranteed in a system with particle-hole symmetry in 
which the zero mode is isolated from the continuous spectrum by a finite
energy gap. Therefore, studying gap-closing surfaces in the parameter
space of the coupling constants characterizing the theory is of
crucial importance in identifying the stability of the Majorana 
modes as well as the phase boundaries between different topological phases.

In this paper, we compute
the conditions for the closing of the gap in 
\mbox{$\Delta^{\ }_{\mathrm{t}}/\Delta^{\ }_{\mathrm{s}}$--$\mu$} 
space by
exploring a one-to-one mapping to the normal-state dispersion
relation, in which a function of the ratio 
$\Delta^{\ }_{\mathrm{t}}/\Delta^{\ }_{\mathrm{s}}$ serves as a
reparameterization of the magnitude of the momenta in the dispersion
relation. Thereto, we show that there are as many lines in
\mbox{$\Delta^{\ }_{\mathrm{t}}/\Delta^{\ }_{\mathrm{s}}$--$\mu$} space
at which the gap closes as there are branches in the dispersion relation. 
But in the case of the Rashba-Dirac dispersion, 
it is possible to go from one 
side of a gap vanishing line to another without crossing it 
by going through the point at infinity
($\Delta^{\ }_{\mathrm{s}}=0$).
Thus, there are 
not two distinct phases separated by a transition 
in this case, but there is a single phase instead.

In the Fermi limit relevant to 2D TRS noncentrosymmetric superconductors,
we find that the conditions for the closing of the spectral gap do
separate two gapful phases. 
These two regimes are those in which either
the singlet or the triplet pairing controls the physics. 
The detailed shape of the phase boundaries is dictated
by the normal-state dispersion relation. The presence of the TRS
spin-orbit coupling leads to interesting effects at certain values of
the chemical potential, for example, re-entrance to the phase
dominated by singlet-pairing physics even when $\Delta^{\
}_{\mathrm{t}}/\Delta^{\ }_{\mathrm{s}}$ is large. We find two
Majorana zero modes bound to an isolated vortex 
in the triplet controlled phase but they
disappear in the singlet controlled phase.  
We find that the vortices
that bind this pair of Majorana zero modes have unit flux, as opposed to
the half-vortices needed in the case of two-dimensional 
$p^{\ }_{x}\pm\text{i}p^{\ }_{y}$ superconductors that break time-reversal
symmetry. The physical reason for this difference is that, when TRS
holds, the spin-resolved pairing amplitudes 
$\Delta^{\ }_{\uparrow\uparrow}$ and
$\Delta^{\ }_{\downarrow\downarrow}$ are not independent, and thus one
cannot introduce vorticity in one but not the other, which is the case
for the half vortices in the $p^{\ }_{x}\pm\text{i}p^{\ }_{y}$
superconductors. Therefore the pair of Majorana fermions that we find for full
vortices in the triplet case is distinct from those found by Read and
Green~\cite{Read00} and Ivanov.~\cite{Ivanov01}
The pair of Majorana fermions that we find
is not robust to a generic weak perturbation 
that breaks translation invariance, for these Majorana fermions
are not related to each other by the operation of time reversal.
This pair of Majorana fermions is thus unrelated to the one introduced
by Qi \emph{et al.\ }in Ref.~\onlinecite{Qi09a}
as a mean to identify the triplet dominated TRS phase as a
nontrivial 2D $\mathbb{Z}^{\ }_{2}$ topological superconducting phase.
We conclude that,
although both superconducting regimes can be distinguished by the even number
of Majorana fermions that an isolated TRS-breaking vortex binds, 
this distinction is not topological, 
for it is not robust to static disorder, for example.

In addition to this interesting interplay 
between the singlet and triplet pair potentials
for the existence of Majorana fermions,
superconductivity for surface states 
in a 3D TRS topological band insulator and noncentrosymmetric materials
has other curious properties, one
of which is the following. Because spin is not a good quantum number,
even if the pair potential contains, say, only the singlet
component, the condensate will nevertheless have triplet correlations. 
For example, the pairing correlation for electrons 
$\langle c^{\ }_{\boldsymbol{k}\uparrow}
         c^{\ }_{-\boldsymbol{k}\uparrow}\rangle\ne 0$ 
(and 
$\langle c^{\ }_{\boldsymbol{k}\downarrow} 
         c^{\ }_{-\boldsymbol{k}\downarrow}\rangle\ne 0$ as well) 
for $\mu\ne 0$, even if $\Delta^{\ }_{\mathrm{t}}=0$. 
One measurable consequence is that, generically,
these superconducting states would lead to detectable Josephson
currents when connected to either conventional $s$-wave or $p$-wave
superconductors, manifesting the fact that they have both types of
correlations (even if only the singlet pair potential
$\Delta^{\ }_{\mathrm{s}}$ is nonzero).

The paper is organized as follows.  
We define the model in Sec.~\ref{sec: Definition}.
We show in Sec.~\ref{sec: Susceptibility}
that the normal-state Pauli magnetic susceptibility has the
remarkable property that it depends on all states in the Fermi sea, not
only on the density of states at the Fermi level, and that it is
anisotropic for the surface states of a 3D TRS topological band insulator. 
The dynamical Pauli susceptibility tensor 
in both the normal and superconducting states
encodes rich magnetoelectric effects that are responsible
for the spin-Hall effect among others.
We study in Sec.~\ref{sec: Superconducting instabilities}
self-consistently the interplay between
the singlet and triplet components to the 
superconducting instabilities of the surface states 
in a 3D TRS topological band insulator or in 2D
noncentrosymmetric materials.
The generic mean-field phase diagram
for a TRS two-band BdG Hamiltonian in the isotropic continuum limit
that interpolates between 
the regimes (\ref{eq: hierarchy A}) and (\ref{eq: hierarchy B})
is constructed in Sec.~\ref{sec: Mean-field phase diagram}.
We find that the Rashba-Dirac limit $t/\alpha=0$, 
that pertains to the surface states of a 3D TRS topological
band insulator, is singular in that
there exists only one single phase in the phase diagram.
In Sec.~\ref{sec: Majorana fermions a},
we construct explicitly the \textit{single} Majorana fermion bound to the core
of an isolated vortex in the superconducting phase
of the surface states of a 3D TRS topological
band insulator.
In Sec.~\ref{sec: Majorana fermions b},
we construct explicitly the \textit{pair} 
of Majorana fermions bound to the core
of an isolated vortex in the triplet dominated superconducting phase
of a 2D noncentrosymmetric superconductor.
We conclude with Sec.~\ref{sec: Discussion}.

\begin{figure}
(a)\includegraphics[angle=0,scale=0.3]{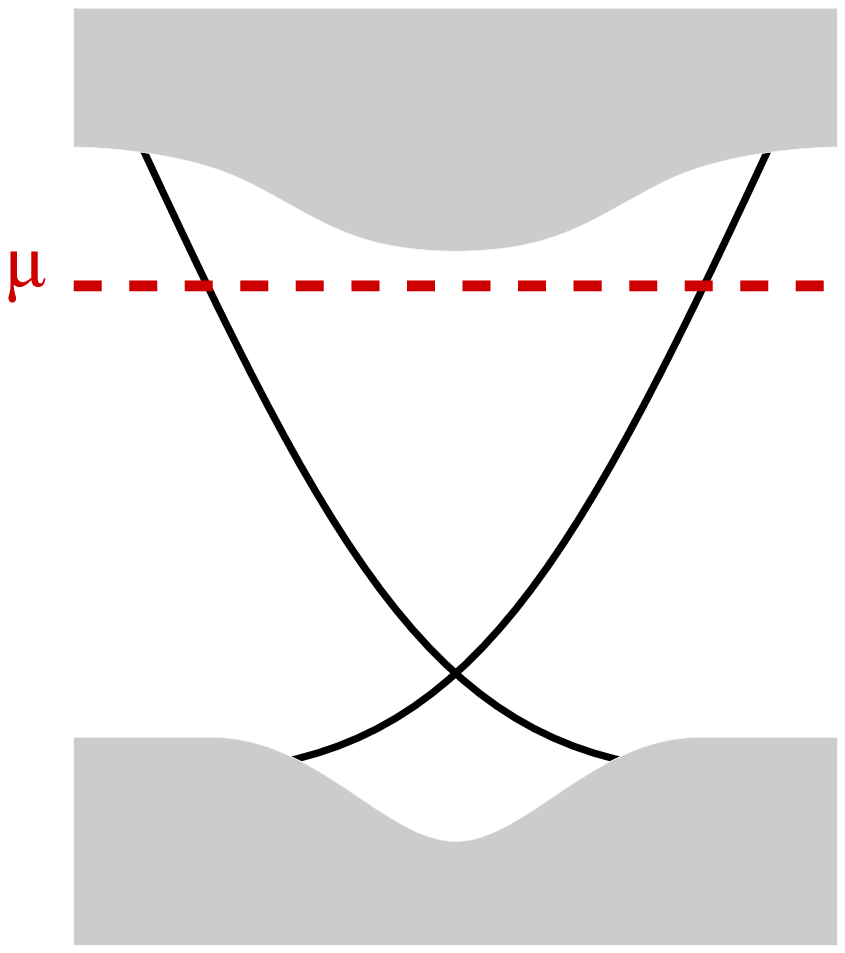}
(b)\includegraphics[angle=0,scale=0.3]{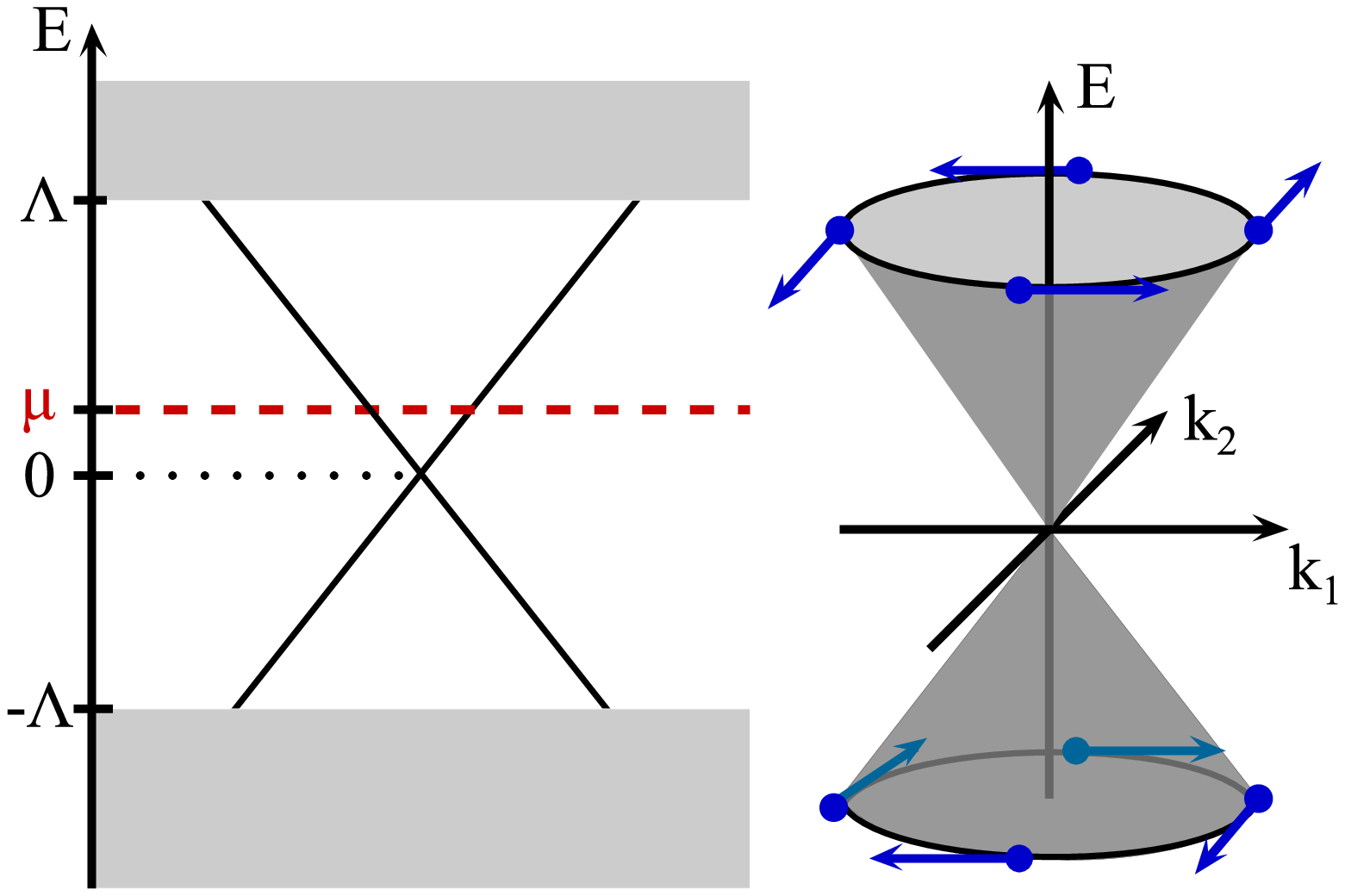}
\caption{(Color online)
(a) Schematic picture of the surface states of the topological insulator 
Bi$_2$Se$_3$ (black lines). 
The chemical potential $\mu$ is far from the Rashba-Dirac (nodal) point and close 
to the conduction-band continuum (upper gray region), 
while the Rashba-Dirac (nodal) point is close to the valence-band continuum 
(lower gray region).
(b) Left: one-dimensional cut of the dispersion of the  
Rashba-Dirac model defined by Eq.~(\ref{eq: def 2x2 Rashba-Dirac H0}). 
In particular, we study the case where $\mu$ is close 
to the Rashba-Dirac (nodal) point 
rather than close to the energy cutoff $\pm\Lambda$ 
that defines the onset of the conduction band and the valence band. 
Right: the expectation values of the electron spins are perpendicular 
to their momenta and oriented in opposite directions for the upper and 
lower cone 
[see Eqs.~(\ref{eq: expectation b}) 
and~(\ref{eq: expectation c}) at $B=0$].
         }
\label{fig:dispersions}
\end{figure}

\section{
Definition
        }
\label{sec: Definition}

\subsection{
Normal state
           }

In the continuum limit and in the single-particle approximation, 
we define the single-node Rashba-Dirac Hamiltonian
\begin{equation}
\mathcal{H}^{\mathrm{sur}}_{0\,\boldsymbol{k}}:=
\hbar v^{\ }_{\mathrm{RD}}\,
\left(
k^{\ }_1
\sigma^{\ }_{2}
-
k^{\ }_2
\sigma^{\ }_{1}
\right),
\qquad
\hbar v^{\ }_{\mathrm{RD}}\,|\boldsymbol{k}|<\Lambda.
\label{eq: def 2x2 Rashba-Dirac H0}
\end{equation}
The Rashba-Dirac velocity is $v^{\ }_{\mathrm{RD}}$
and we restrict the momentum $\hbar\boldsymbol{k}$
by the cutoff $\Lambda/v^{\ }_{\mathrm{RD}}$ beyond which
the surface states of a TRS topological band insulator
merge into the bulk states. 
The two-dimensional momentum
$\hbar\boldsymbol{k}=\hbar(k^{\ }_{1},k^{\ }_{2})$
couples to the Pauli matrices
$\boldsymbol{\sigma}=(\sigma^{\ }_{1},\sigma^{\ }_{2})$.
These Pauli matrices act on the internal space of the 
spin-1/2 degrees of freedom carried by the surface electron (hole)
in the laboratory frame of reference.
This coupling between the electron (hole) crystal 
wave vector and the spin of the electron (hole)
prevents conservation of the electron (hole) spin.
However, TRS is conserved so that the linear dispersion
that follows from 
Eq.~(\ref{eq: def 2x2 Rashba-Dirac H0})
is twofold Kramers degenerate.

For the surface states of Bi$_2$Se$_3$,
the Rashba-Dirac velocity is measured to be
$v^{\ }_{\mathrm{RD}}\approx 5.1\times 10^{5}\,\hbox{m s}^{-1}$.%
~\cite{Xia09}
Furthermore, the Rashba-Dirac energy
(the energy measured at the Rashba-Dirac point) 
$\varepsilon^{\ }_{\mathrm{RD}}\approx0.3\,\hbox{eV}$
is close to the insulating band gap of
$0.35\,\hbox{eV}$ 
for the bulk states in which the surface states merge.%
~\cite{Xia09}
Hence, the superconducting gap 
$\Delta\approx 3\times10^{-4}\,\hbox{eV}$
in intercalated Cu$_x$Bi$_2$Se$_3$
is minute compared to $\varepsilon^{\ }_{\mathrm{RD}}$ 
in Bi$_2$Se$_3$.
If there are Rashba-Dirac surface states in Cu$_x$Bi$_2$Se$_3$
involved in pairing correlations or if there
are Rashba-Dirac surface states in Bi$_2$Se$_3$
involved in pairing correlations induced by the proximity
to superconducting Cu$_x$Bi$_2$Se$_3$, 
they are likely to be far away from the Rashba-Dirac point. 
On the other hand, we take the point of view
that it is only a matter of time before a way is found to tune
the chemical potential of the TRS topological surface states
through the Rashba-Dirac point
(substituted Cu$_y$Bi$_{2-y}$Se$_3$ might be a candidate). 
Hence, one goal of this paper is to characterize
pairing correlations among the surface states of
a TRS topological band insulator upon tuning the
Fermi energy through the Rashba-Dirac point 
(see Fig.~\ref{fig:dispersions}).

We shall compare our study of Eq.~(\ref{eq: def 2x2 Rashba-Dirac H0})
with that of the two-dimensional Rashba tight-binding model
\begin{subequations}
\label{eq: def 2x2 Rashba on lattice}
\begin{equation}
\mathcal{H}^{2\text{D}}_{0\,\boldsymbol{k}}:=
\varepsilon^{\ }_{\boldsymbol{k}}
\sigma^{\ }_{0}
+
\boldsymbol{g}^{\ }_{\boldsymbol{k}}
\cdot
\boldsymbol{\sigma},
\qquad
\boldsymbol{k}\in\hbox{ BZ}.
\label{eq: def 2x2 Rashba on lattice a}
\end{equation}
Here, $\sigma^{\ }_{0}$ is the unit $2\times2$ matrix
in spin space and the wave vector $\boldsymbol{k}$
is restricted to the first Brillouin zone (BZ).
It describes the hopping on a square lattice with the
SRS dispersion
\begin{equation}
\varepsilon^{\ }_{\boldsymbol{k}}=
-
2t
\left(
\cos k^{\ }_{1}
+
\cos k^{\ }_{2}
\right),
\qquad
t\geq0,
\label{eq: def 2x2 Rashba on lattice b}
\end{equation}
and with the Rashba spin-orbit coupling
\begin{equation}
\boldsymbol{g}^{\ }_{\boldsymbol{k}}=
\alpha
\begin{pmatrix}
-
\sin k^{\ }_{2}
\\
\sin k^{\ }_{1}
\end{pmatrix},
\label{eq: def 2x2 Rashba on lattice c}
\end{equation}
\end{subequations}
say. Our convention throughout this paper will be that
$\alpha\geq0$.

An important difference between
the surface states of a 3D TRS topological band insulator
and the Rashba tight-binding states
is that the surface states span an odd number less of Fermi surfaces.
This is a manifestation 
of the fermion doubling that occurs when attempting
to regularize a $D$-dimensional single Rashba-Dirac cone 
by a $D$-dimensional tight-binding model.
The fermion doubling can only be eliminated by the addition
of the Wilson term
\begin{equation}
\mathcal{H}^{\text{W}}_{0\,\boldsymbol{k}}:=
t^{\ }_{\text{W}}
\left(
2
-
\cos k^{\ }_{1}
-
\cos k^{\ }_{2}
\right)
\sigma^{\ }_{3},
\quad
t^{\ }_{\text{W}}\gg
t
+
\alpha,
\end{equation}
to the tight-binding Hamiltonian~(\ref{eq: def 2x2 Rashba on lattice a})
at the cost of breaking TRS.

Hamiltonian~(\ref{eq: def 2x2 Rashba-Dirac H0})
is scale invariant. It then follows that the density
of states per unit area is proportional to the 
absolute value of the chemical potential $\mu$
and vanishes at the Rashba-Dirac point $\mu=0$.
The effects of this scale invariance on
charge transport, including the orbital effects of a magnetic field, 
are identical to those in graphene in the single
Rashba-Dirac cone approximation, if the Zeeman
coupling to a magnetic field is ignored. 
In Sec.~\ref{sec: Susceptibility}, we are going to study
the inherently strong effects of the spin-orbit coupling on
the Pauli magnetic susceptibility. However, before doing so,
we want to include the possibility of a superconducting instability
that we first treat at the mean-field level without imposing the condition
of self-consistency.

\label{sec: Sc state}

\subsection{
Mean-field superconducting state
           }
We rewrite the continuum Hamiltonian 
(\ref{eq: def 2x2 Rashba-Dirac H0})
or the lattice Hamiltonian 
(\ref{eq: def 2x2 Rashba on lattice})
in the language of second quantization.
For simplicity, we choose a tight-binding notation.
Reverting notation to the continuum is straightforward.
We thus introduce the spinor 
$\psi^{\dag}_{\boldsymbol{k}}=
\left(
c^{\dag}_{\boldsymbol{k}\uparrow}, 
c^{\dag}_{\boldsymbol{k}\downarrow}
\right)$ 
for electrons in the spin basis of the laboratory frame
of reference and the spinor 
$\tilde{\psi}^{\dag}_{\boldsymbol{k}}=
\left(
a^{\dag}_{\boldsymbol{k}+}, 
a^{\dag}_{\boldsymbol{k}-}
\right)$ 
in the helicity basis defined below. This gives
\begin{subequations}
\begin{equation}
\begin{split}
&
H^{\ }_{0}=
\sum_{\boldsymbol{k}\in\mathrm{BZ}}
\psi^{\dag}_{\boldsymbol{k}} 
\mathcal{H}^{\ }_{0;\boldsymbol{k}} 
\psi^{\ }_{\boldsymbol{k}}=
\sum_{\boldsymbol{k}\in\mathrm{BZ}}
\tilde{\psi}^{\dag}_{\boldsymbol{k}}
\tilde{\mathcal{H}}_{0;\boldsymbol{k}} 
\tilde{\psi}^{\ }_{\boldsymbol{k}},
\\
&
\mathcal{H}^{\ }_{0;\boldsymbol{k}}=
\left(
\varepsilon^{\ }_{\boldsymbol{k}}
-
\mu
\right)
\sigma^{\ }_{0}
+ 
\boldsymbol{g}^{\ }_{\boldsymbol{k}}
\cdot
\boldsymbol{\sigma},
\\
&
\tilde{\mathcal{H}}_{0;\boldsymbol{k}}=
\left(\begin{array}{cc}
\xi^{\ }_{\boldsymbol{k}+}
&
0
\\
0
&
\xi^{\ }_{\boldsymbol{k}-}
\end{array}
\right).
\label{eq: def second quantization}
\end{split}
\end{equation}
The single-particle dispersion is here given by%
~\cite{footnote: alpha to-alpha}
\begin{equation}
\xi^{\ }_{\boldsymbol{k}\pm}=
\varepsilon^{\ }_{\boldsymbol{k}}
-
\mu
\pm
|\boldsymbol{g}^{\ }_{\boldsymbol{k}}|,
\label{eq:NS single particle dispersion}
\end{equation}
while the transformation between the
laboratory basis and the helicity basis is given 
by the unitary $2\times2$ matrix
\begin{equation}
\begin{split}
&
\Pi^{\ }_{\boldsymbol{k}}\equiv
\frac{1}{\sqrt{2}}
\left(
\begin{array}{cc}
1
&
1
\\ 
e^{\text{i}\varphi^{\ }_{\boldsymbol{k}}}
&
-
e^{\text{i}\varphi^{\ }_{\boldsymbol{k}}}
\end{array}
\right)\\
&
\hphantom{\Pi^{\ }_{\boldsymbol{k}}}
:=
\frac{1}{\sqrt{2}}
\left(
\begin{array}{cc}
1
&
1
\\
\frac{
g^{\ }_{\boldsymbol{k}1}
+
\text{i}
g^{\ }_{\boldsymbol{k}2}}
     {
|\boldsymbol{g}^{\ }_{\boldsymbol{k}}|
     }
&
-
\frac{
g^{\ }_{\boldsymbol{k}1}
+
\text{i}
g^{\ }_{\boldsymbol{k}2}
     }
     {
|\boldsymbol{g}^{\ }_{\boldsymbol{k}}|}
\end{array}
\right),
\label{eq: def trsf to helicity basis}
\end{split}
\end{equation}
whereby
\begin{equation}
\begin{split}
&
\psi^{\dag}_{\boldsymbol{k}}=
\tilde{\psi}^{\dag}_{\boldsymbol{k}}
\Pi^{\dag}_{\boldsymbol{k}},
\
\psi^{\ }_{\boldsymbol{k}}=
\Pi^{\ }_{\boldsymbol{k}} 
\tilde{\psi}^{\ }_{\boldsymbol{k}},
\
\tilde{\mathcal{H}}^{\ }_{0;\boldsymbol{k}}=
\Pi^{\dag}_{\boldsymbol{k}}
\mathcal{H}^{\ }_{0;\boldsymbol{k}}
\Pi^{\ }_{\boldsymbol{k}}.
\label{eq:Spinor Trafo}
\end{split}
\end{equation}
\end{subequations}

The (mean-field) BdG Hamiltonian is defined by
\begin{subequations}
\begin{equation}
\begin{split}
H:=&\,
\sum_{\boldsymbol{k}\in\text{BZ}}
\Psi^{\dag}_{\boldsymbol{k}}
\left(
\begin{array}{cc}
\mathcal{H}^{\ }_{0;\boldsymbol{k}}
& 
\Delta^{\ }_{\boldsymbol{k}}
\\
\Delta^{\dag}_{\boldsymbol{k}}
&
-
\mathcal{H}^{T}_{0;-\boldsymbol{k}}
\end{array}
\right)
\Psi^{\ }_{\boldsymbol{k}}
\\
=&\,
\sum_{\boldsymbol{k}\in\text{BZ}}
\Phi^{\dag}_{\boldsymbol{k}}
\left(
\begin{array}{cc}
\mathcal{H}^{\ }_{0;\boldsymbol{k}}
& 
\Delta^{\ }_{\boldsymbol{k}}
\left(
-\text{i}\sigma^{\ }_{2}\right)
\\
\text{i}\sigma^{\ }_{2}
\Delta^{\dag}_{\boldsymbol{k}}
&
-\sigma^{\ }_{2}
\mathcal{H}^{T}_{0;-\boldsymbol{k}}
\sigma^{\ }_{2}	
\end{array}
\right)
\Phi^{\ }_{\boldsymbol{k}}
\\
=&\,
\sum_{\boldsymbol{k}\in\mathrm{BZ}}
\tilde{\Phi}^{\dag}_{\boldsymbol{k}}
\begin{pmatrix}
\tilde{\mathcal{H}}^{\ }_{0;\boldsymbol{k}}
& 
\tilde{\Delta}^{\ }_{\boldsymbol{k}}
\\
\tilde{\Delta}^{\dag}_{\boldsymbol{k}}
&
-
\tilde{\mathcal{H}}^{T}_{0;-\boldsymbol{k}}
\end{pmatrix}
\tilde{\Phi}^{\ }_{\boldsymbol{k}}
\end{split}
\label{eq:BdG Hamiltonian}
\end{equation}
where the bispinors 
$\Psi^{\dag}_{\boldsymbol{k}}$, 
$\Phi^{\dag}_{\boldsymbol{k}}$,
and 
$\tilde{\Phi}^{\dag}_{\boldsymbol{k}}$ 
are given by
\begin{equation}
\begin{split}
&
\Psi^{\dag}_{\boldsymbol{k}}=
\left(
\psi^{\dag}_{\boldsymbol{k}},
\psi^{\   }_{-\boldsymbol{k}}
\right)=
\left(
 c^{\dag}_{ \boldsymbol{k}\uparrow}, 
 c^{\dag}_{ \boldsymbol{k}\downarrow},
 c^{\   }_{-\boldsymbol{k}\uparrow},
 c^{\   }_{-\boldsymbol{k}\downarrow}
\right),
\\
&
\Phi^{\dag}_{\boldsymbol{k}}=
\left(
\psi^{\dag}_{\boldsymbol{k}},
\text{i}\sigma^{\ }_{2}
\psi^{\   }_{-\boldsymbol{k}}
\right)=
\left(
 c^{\dag}_{ \boldsymbol{k}\uparrow}, 
 c^{\dag}_{ \boldsymbol{k}\downarrow},
 c^{\   }_{-\boldsymbol{k}\downarrow},
-c^{\   }_{-\boldsymbol{k}\uparrow}
\right),
\\
&
\tilde{\Phi}^{\dag}_{\boldsymbol{k}}=
\left(
 a^{\dag}_{\boldsymbol{k}+}, 
 a^{\dag}_{\boldsymbol{k}-},
 e^{\text{i}\varphi^{\ }_{-\boldsymbol{k}}}
 a^{\ }_{-\boldsymbol{k}+},
-
 e^{\text{i}\varphi^{\ }_{-\boldsymbol{k}}}
 a^{\ }_{-\boldsymbol{k}-}
\right),
\end{split}
\label{eq: def bispinors}
\end{equation}
\end{subequations}
respectively.
We have chosen to construct the bispinors 
$\Phi^{\dag}_{\boldsymbol{k}}$ 
and 
$\tilde{\Phi}^{\dag}_{\boldsymbol{k}}$ 
from the spinors 
$\psi^{\dag}_{\boldsymbol{k}}$ 
and
$\tilde{\psi}^{\dag}_{\boldsymbol{k}}$,
respectively, and their time-reversed partners. 
Thereby, we have to take care of the action 
of the time-reversal operation $\mathcal{T}$ 
on the laboratory and the helicity single-particle states 
labeled by the wave vector $\boldsymbol{k}$ and the
indices $s=\uparrow,\downarrow$ and
$\lambda=\pm1$, respectively. 
For the laboratory basis, it is
\begin{equation}
\mathcal{T}
 \left|\right. \boldsymbol{k}\uparrow  \left.\right\rangle=
+\left|\right.-\boldsymbol{k}\downarrow\left.\right\rangle,
\quad
\mathcal{T}
 \left|\right. \boldsymbol{k}\downarrow  \left.\right\rangle=
-\left|\right.-\boldsymbol{k}\uparrow    \left.\right\rangle,
\label{eq: def TR on laboratory states}
\end{equation}
i.e., it is off-diagonal in the laboratory spin basis.
For the helicity basis,
\begin{subequations}
\begin{equation}
\begin{split}
&
\left|\right. \boldsymbol{k}+\left.\right\rangle=
\frac{1}{\sqrt{2}}
\left(
\left|\right. \boldsymbol{k}\uparrow  \left.\right\rangle
+
e^{+\text{i}\varphi^{\ }_{\boldsymbol{k}}}
\left|\right. \boldsymbol{k}\downarrow\left.\right\rangle
\right),
\\
&
\left|\right. \boldsymbol{k}-\left.\right\rangle=
\frac{1}{\sqrt{2}}
\left(
\left|\right. \boldsymbol{k}\uparrow  \left.\right\rangle
-
e^{+\text{i}\varphi^{\ }_{\boldsymbol{k}}}
\left|\right. \boldsymbol{k}\downarrow\left.\right\rangle
\right),
\end{split}
\end{equation}
together with 
\begin{equation}
e^{\text{i}\varphi^{\ }_{\boldsymbol{k}}}=
-
e^{\text{i}\varphi^{\ }_{-\boldsymbol{k}}}
\end{equation}
imply that it is
\begin{equation}
\begin{split}
&
\mathcal{T}
\left|\right.\boldsymbol{k}\lambda\left.\right\rangle=
\lambda
e^{-\text{i}\varphi^{\ }_{-\boldsymbol{k}}}
\left|\right.-\boldsymbol{k}\lambda\left.\right\rangle,
\end{split}
\label{eq: def TR on helicity states}
\end{equation}
\end{subequations}
i.e., it is diagonal in the helicity internal space
but with the wave vector and helicity-dependent 
eigenvalue
$\lambda e^{\text{i}\varphi^{\ }_{\boldsymbol{k}}}$
that is odd under the inversion 
$\boldsymbol{k}\to-\boldsymbol{k}$.
Hence, the bispinors 
$\Phi^{\dag}_{\boldsymbol{k}}$ 
and 
$\tilde{\Phi}^{\dag}_{\boldsymbol{k}}$ 
follow.

We parameterize the $2\times2$ pair-potential matrix by
\begin{subequations}
\begin{equation}
\Delta^{\ }_{\boldsymbol{k}}=
\left(
\Delta^{\ }_{\text{s},\boldsymbol{k}}
\sigma^{\ }_{0}
+
\boldsymbol{d}^{\ }_{\boldsymbol{k}}
\cdot
\boldsymbol{\sigma}
\right)
(\text{i}\sigma^{\ }_{2})
\end{equation}
in the laboratory frame for the spin degrees of freedom.
PHS, which embodies Fermi statistics within the BdG formulation, 
demands that it is an antisymmetric operator, i.e.,
\begin{equation}
\Delta^{\ }_{\text{s},\boldsymbol{k}}=
\Delta^{\ }_{\text{s},-\boldsymbol{k}},
\qquad
\boldsymbol{d}^{\ }_{\boldsymbol{k}}=
-
\boldsymbol{d}^{\ }_{-\boldsymbol{k}}.
\end{equation}
TRS imposes the conditions
\begin{equation}
\Delta^{\ }_{\text{s},\boldsymbol{k}}=
\Delta^{* }_{\text{s},-\boldsymbol{k}},
\qquad
\boldsymbol{d}^{\ }_{\boldsymbol{k}}=
-
\boldsymbol{d}^{* }_{-\boldsymbol{k}}.
\end{equation}
\end{subequations}

Throughout this paper, we consider Cooper pairs
made of time-reversed helicity single-particle
states from Eq.~(\ref{eq: def TR on helicity states}). Hence,
we take the $2\times2$ pair-potential matrix
\begin{subequations}
\begin{equation}
\tilde{\Delta}^{\ }_{\boldsymbol{k}}=
\begin{pmatrix}
\tilde{\Delta}^{\ }_{\boldsymbol{k}+}
&
0
\\
0
&
\tilde{\Delta}^{\ }_{\boldsymbol{k}-}
\end{pmatrix}
\label{Deltatildediagonal}
\end{equation}
to be diagonal in the helicity basis, and
it then follows that 
\begin{equation}
\tilde{\Delta}^{\ }_{\boldsymbol{k}+}=
\tilde{\Delta}^{* }_{-\boldsymbol{k}+},
\qquad
\tilde{\Delta}^{\ }_{\boldsymbol{k}-}=
\tilde{\Delta}^{* }_{-\boldsymbol{k}-},
\end{equation}
\end{subequations}
as a consequence of TRS.
Furthermore, we find with the help of
Eq.~\eqref{eq: def bispinors}
the $4\times4$ Hermitian matrix
(the complex notation 
$z=x+\text{i}y$
and
$\bar{z}=x-\text{i}y$
is occasionally used)
\begin{subequations}
\begin{equation}
\mathcal{H}^{\ }_{\boldsymbol{k}}=
\begin{pmatrix}
\varepsilon^{\ }_{\boldsymbol{k}}
-
\mu
&
\bar{g}^{\ }_{\boldsymbol{k}}
&
\Delta^{\ }_{\mathrm{s},\boldsymbol{k}}
&
\Delta^{\ }_{\mathrm{t},\boldsymbol{k}}
e^{-\text{i}\varphi^{\ }_{\boldsymbol{k}}}
\\
g^{\ }_{\boldsymbol{k}}
&
\varepsilon^{\ }_{\boldsymbol{k}}
-
\mu
&
\Delta^{\ }_{\mathrm{t},\boldsymbol{k}}
e^{+\text{i}\varphi^{\ }_{\boldsymbol{k}}}
&
\Delta^{\ }_{\mathrm{s},\boldsymbol{k}}
\\
\text{c.c.}
&
\text{c.c.}
&
-
\varepsilon^{\ }_{\boldsymbol{k}}
+
\mu
&
-
\bar{g}^{\ }_{\boldsymbol{k}}
\\
\text{c.c.}
&
\text{c.c.}
&
-
g^{\ }_{\boldsymbol{k}}
&
-
\varepsilon^{\ }_{\boldsymbol{k}}
+
\mu
\end{pmatrix},
\label{eq: BdG in terms tilde tilde pairings}
\end{equation}
where we recall that 
$\varphi^{\ }_{\boldsymbol{k}}:=\arg\, g^{\ }_{\boldsymbol{k}}$
and
\begin{equation}
\begin{split}
&
\Delta^{\ }_{\mathrm{s},\boldsymbol{k}}=
\frac{1}{2}
\left(
\tilde{\Delta}^{\ }_{\boldsymbol{k}+}
+
\tilde{\Delta}^{\ }_{\boldsymbol{k}-}
\right)=
\Delta^{* }_{\mathrm{s},-\boldsymbol{k}},
\\
&
\Delta^{\ }_{\mathrm{t},\boldsymbol{k}}=
\frac{1}{2}
\left(
\tilde{\Delta}^{\ }_{\boldsymbol{k}+}
-
\tilde{\Delta}^{\ }_{\boldsymbol{k}-}
\right)=
\Delta^{* }_{\mathrm{t},-\boldsymbol{k}},
\\
&
\boldsymbol{d}^{\ }_{\boldsymbol{k}}=
\frac{1}{2}
\left(
\tilde{\Delta}^{\ }_{\boldsymbol{k}+}
-
\tilde{\Delta}^{\ }_{\boldsymbol{k}-}
\right)
\frac{
\boldsymbol{g}^{\ }_{\boldsymbol{k}}
     }
     {
|\boldsymbol{g}_{\boldsymbol{k}}|
     }=
-\boldsymbol{d}^{* }_{-\boldsymbol{k}},
\label{tilde tilde to spin trafo}
\end{split}
\end{equation}
in the $\Phi^{\ }_{\boldsymbol{k}}$ 
representation of Eq.~(\ref{eq:BdG Hamiltonian}).

The fact that the vector $\boldsymbol{d}^{\ }_{\boldsymbol{k}}$ 
is parallel to 
$\boldsymbol{g}^{\ }_{\boldsymbol{k}}$ 
is a consequence of our assumption that Cooper pairs
are made of time-reversed helicity single-particle states,
i.e., Eq.~(\ref{Deltatildediagonal}). This assumption
is justified if the pairing interaction preserves the
symmetry of the noninteracting Hamiltonian. 
Following the literature on noncentrosymmetric superconductors,%
~\cite{Sigrist07} we are thus assuming that the
symmetry of the noninteracting Hamiltonian is preserved by the 
self-consistent inclusion of the pairing interaction.

We also demand that Hamiltonian%
~(\ref{eq: BdG in terms tilde tilde pairings})
is single valued in the BZ. This restricts the triplet
pairing $\Delta^{\ }_{\text{t},\boldsymbol{k}}$
to vanish at least as fast as 
$|\boldsymbol{g}^{\ }_{\boldsymbol{k}}|$,
\begin{equation}
\lim_{|\boldsymbol{g}^{\ }_{\boldsymbol{k}}|\to0}
\frac{
|\Delta^{\ }_{\text{t},\boldsymbol{k}}|
     }
     {
|\boldsymbol{g}^{\ }_{\boldsymbol{k}}|
     }
<
c
\end{equation}
for some number $c$ larger than or equal to 0.
\end{subequations}
With our choice of gauge,
$\tilde{\Delta}^{\ }_{\boldsymbol{k}+}$
and
$\tilde{\Delta}^{\ }_{\boldsymbol{k}-}$
or, equivalently,
$\Delta^{\ }_{\text{s},\boldsymbol{k}}$
and
$\Delta^{\ }_{\text{t},\boldsymbol{k}}$
are real valued.  
In Sec.~\ref{sec: Majorana fermions}, 
where we study TRS-breaking vortices, 
we revert instead to complex order parameters to
accommodate twists in the phases of the singlet and triplet
pair potentials.
Finally, we observe
that the pair-potential eigenvalues
\begin{equation}
\tilde{\Delta}^{\ }_{\boldsymbol{k}\lambda}=
\Delta^{\ }_{\text{s},\boldsymbol{k}}
+
\lambda
\boldsymbol{d}^{\ }_{\boldsymbol{k}}
\cdot
\frac{\boldsymbol{g}^{\ }_{\boldsymbol{k}}}
{\left|\boldsymbol{g}_{\boldsymbol{k}}\right|},
\qquad
\lambda=\pm,
\label{eq: double tilde representation pairings}
\end{equation}
transform according to the same
irreducible representation of the space group.
For example, in the isotropic continuum limit with
$s$-wave pairing
they are functions of $|\boldsymbol{k}|$ only.
    
The BdG Hamiltonian~(\ref{eq:BdG Hamiltonian}) is of the form
\begin{equation}
\begin{split}
&
H\equiv
\sum_{\lambda=\pm}
H^{\ }_{\lambda}:=
\sum_{\lambda=\pm}
\sum_{\boldsymbol{k}\in\text{BZ}}
H^{\ }_{\boldsymbol{k}\lambda},
\\
&
H^{\ }_{\boldsymbol{k}\lambda}=
\xi^{\ }_{\boldsymbol{k}\lambda}
a^{\dag}_{\boldsymbol{k}\lambda}
a^{\   }_{\boldsymbol{k}\lambda}
+
\lambda
\tilde{\Delta}^{\ }_{\boldsymbol{k}\lambda}
\left(
e^{
\text{i}\varphi^{\ }_{-\boldsymbol{k}}
  }
a^{\   }_{-\boldsymbol{k}\lambda}
a^{\   }_{ \boldsymbol{k}\lambda}
+
\mathrm{H.c.}
\right).
\end{split}
\label{eq: MF H}
\end{equation}
The mean-field ground state is the state
$|\Upsilon^{\ }_{\mathrm{mf}}\rangle$
annihilated by 
$H$. 
It is obtained as the direct product
$|\Upsilon^{\ }_{\mathrm{mf}}\rangle=
|\Upsilon^{+}_{\mathrm{mf}}\rangle
\otimes
|\Upsilon^{-}_{\mathrm{mf}}\rangle$, 
where
$|\Upsilon^{\lambda}_{\mathrm{mf}}\rangle$
is annihilated by $H^{\ }_{\lambda}$ for each of the helicities
$\lambda=\pm$.

Each 
helicity supports quasiparticles obeying the 
PHS (relative to the chemical potential)
dispersion
$\pm E^{\ }_{\boldsymbol{k}\lambda}$
with 
\begin{equation}
E^{\ }_{\boldsymbol{k}\lambda}=
\sqrt{
\xi^{2}_{\boldsymbol{k}\lambda}
+
\tilde{\Delta}^{2 }_{\boldsymbol{k}\lambda}
     },
\qquad
\lambda=\pm.
\label{eq:Sc single particle dispersion}
\end{equation}
However, the ground states
$|\Upsilon^{+}_{\mathrm{mf}}\rangle$
and
$|\Upsilon^{-}_{\mathrm{mf}}\rangle$
are not independent since they are tied to
each other by TRS. Indeed, TRS implies that the relative phase of the
pairing potentials $\tilde{\Delta}^{\ }_{\boldsymbol{k}\lambda}$
with the helicities $\lambda=\pm$
is locked to be 0 or $\pi$, as follows from
the transformation law%
~(\ref{eq: def TR on helicity states}), i.e.,
\begin{equation}
\mathcal{T}
a^{\dag}_{\boldsymbol{k}\lambda}
\mathcal{T}^{-1}=
\lambda
e^{
-\text{i}\varphi^{\ }_{\boldsymbol{k}}
  }
a^{\dag}_{-\boldsymbol{k}\lambda},
\quad
\mathcal{T}
a^{\   }_{\boldsymbol{k}\lambda}
\mathcal{T}^{-1}=
\lambda
e^{
+\text{i}\varphi^{\ }_{-\boldsymbol{k}}
  }
a^{\   }_{-\boldsymbol{k}\lambda}.
\end{equation}

To construct $|\Upsilon^{\ }_{\mathrm{mf}}\rangle$, we perform
a Bogoliubov transformation 
for each helicity index $\lambda=\pm$ independently.
Thus, for each helicity $\lambda=\pm$, we define
\begin{subequations}
\begin{equation}
\gamma^{\ }_{\boldsymbol{k}\lambda}:= 
U^{\ }_{\boldsymbol{k}\lambda}\,
a^{\ }_{\boldsymbol{k}\lambda}
-
V^{\ }_{\boldsymbol{k}\lambda}\,
a^{\dagger}_{\boldsymbol{-k}\lambda}
\end{equation}
with the complex-valued coefficients
$U^{\ }_{\boldsymbol{k}\lambda}$
and
$V^{\ }_{\boldsymbol{k}\lambda}$,
\begin{equation}
\begin{split}
&
|U^{\ }_{\boldsymbol{k}\lambda}|^2 := 
\frac{1}{2}
\left(
1
+
\frac{
\xi^{\ }_{\boldsymbol{k}\lambda}
     }
     {
E^{\ }_{\boldsymbol{k}\lambda}
     }
\right),
\\
&
|V^{\ }_{\boldsymbol{k}\lambda}|^2:= 
\frac{1}{2}
\left(
1
-
\frac{
\xi^{\ }_{\boldsymbol{k}\lambda}
     }
     {
E^{\ }_{\boldsymbol{k}\lambda}
     }
\right),
\\
&
\frac{
U^{\ }_{\boldsymbol{k}\lambda}
     }
     {
V^{\ }_{\boldsymbol{k}\lambda}
     }=
-
\frac{
\lambda 
e^{-\text{i}\varphi^{\ }_{-\boldsymbol{k}}}
\tilde{\Delta}^{\ }_{\boldsymbol{k}\lambda}
     }
     {
E^{\ }_{\boldsymbol{k}\lambda}
-
\xi^{\ }_{\boldsymbol{k}\lambda}
     }.
\end{split}
\end{equation}
\end{subequations}
Under this transformation
\begin{equation}
H=
\sum_{\boldsymbol{k}\in\text{BZ}}
\sum_{\lambda=\pm}
E^{\ }_{\boldsymbol{k}\lambda}\,
\gamma^{\dag}_{\boldsymbol{k}\lambda}
\gamma^{\   }_{\boldsymbol{k}\lambda}.
\end{equation}
The mean-field ground state is then
\begin{subequations}
\label{eq: MF GS done}
\begin{equation}
|\Upsilon^{\ }_{\mathrm{mf}}\rangle=
\prod_{\lambda=\pm}
\prod_{\boldsymbol{k}}
\left(
U^{\ }_{\boldsymbol{k}\lambda}
+
V^{\ }_{\boldsymbol{k}\lambda}
a^{\dag}_{ \boldsymbol{k}\lambda}
a^{\dag}_{-\boldsymbol{k}\lambda}
\right)
|0\rangle 
\end{equation}
provided
\begin{equation}
\gamma^{\ }_{\boldsymbol{k}\lambda}
|\Upsilon^{\ }_{\mathrm{mf}}\rangle=
0
\end{equation} 
\end{subequations}
holds for all $\boldsymbol{k}$ and all $\lambda=\pm$.

By construction, the mean-field ground state~(\ref{eq: MF GS done})
is TRS. SRS is, however, broken. Consequently,
\begin{equation}
\begin{split}
\langle\Upsilon^{\ }_{\mathrm{mf}}|
c^{\ }_{-\boldsymbol{k}\uparrow}
c^{\ }_{ \boldsymbol{k}\uparrow}
|\Upsilon^{\ }_{\mathrm{mf}}\rangle
=&\,
\frac{
e^{-\text{i}\varphi^{\ }_{-\boldsymbol{k}}}
     }
     {
4
     }
\left( 
\frac{
\tilde{\Delta}^{\ }_{\boldsymbol{k}+}
     }
     {
E^{\ }_{\boldsymbol{k}+}
     }
-
\frac{
\tilde{\Delta}^{\ }_{\boldsymbol{k}-}
     }
     {
E^{\ }_{\boldsymbol{k}-}
     } 
\right),
\\
\langle\Upsilon^{\ }_{\mathrm{mf}}|
c^{\ }_{-\boldsymbol{k}\downarrow}
c^{\ }_{ \boldsymbol{k}\downarrow}
|\Upsilon^{\ }_{\mathrm{mf}}\rangle
=&\,
\frac{
e^{\text{i}\varphi^{\ }_{\boldsymbol{k}}}
     }
     {
4
     }
\left( 
\frac{
\tilde{\Delta}^{\ }_{\boldsymbol{k}+}
     }
     {
E^{\ }_{\boldsymbol{k}+}
     }
-
\frac{
\tilde{\Delta}^{\ }_{\boldsymbol{k}-}
     }
     {
E^{\ }_{\boldsymbol{k}-}
     } 
\right),
\\
\langle\Upsilon^{\ }_{\mathrm{mf}}|
c^{\ }_{-\boldsymbol{k}\uparrow}
c^{\ }_{ \boldsymbol{k}\downarrow}
|\Upsilon^{\ }_{\mathrm{mf}}\rangle
=&\,
\frac{1}{4}
\left( 
\frac{
\tilde{\Delta}^{\ }_{\boldsymbol{k}+}
     }
     {
E^{\ }_{\boldsymbol{k}+}
     }
+
\frac{
\tilde{\Delta}^{\ }_{\boldsymbol{k}-}
     }
     {
E^{\ }_{\boldsymbol{k}-}
     } 
\right),
\end{split}
\end{equation}
are generically nonvanishing (one exception is the Rashba-Dirac limit 
$\varepsilon^{\ }_{\boldsymbol{k}}=0$ at the Rashba-Dirac point $\mu=0$)
even though the pair potential may be purely singlet when
\begin{equation}
\tilde{\Delta}^{\ }_{\boldsymbol{k}+}=
\tilde{\Delta}^{\ }_{\boldsymbol{k}-}
\end{equation} 
or purely triplet when
\begin{equation}
\tilde{\Delta}^{\ }_{\boldsymbol{k}+}=
-
\tilde{\Delta}^{\ }_{\boldsymbol{k}-}.
\end{equation}
In a superconducting state that preserves SRS,
the ground state has no spin correlations 
other than that of the pair condensate.

\section{
Susceptibility
        }
\label{sec: Susceptibility}

\subsection{
Static and uniform Pauli magnetic susceptibility at $T=0$ in the normal state
           }

We are after the Pauli magnetization per electron 
induced by the Zeeman coupling 
$
\propto
-
B^{\ }_{1}
\sigma^{\ }_{1}
-
B^{\ }_{2}
\sigma^{\ }_{2}
-
B^{\ }_{3}
\sigma^{\ }_{3}
$,
where it is understood that 
$\sigma^{\ }_{3}$ is the third Pauli matrix
and the in-plane components of the magnetic field
are $B^{\ }_{1}$ and $B^{\ }_{2}$ while the out-of-plane 
component  $B^{\ }_{3}$ is taken along the spin quantization axis
in the laboratory frame of reference.

To obtain the Pauli magnetization
per electron at $T=0$, we start from
Eq.~(\ref{eq: def 2x2 Rashba-Dirac H0}) 
with the Zeeman coupling added
\begin{equation}
\mathcal{H}^{\mathrm{sur}}_{\boldsymbol{B}\,\boldsymbol{k}}:=
\left(\hbar v^{\ }_{\mathrm{RD}}
k^{\ }_1
-B^{\ }_{2}
\right)
\sigma^{\ }_{2}
-
\left(
\hbar v^{\ }_{\mathrm{RD}}
k^{\ }_2
+B^{\ }_{1}
\right)
\sigma^{\ }_{1}
-B^{\ }_{3}
\sigma^{\ }_{3},
\label{eq: def 2x2 Rashba-Dirac HB}
\end{equation}
compute the expectation value of the spin operator
$\hbar\sigma^{\ }_{1,2,3}/2$ for all the Bloch states,
and sum these expectation values up to the chemical potential
$\mu$. The Pauli susceptibility
per electron then follows by differentiation
with respect to $B^{\ }_{1,2,3}$ followed
by setting $B^{\ }_{1,2,3}=0$.
We set $\hbar=v^{\ }_{\mathrm{RD}}=1$ to simplify
notation.

As long as $B^{2}_{1}+B^{2}_{2}>0$, 
the eigenvalue
\begin{subequations}
\begin{equation}
\xi^{\ }_{\pm}(\boldsymbol{k})=
-
\mu
\pm
\sqrt{
\left(
k^{\ }_{1}
-
B^{\ }_{2}
\right)^{2}
+
\left(
k^{\ }_{2}
+
B^{\ }_{1}
\right)^{2}
+
B^{2}_{3}
     }
\end{equation}
has the eigenstate
\begin{equation}
\Psi^{\ }_{\pm}(\boldsymbol{k})=
\frac{1}{\mathcal{N}^{\ }_{\pm}(\boldsymbol{k})}
\begin{pmatrix}
-
k^{\ }_{2}
-
\mathrm{i}
k^{\ }_{1}
- 
B^{\ }_{1}
+
\mathrm{i}
B^{\ }_{2}
\\
\xi^{\ }_{\pm}(\boldsymbol{k})
+
\mu
+
B^{\ }_{3}
\end{pmatrix}
\end{equation}
with the normalization
\begin{equation}
\mathcal{N}^{\ }_{\pm}(\boldsymbol{k}):=
\sqrt{
2
\left[
\xi^{\ }_{\pm}(\boldsymbol{k})
+
\mu
\right]
\left[
\xi^{\ }_{\pm}(\boldsymbol{k})
+
\mu
+
B^{\ }_{3}
\right]
     }.
\end{equation}
\end{subequations}
We observe that the effect of in-plane magnetic fields
is to translate the Fermi sea. The spin expectation values
in the Bloch states are
\begin{subequations}
\begin{equation}
\begin{split}
\Psi^{\dag}_{\pm}(\boldsymbol{k})
\sigma^{\ }_{3}
\Psi^{\   }_{\pm}(\boldsymbol{k})
=&\,
\mp
\frac{
B^{\ }_{3}
     }
     {
\sqrt{
B^{2}_{3}
+
\left|\bar{k}+B\right|^2
     }
     } 
\\
=&\,
\mp
\frac{B^{\ }_{3}}{|\boldsymbol{k}|}
[F\left(B/\bar{k},\bar{B}/k\right)+\cdots],
\end{split}
\label{eq: expectation a}
\end{equation}
\begin{equation}
\begin{split}
\left.
\Psi^{\dag}_{\pm}(\boldsymbol{k})
\sigma
\Psi^{\   }_{\pm}(\boldsymbol{k})
\right|^{\ }_{B^{\ }_{3}=0}
=&\,
-
\frac{
2
\left(
\bar{k}
+B
\right)
\left(
\xi^{\ }_{\pm}(\boldsymbol{k})
+
\mu
\right)
     }
     {
|
\bar{k}
+
B
|^{2}
+
\left(
\xi^{\ }_{\pm}(\boldsymbol{k})
+
\mu
\right)^2
     }
\\
=&\,
\mp
\frac{
\bar{k}
+
B
     }
     {
|
\bar{k}
+
B
|
     }
\\
=&\,
\mp
\frac{\bar{k}+B}{|\boldsymbol{k}|}
[F\left(B/\bar{k},\bar{B}/k\right)+\cdots],
\end{split}
\label{eq: expectation b}
\end{equation}
and 
\begin{equation}
\begin{split}
\left.
\Psi^{\dag}_{\pm}(\boldsymbol{k})
\bar{\sigma}
\Psi^{\   }_{\pm}(\boldsymbol{k})
\right|^{\ }_{B^{\ }_{3}=0}
=&\,
-
\frac{
2
\left(
k
+
\bar{B}
\right)
\left(
\xi^{\ }_{\pm}(\boldsymbol{k})
+
\mu
\right)
     }
     {
|
\bar{k}
+
B
|^{2}
+
\left(
\xi^{\ }_{\pm}(\boldsymbol{k})
+
\mu
\right)^2
     }
\\
=&\,
\mp
\frac{
k
+
\bar{B}
     }
     {
|
k
+
\bar{B}
|
     }
\\
=&\,
\mp
\frac{k+\bar{B}}{|\boldsymbol{k}|}
[F\left(\bar{B}/k,B/\bar{k}\right)+\cdots].
\end{split}
\label{eq: expectation c}
\end{equation}
\end{subequations}
Here, we have introduced the complex notations
\begin{subequations}
\begin{equation}
\sigma=
\sigma^{\ }_{1}
+
\mathrm{i}
\sigma^{\ }_{2},
\qquad
\bar\sigma=
\sigma^{\ }_{1}
-
\mathrm{i}
\sigma^{\ }_{2},
\end{equation}
for the Pauli matrices, 
\begin{equation}
k=
k^{\ }_{2}
+
\mathrm{i}
k^{\ }_{1},
\qquad
\bar{k}=
k^{\ }_{2}
-
\mathrm{i}
k^{\ }_{1},
\end{equation}
for the momenta, and
\begin{equation}
B=
B^{\ }_{1}
+
\mathrm{i}
B^{\ }_{2},
\qquad
\bar{B}=
B^{\ }_{1}
-
\mathrm{i}
B^{\ }_{2},
\end{equation}
for the in-plane components of the magnetic field. 
We have also introduced
the real-valued function 
\begin{equation}
F\left(z,\bar{z}\right):=
1
-
\frac{1}{2}
\left(
z
+
\bar{z}
\right)
\end{equation}
that comes about to first order in an expansion in powers of
the components of the magnetic field.
\end{subequations}
The magnetization per electron
is obtained by integrating
over all single-particle energies up to the
chemical potential $\mu$. 
We conclude that the Pauli magnetic susceptibility tensor
per electron is
\begin{equation}
\chi^{\ }_{ab}\propto
\delta^{\ }_{ab}
\times
\begin{cases}
\hphantom{2}
 \pi
\left(
\Lambda
-
|\mu|
\right),& 
\hbox{  if $a=1,2$,}
\\
&
\\
2\pi
\left(
\Lambda
-
|\mu|
\right),
&
\hbox{  if $a=3$,}
\end{cases}
\label{eq: main result Pauli susc}
\end{equation}
in the noninteracting approximation and at $T=0$.

The Pauli magnetic susceptibility per electron%
~(\ref{eq: main result Pauli susc})
also holds for the Rashba tight-binding Hamiltonian%
~(\ref{eq: def 2x2 Rashba on lattice})
with minor modifications provided the limit
$t/\alpha\to0$ is taken,
($N$ is the number of lattice sites)
\begin{equation}
\chi^{\ }_{ab}=
\delta^{\ }_{ab}
\times
\begin{cases}
\frac{1/2}{N}
\sum\limits_{|\mu|<|\boldsymbol{g}^{\ }_{\boldsymbol{k}}|}
\frac{1}{|\boldsymbol{g}^{\ }_{\boldsymbol{k}}|},
& 
\hbox{  if $a=1,2$,}
\\
&
\\
\frac{1}{N}
\sum\limits_{|\mu|<|\boldsymbol{g}^{\ }_{\boldsymbol{k}}|}
\frac{1}{|\boldsymbol{g}^{\ }_{\boldsymbol{k}}|},
&
\hbox{  if $a=3$.}
\end{cases}
\end{equation}
In the opposite limit $\alpha/t\to0$,
we cannot use the lattice counterpart to
Eq.~(\ref{eq: expectation a})
to compute $\chi^{\ }_{33}$,
since our choice for the spinor representation is singular in this limit.
We can however use the lattice counterparts to Eqs.%
~(\ref{eq: expectation b})
and 
(\ref{eq: expectation c})
to compute $\chi^{\ }_{11}$ 
and $\chi^{\ }_{22}$.
By isotropy, we then recover the conventional
Pauli magnetic susceptibility
\begin{equation}
\chi^{\ }_{ab}\propto
\delta^{\ }_{ab}
\nu^{\ }_{\mathrm{F}}(\mu)
\end{equation}
where $\nu^{\ }_{\mathrm{F}}(\mu)$ is the 
density of states per electron and per spin
of the dispersion $\varepsilon^{\ }_{\boldsymbol{k}}$.
This result remains true to first order in $\alpha/t$.

\subsection{
Dynamical Pauli susceptibility in the superconducting state
           }

Another remarkable consequence of the spin-orbit coupling
is that charge-density and spin-density fluctuations are coupled, 
both in the normal and in the superconducting state.%
~\cite{Dyakonov71,Edelstein89,Hirsch99,Raghu09}
The spin-Hall effect is a consequence of this coupling.%
~\cite{Dyakonov71,Hirsch99}
To quantify this statement, we introduce the susceptibility tensor 
in the superconducting state
\begin{subequations}
\begin{equation}
\begin{split}
\left(\hat{\chi}^{\ }_{00}\right)^{\ }_{q}
&=-\frac{1}{\beta N}\sum_{k}\text{tr}\left[G^{\ }_{0;k}X^{\ }_{03}G^{\ }_{0;k+q}X^{\ }_{03}\right],\\
\left(\hat{\chi}^{\ }_{0d}\right)^{\ }_{q}
&=-\frac{1}{\beta N}\sum_{k}\text{tr}\left[G^{\ }_{0;k}X^{\ }_{03}G^{\ }_{0;k+q}X^{\ }_{d0}\right],\\
\left(\hat{\chi}^{\ }_{b0}\right)^{\ }_{q}
&=-\frac{1}{\beta N}\sum_{k}\text{tr}\left[G^{\ }_{0;k}X^{\ }_{b0}G^{\ }_{0;k+q}X^{\ }_{03}\right],\\
\left(\hat{\chi}^{\ }_{bd}\right)^{\ }_{q}
&=-\frac{1}{\beta N}\sum_{k}\text{tr}\left[G^{\ }_{0;k}X^{\ }_{b0}G^{\ }_{0;k+q}X^{\ }_{d0}\right],
\end{split}
\label{eq: def Fourier components result for chimunu}
\end{equation}
where the indices $b$ and $d$ run over the values 1,2, and 3, 
and 
\begin{equation}
X^{\ }_{\mu\nu}:=
\sigma^{\ }_{\mu}\otimes\tau^{\ }_{\nu},
\qquad
\mu,\nu=0,1,2,3,
\label{eq: def of Xmunu}
\end{equation}
with the unit $2\times2$ matrix $\tau^{\ }_{0}$ and
the Pauli matrices $\boldsymbol{\tau}$ acting on the 
particle-hole two-dimensional subspace.
According to the mean-field Hamiltonian in the superconducting state~\eqref{eq:BdG Hamiltonian}, 
the single-particle Green's function is 
\begin{equation}
\begin{split}
G^{\ }_{0;k}:=&
\left[
	-\mathrm{i}\omega^{\ }_{n}X^{\ }_{00}
	+\left(\varepsilon^{\ }_{\bs{k}}-\mu\right)X^{\ }_{03}
	+g^{\ }_{\bs{k}1}X^{\ }_{13}+g^{\ }_{\bs{k}2}X^{\ }_{23}\right.\\
	&
	\left.
	+\Delta^{\ }_{s,\bs{k}}X^{\ }_{01}
	+\Delta^{\ }_{t,\bs{k}}\left(\hat{g}_{\bs{k}1}X^{\ }_{11}+\hat{g}_{\bs{k}2}X^{\ }_{21}\right)
\right]^{-1}
.
\end{split}
\label{eq: def G0k}
\end{equation}
\end{subequations}
Our notation applies to a lattice made of $N$ sites,
periodic boundary conditions are assumed,
$\beta$ is the inverse temperature 
(the Boltzmann constant is set to unity), 
finally
$q=(\text{i}\varpi^{\ }_{l},\boldsymbol{q})$
and 
$k=(\text{i}\omega^{\ }_{n},\boldsymbol{k})$
are three vectors with bosonic and fermionic Matsubara
frequencies, respectively, while $\boldsymbol{q}$
and $\boldsymbol{k}$ belong to the first BZ.
It is straightforward to modify this notation for the
case of the continuum limit.

After performing the summation over the fermionic Matsubara
frequencies in Eq.~(\ref{eq: def Fourier components result for chimunu})
and with some additional lengthy algebra,
the dynamical Pauli susceptibility tensor simplifies to
\begin{subequations}
\begin{equation}
\begin{split}
\left(\hat{\chi}^{\ }_{\mu\nu}\right)^{\ }_{q}=
\frac{1}{4N}
&\sum_{\boldsymbol{k}}
\sum_{\lambda,\lambda',\gamma,\gamma'}
\left(
\Gamma^{\lambda,\lambda'}_{\mu\nu}
\right)^{\ }_{\boldsymbol{k},\boldsymbol{q}}
\left(
C^{\lambda,\lambda',\gamma,\gamma'}_{\mu\nu}
\right)^{\ }_{\boldsymbol{k},\boldsymbol{q}}\\
&\times
\frac{
f^{\ }_{\text{FD}}
(\gamma E^{\ }_{\boldsymbol{k}\lambda})
-
f^{\ }_{\text{FD}}
(\gamma'E^{\ }_{\boldsymbol{k}+\boldsymbol{q}\lambda'})
     }
     {
\gamma E^{\ }_{\boldsymbol{k}\lambda}
-
\gamma'E^{\ }_{\boldsymbol{k}+\boldsymbol{q}\lambda'}
+
\text{i}\varpi^{\ }_{l}
     },
\end{split}     
\label{dynsuscept}
\end{equation}
where $\gamma,\gamma'=\pm$ and $\mu,\nu=0,1,2,3$, 
the single-particle dispersion in the
superconducting state $E^{\ }_{\boldsymbol{k}\lambda}$ is
defined in Eq.~\eqref{eq:Sc single particle dispersion}
for the helicities $\lambda=\pm$, while
\begin{equation}
f^{\ }_{\text{FD}}(z)=
\frac{1}{e^{\beta z}+1}
\end{equation}
is the Fermi-Dirac function.
The vertex is given by
($\boldsymbol{\hat{g}}^{\ }_{\boldsymbol{k}}\equiv
  \boldsymbol{g}^{\ }_{\boldsymbol{k}}/|\boldsymbol{g}^{\ }_{\boldsymbol{k}}|$)
\begin{equation}
\begin{split}
\left(
\Gamma^{\lambda,\lambda'}_{00}
\right)^{\ }_{\boldsymbol{k},\boldsymbol{q}}=&\,
1
+
\lambda\lambda'
\boldsymbol{\hat{g}}^{\ }_{\boldsymbol{k}}
\cdot
\boldsymbol{\hat{g}}^{\ }_{\boldsymbol{k}+\boldsymbol{q}},
\\
\left(
\Gamma^{\lambda,\lambda'}_{b0}
\right)^{\ }_{\boldsymbol{k},\boldsymbol{q}}=&\,
\lambda'
\hat{g}^{\ }_{b;\boldsymbol{k}+\boldsymbol{q}}
+
\lambda
\hat{g}^{\ }_{b;\boldsymbol{k}}
+
\lambda\lambda'
\text{i}
\epsilon_{abc}
\hat{g}^{\ }_{a;\boldsymbol{k}}
\hat{g}^{\ }_{c;\boldsymbol{k}+\boldsymbol{q}},
\\
\left(
\Gamma^{\lambda,\lambda'}_{0d}
\right)^{\ }_{\boldsymbol{k},\boldsymbol{q}}=&\,
\lambda'
\hat{g}^{\ }_{d;\boldsymbol{k}+\boldsymbol{q}}
+
\lambda
\hat{g}^{\ }_{d;\boldsymbol{k}}
-
\lambda\lambda'
\text{i}
\epsilon_{adc}
\hat{g}^{\ }_{a;\boldsymbol{k}}
\hat{g}^{\ }_{c;\boldsymbol{k}+\boldsymbol{q}},
\\
\left(
\Gamma^{\lambda,\lambda'}_{bd}
\right)^{\ }_{\boldsymbol{k},\boldsymbol{q}}=&\,
\delta^{\ }_{bd}
+
\lambda\lambda'
\hat{g}^{\ }_{a;\boldsymbol{k}}
f^{ac}_{bd}
\hat{g}^{\ }_{c;\boldsymbol{k}+\boldsymbol{q}}
\\
&\,
+
\text{i}\epsilon^{\ }_{abd}
\left(
\lambda
\hat{g}^{\ }_{a;\boldsymbol{k}}
-
\lambda'
\hat{g}^{\ }_{a;\boldsymbol{k}+\boldsymbol{q}}
\right),
\end{split}
\end{equation}
where the tensor $f^{ac}_{bd}$ is defined by
\begin{equation}
\begin{split}
f^{ac}_{bd}:=&\,
\delta^{\ }_{ab}
\delta^{\ }_{cd}
-
\delta^{\ }_{ac}
\delta^{\ }_{bd}
+
\delta^{\ }_{ad}
\delta^{\ }_{bc}
=
f^{ac}_{db}.
\end{split}
\label{eq: def tensor f ...}
\end{equation}
Finally, the coherence factors are given by
\begin{equation}
\begin{split}
\left(
C^{\lambda,\lambda',\gamma,\gamma'}_{00}
\right)^{\ }_{\boldsymbol{k},\boldsymbol{q}}
&=
1+\gamma\gamma'\frac{\xi_{\bs{k},\lambda}\xi_{\bs{k}+\bs{q},\lambda'}-
\tilde{\Delta}_{\bs{k},\lambda}\tilde{\Delta}_{\bs{k}+\bs{q},\lambda'}}
{E_{\bs{k}+\bs{q},\lambda'}E_{\bs{k},\lambda}},\\
\left(
C^{\lambda,\lambda',\gamma,\gamma'}_{b0}
\right)^{\ }_{\boldsymbol{k},\boldsymbol{q}}
&=
\left(
C^{\lambda,\lambda',\gamma,\gamma'}_{0d}
\right)^{\ }_{\boldsymbol{k},\boldsymbol{q}}
=
\gamma\frac{\xi_{\bs{k},\lambda}}{E_{\bs{k},\lambda}}+\gamma'\frac{\xi_{\bs{k}+\bs{q},\lambda'}}
{E_{\bs{k}+\bs{q},\lambda'}},\\
\left(
C^{\lambda,\lambda',\gamma,\gamma'}_{bd}
\right)^{\ }_{\boldsymbol{k},\boldsymbol{q}}
&=
1+\gamma\gamma'\frac{\xi_{\bs{k},\lambda}\xi_{\bs{k}+\bs{q},\lambda'}+
\tilde{\Delta}_{\bs{k},\lambda}\tilde{\Delta}_{\bs{k}+\bs{q},\lambda'}}
{E_{\bs{k}+\bs{q},\lambda'}E_{\bs{k},\lambda}},
\label{eq:}
\end{split}
\end{equation}
\end{subequations}
where the single-particle dispersion in the normal state 
$\xi^{\ }_{\bs{k},\lambda}$ and the superconducting pair potentials 
$\tilde{\Delta}^{\ }_{\bs{k},\lambda}$ are defined
in Eq.~\eqref{eq:NS single particle dispersion} and in 
Eq.~\eqref{Deltatildediagonal}, respectively.
The dynamical Pauli susceptibility of the normal state is obtained 
by taking the limit 
$\tilde{\Delta}_{\bs{k},\lambda}\rightarrow0$, 
$\lambda=\pm$, 
supplemented by the replacements $\gamma E^{\ }_{\bs{k}\lambda}\rightarrow E^{\ }_{\bs{k}\lambda}$ and
$\gamma' E^{\ }_{\bs{k+q}\lambda'}\rightarrow E^{\ }_{\bs{k+q}\lambda'}$
so as to remove the particle-hole symmetry.

We then recover the result from Ref.~\onlinecite{Takimoto09}.
If we furthermore set $\varpi^{\ }_{l}=0$ in Eq.~\eqref{dynsuscept}, 
we obtain the static susceptibility of the normal state.
At the Rashba-Dirac point, 
i.e., for $\xi^{\ }_{\boldsymbol{k}\lambda}=
\lambda |\boldsymbol{g}^{\ }_{\boldsymbol{k}}|$, 
the following components of the static susceptibility vanish:
$\hat{\chi}^{\ }_{01}$, 
$\hat{\chi}^{\ }_{02}$, 
$\hat{\chi}^{\ }_{10}$, 
$\hat{\chi}^{\ }_{20}$, 
$\hat{\chi}^{\ }_{13}$,
$\hat{\chi}^{\ }_{23}$, 
$\hat{\chi}^{\ }_{31}$, 
and
$\hat{\chi}^{\ }_{32}$.

\begin{figure}
\includegraphics[angle=0,scale=0.3]{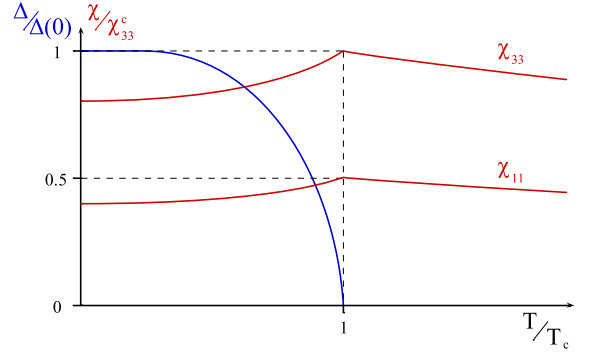}
\caption{
(Color online)
Temperature dependence of the selfconsistent superconducting gap 
(red, normalized by the value at zero temperature) and
the in-plane and out-of-plane Pauli magnetic susceptibility 
(red, normalized by the maximum of the out-of-plane susceptibility)
for 
$\mu=5\omega^{\ }_{\text{D}}$ and $|V|\nu(\omega^{\ }_{\text{D}})=1$. 
Here, $\omega^{\ }_{\text{D}}$ is the Debye
cutoff used for the gap equations.
         }
\label{fig:static susceptibilitiy}
\end{figure}

We now consider the isotropic continuum limit%
~(\ref{eq: def 2x2 Rashba-Dirac H0}) (the Rashba-Dirac limit)
together with an attractive contact density-density interaction 
$-|V|\delta(\boldsymbol{r}-\boldsymbol{r}')$, 
which induces purely singlet pairing $\Delta^{\ }_{\text{s}}$ 
as we will show in Sec. \ref{subsec: Density-density interaction}.
The gap equation at the chemical potential $\mu$ and the 
inverse temperature $\beta$ is
\begin{subequations}
\begin{equation}
1=
|V|\nu(\omega^{\ }_{\text{D}})
\sum_{\lambda=\pm}
\int\limits_{\mu-\omega^{\ }_{\text{D}}}^{\mu+\omega^{\ }_{\text{D}}}
\frac{
d\varepsilon\,
\nu(\varepsilon)
     }
     {
\nu(\omega^{\ }_{\text{D}})
     }
\frac{
\tanh\beta E^{\ }_{\lambda}(\varepsilon)/2
     }
     {
2E^{\ }_{\lambda}(\varepsilon)
     }
\end{equation}
where the ``Debye'' energy cutoff $\omega^{\ }_{\text{D}}$ 
has been introduced, 
\begin{equation}
E^{\ }_{\lambda}(\varepsilon):=
\sqrt{(\varepsilon-\lambda\mu)^{2}+\Delta^{2}_{\text{s}}},
\end{equation}
and
\begin{equation}
\nu(\varepsilon):=
\frac{|\varepsilon|}{2\pi(\hbar v^{\ }_{\mathrm{RD}})^{2}}
\end{equation}
\end{subequations}
is the Rashba-Dirac density of states per unit area.
The temperature dependence 
of the static Pauli magnetic susceptibility for an
out-of-plane uniform applied magnetic field
is then given by
$\chi^{\ }_{33}=2\chi^{\ }_{11}=2\chi^{\ }_{22}$ with
\begin{equation}
\begin{split}
\chi^{\ }_{33}\,\propto&\,
\sum_{\lambda=\pm}
\int \frac{d^{2}\boldsymbol{k}}{(2\pi)^{2}}
\frac{
E^{2 }_{\boldsymbol{k}\lambda}
+
\xi^{\ }_{\boldsymbol{k}, \lambda}
\xi^{\ }_{\boldsymbol{k},-\lambda}
+
\Delta^{2}_{\text{s}}(T)
     }
     {
E^{\ }_{\boldsymbol{k}\lambda}
     } 
\\
&\,\times
\frac{
1
     }
     {
\xi^{2 }_{\boldsymbol{k}, \lambda}
-
\xi^{2 }_{\boldsymbol{k},-\lambda}
     }
\tanh
\frac{
\beta E^{\ }_{\boldsymbol{k}\lambda}
     }
     {
2
     }.
\end{split}
\end{equation}
We plot the temperature dependence of the
self-consistent pair potential $\Delta^{\ }_{\text{s}}(T)$ 
and of $\chi^{\ }_{33}(T)$ in Fig.~\ref{fig:static susceptibilitiy}.
First, we note that $\chi^{\ }_{33}(T=0)\neq 0$.
This is a direct consequence of the spin-orbit coupling.\cite{Edelstein89}$^{-}$\cite{Frigeri04}
Second, we note that
$\chi^{\ }_{33}(T)$ decreases as a function of temperature beyond the
critical temperature, i.e., when $T>T^{\ }_{\text{c}}$.
Although the finite value of
$\chi^{\ }_{33}(T=0)$ is typical of noncentrosymmetric
superconductors,\cite{Edelstein89}$^{-}$\cite{Frigeri04}
$\chi^{\ }_{33}(T>T^{\ }_{\text{c}})$ only saturates
to a value proportional to the density of state at the Fermi level
in the regime~(\ref{eq: hierarchy A}).
In the regime~(\ref{eq: hierarchy B}), 
the decrease of $\chi^{\ }_{33}(T>T^{\ }_{\text{c}})$
can be understood with the help of
Eq.~(\ref{eq: main result Pauli susc})
if $\mu$ is substituted for $T$ to mimic thermal population.
Indeed, Eq.~(\ref{eq: main result Pauli susc})
implies that the normal-state $\chi^{\ }_{33}(\mu)$ at $T=0$
increases with $\mu$ if the Fermi level 
is below the Rashba-Dirac point, but decreases with $\mu$
if the Fermi level is above the Rashba-Dirac point, for
the states above the Rashba-Dirac point give a contribution 
that cancels part of the susceptibility coming from the states 
below the Rashba-Dirac point.

\section{
Superconducting instabilities
        }
\label{sec: Superconducting instabilities}

\subsection{
Density-density interaction
           }
\label{subsec: Density-density interaction}

To obtain the BdG Hamiltonian~(\ref{eq:BdG Hamiltonian})
self-consistently, we consider first a
density-density interaction given by
\begin{equation}
\begin{split}
&
H^{\ }_{V}:=
\frac{1}{2}
\sum_{\boldsymbol{q}}
V^{\ }_{ \boldsymbol{q}}
\rho^{\ }_{ \boldsymbol{q}}
\rho^{\ }_{-\boldsymbol{q}},
\qquad
\rho^{\ }_{\boldsymbol{q}}:=
\sum_{\boldsymbol{k},s=\uparrow,\downarrow}
c^{\dag}_{\boldsymbol{k}+\boldsymbol{q}s}
c^{\ }_{\boldsymbol{k}s}
\end{split}
\end{equation}
where $V^{\ }_{\boldsymbol{q}}$ is an even function of momentum.
Normal ordering yields
\begin{equation}
\begin{split}
H^{\ }_{V}=&\,
\frac{1}{2}
\sum_{\boldsymbol{q}}
V^{\ }_{ \boldsymbol{q}}
\sum_{\boldsymbol{k},\boldsymbol{k}'}
\sum_{s,s'}
c^{\dag}_{\boldsymbol{k}+\boldsymbol{q}s}
c^{\dag}_{\boldsymbol{k}'-\boldsymbol{q}s'}
c^{\   }_{\boldsymbol{k}'s'}
c^{\   }_{\boldsymbol{k}s}
\\
&\,
+
\frac{1}{2}
\sum_{\boldsymbol{q}}
V^{\ }_{ \boldsymbol{q}}
\sum_{\boldsymbol{k}}
\sum_{s}
c^{\dag}_{ \boldsymbol{k}s}
c^{\   }_{ \boldsymbol{k}s}.
\end{split}
\label{eq: normal ordering}
\end{equation}
After renormalization of the chemical potential and
restriction of the normal-ordered interaction 
to the scattering of Cooper pairs
with vanishing center-of-mass momentum,
we obtain the reduced Hamiltonian 
\begin{equation}
H^{\text{red}}_{V}=
\frac{1}{2}
\sum_{\boldsymbol{k},\boldsymbol{p}}
V^{\ }_{\boldsymbol{k}-\boldsymbol{p}}
\sum_{s,s'}
c^{\dag}_{ \boldsymbol{k}s}
c^{\dag}_{-\boldsymbol{k}s'}
c^{\   }_{-\boldsymbol{p}s'}
c^{\   }_{ \boldsymbol{p}s}.
\label{eq: reduced BCS}
\end{equation}
We show in Appendix~\ref{appsec: proofs}
that the reduced interaction%
~(\ref{eq: reduced BCS})
has the representation
\begin{equation}
\begin{split}
H^{\text{red}}_{V}=&\,
\frac{1}{2}
\sum_{\boldsymbol{k},\boldsymbol{p}}
\sum_{\lambda,\lambda'=\pm1}
V^{\ }_{\boldsymbol{k}-\boldsymbol{p}}
e^{\text{i}
\left(\varphi^{\ }_{\boldsymbol{p}}-\varphi^{\ }_{\boldsymbol{k}}\right)}
\\
&\,
\times
\left[
\cos
\left(
\varphi^{\ }_{\boldsymbol{p}}
-
\varphi^{\ }_{\boldsymbol{k}}
\right)
+
\lambda\lambda'
\right]
a^{\dag}_{ \boldsymbol{k}\lambda}
a^{\dag}_{-\boldsymbol{k}\lambda}
a^{\   }_{-\boldsymbol{p}\lambda'}
a^{\   }_{ \boldsymbol{p}\lambda'}
\\
&\,
+
\cdots.
\end{split}
\label{eq: transformed V}
\end{equation}
The terms $\cdots$ that were omitted involve pairs
of creation or of annihilation operators of opposite
helicities. We ignore these terms because we are going 
to perform a mean-field approximation for Cooper pairs
made of time-reversed helicity single-particle
states from Eq.~(\ref{eq: def TR on helicity states}).

We define the mean-field 
superconducting order parameters to be
\begin{subequations}
\label{eq: def sc order and pairing potentials}
\begin{equation}
\tilde{\delta}^{\ }_{\boldsymbol{k}\lambda}:=
\lambda
e^{\text{i}\varphi^{\ }_{-\boldsymbol{k}}}
\left\langle
a^{\ }_{-\boldsymbol{k}\lambda}
a^{\ }_{ \boldsymbol{k}\lambda}
\right\rangle^{\ }_{\beta,\mu}=
+
\tilde{\delta}^{\ }_{-\boldsymbol{k}\lambda}
\label{eq: def sc order and pairing potentials a}
\end{equation}
where $\lambda=\pm$.
The angular bracket represents the statistical averaging
at inverse temperature $\beta$ and chemical potential $\mu$.
We also define the mean-field helicity pairing potentials to be
\begin{equation}
\begin{split}
\tilde{\Delta}^{\ }_{\boldsymbol{k}\lambda}:=&\,
\frac{1}{2}
\sum_{\boldsymbol{p},\lambda'}
V^{\ }_{\boldsymbol{k}-\boldsymbol{p}}
\left[
\lambda\lambda'
\cos
\left(
\varphi^{\ }_{\boldsymbol{p}}
-
\varphi^{\ }_{\boldsymbol{k}}
\right)
+
1
\right]
\tilde{\delta}^{\ }_{\boldsymbol{p}\lambda'}
\\
=&\,
\tilde{\Delta}^{\ }_{-\boldsymbol{k}\lambda}
\end{split}
\label{eq: def sc order and pairing potentials b}
\end{equation}
\end{subequations}
where $\lambda=\pm$.The mean-field superconducting order parameter%
~(\ref{eq: def sc order and pairing potentials a})
and the pairing potentials%
~(\ref{eq: def sc order and pairing potentials b})
enter the (mean-field) BdG Hamiltonian
of the form~(\ref{eq: MF H}) and obey the self-consistent conditions
\begin{subequations}
\begin{equation}
\tilde{\delta}^{\ }_{\boldsymbol{p}\lambda}
=
-
\frac{
\tilde{\Delta}^{\ }_{\boldsymbol{p}\lambda}
     }
     {
2 E^{\ }_{\boldsymbol{p}\lambda}
     }
\tanh(\beta E^{\ }_{\boldsymbol{p}\lambda}/2)
\label{eq:selfconsistent conditions}
\end{equation}
where
the single-particle dispersion in the
superconducting state $E^{\ }_{\boldsymbol{p}\lambda}$ is
defined in Eq.~\eqref{eq:Sc single particle dispersion}.
\end{subequations}

If the pairing interaction is independent of momentum 
(i.e., a contact interaction in space), 
the summation over $\boldsymbol{p}$ on the right-hand side of 
Eq.~\eqref{eq: def sc order and pairing potentials b}
cancels the dependence on $\lambda$. 
Hence, both order parameters are then equal
$\tilde{\Delta}^{\ }_{\boldsymbol{k}+}=
 \tilde{\Delta}^{\ }_{\boldsymbol{k}-}$
and we can see from the transformation%
~\eqref{tilde tilde to spin trafo}
that the pairing potential will be of pure spin-singlet nature.
Observe that this result is independent of the noninteracting part 
of the Hamiltonian, and thus valid for both models 
(\ref{eq: def 2x2 Rashba-Dirac H0})
and 
(\ref{eq: def 2x2 Rashba on lattice}). 
It was also found in the context of 3D noncentrosymmetric superconductors 
in Ref.~\onlinecite{Samokhin08}.

We have also solved self-consistently the gap equation 
with the Dirac dispersion 
($\varepsilon^{\ }_{\boldsymbol{k}}\equiv0$) 
for a pairing interaction that is isotropic in momentum space
$V^{\ }_{\bs{q}}=V^{\ }_{|\bs{q}|}$. 
When the chemical potential is much larger than 
the transition temperature $|\mu\beta|\gg1$, 
we have found that the triplet component never exceeds 
the singlet component of the superconducting pairing potential 
if the pairing interaction $V^{\ }_{|\bs{q}|}$ 
never changes sign as a function of $|\bs{q}|$. The latter is true
for most of the commonly used model interactions, 
except Cooper pairing mediated by the Friedel oscillations
induced by the screening of the Coulomb repulsive interaction, 
for example.\cite{Kohn65,Braginskii90}

The density-density interaction as considered here might provide a model 
for the pairing interaction recently discovered at the superconducting 
interfaces in
LaAlO$_3$/SrTiO$_3$ (Ref.~\onlinecite{Reyren07})
or in  some electrolyte/SrTiO$_3$ (Ref.~\onlinecite{Ueno08}) 
that feature a low density and high mobility of the charge carriers.

\subsection{
Heisenberg interaction
           }
\label{subsec: Heisenberg interaction}

As a second example, 
we study the SU(2) preserving spin-density-spin-density interaction 
\begin{equation}
\begin{split}
&
H^{\ }_{\mathrm{H}}:=
\frac{1}{2}
\sum_{\boldsymbol{q}}
J^{\ }_{ \boldsymbol{q}}
\boldsymbol{S}^{\ }_{ \boldsymbol{q}}
\cdot
\boldsymbol{S}^{\ }_{-\boldsymbol{q}},
\quad
\boldsymbol{S}^{\ }_{\boldsymbol{q}}:=
\frac{1}{2}
\sum_{\boldsymbol{k};s,s'}
c^{\dag}_{\boldsymbol{k}+\boldsymbol{q}s}
\boldsymbol{\sigma}_{s,s'}
c^{\ }_{\boldsymbol{k}s'}
\end{split}
\end{equation}
where $J^{\ }_{ \boldsymbol{q}}$ is an even function of momentum.
Proceeding in the same way as in Sec.%
~\ref{subsec: Density-density interaction}, 
we obtain the reduced Hamiltonian for the scattering of Cooper pairs
with vanishing center-of-mass momentum
\begin{equation}
\begin{split}
H^{\text{red}}_{\mathrm{H}}=&\,
\frac{1}{8}
\sum_{\boldsymbol{k},\boldsymbol{p}}
\sum_{s^{\ }_1,s^{\ }_2,s^{\ }_3,s^{\ }_4}
J^{\ }_{\boldsymbol{k}-\boldsymbol{p}}
\boldsymbol{\sigma}_{s^{\ }_1,s^{\ }_4}\cdot
\boldsymbol{\sigma}_{s^{\ }_2,s^{\ }_3}\\
&\,
\times
c^{\dag}_{ \boldsymbol{k}s^{\ }_1}
c^{\dag}_{-\boldsymbol{k}s^{\ }_2}
c^{\   }_{-\boldsymbol{p}s^{\ }_3}
c^{\   }_{ \boldsymbol{p}s^{\ }_4}.
\end{split}
\label{eq: reduced BCS_Heisenberg}
\end{equation}
As is shown in Appendix~\ref{appsec: proofs}
the reduced interaction%
~(\ref{eq: reduced BCS_Heisenberg})
has the following representation
in the helicity basis
\begin{equation}
\begin{split}
H^{\text{red}}_{\text{H}}=&\,
\frac{1}{16}
\sum_{\boldsymbol{k},\boldsymbol{p}}
\sum_{\lambda,\lambda'=\pm1}
J^{\ }_{\boldsymbol{k}-\boldsymbol{p}}
e^{
\text{i}
\left(
\varphi^{\ }_{\boldsymbol{p}}
-
\varphi^{\ }_{\boldsymbol{k}}
\right)
  }
\\
&\,
\times
\left[
\cos
\left(
\varphi^{\ }_{\boldsymbol{p}}
-
\varphi^{\ }_{\boldsymbol{k}}
\right)
-
3
\lambda
\lambda'
\right]
a^{\dag}_{ \boldsymbol{k}\lambda}
a^{\dag}_{-\boldsymbol{k}\lambda}
a^{\   }_{-\boldsymbol{p}\lambda'}
a^{\   }_{ \boldsymbol{p}\lambda'}
\\
&\,
+
\cdots.
\end{split}
\label{eq: transformed Heisenberg}
\end{equation} 
The terms $\cdots$ that were omitted, just as in 
Sec.~\ref{subsec: Density-density interaction}
where we studied density-density interactions, 
involve pairs of creation or of annihilation operators 
of opposite helicities. We ignore these terms because we are going 
to perform a mean-field
approximation with pairs of time-reversed helicity
single-particle states from Eq.~(\ref{eq: def TR on helicity states}).

Again, we define the mean-field 
superconducting order parameters to be
\begin{subequations}
\label{eq: def sc order and pairing potentials if Heis}
\begin{equation}
\tilde{\delta}^{\ }_{\boldsymbol{k}\lambda}:=
\lambda
e^{\text{i}\varphi^{\ }_{-\boldsymbol{k}}}
\left\langle
a^{\ }_{-\boldsymbol{k}\lambda}
a^{\ }_{ \boldsymbol{k}\lambda}
\right\rangle^{\ }_{\beta,\mu}=
+
\tilde{\delta}^{\ }_{-\boldsymbol{k}\lambda}
\label{eq: def sc order and pairing potentials if Heis a}
\end{equation}
where $\lambda=\pm$ and
the angular bracket represents the statistical averaging
at inverse temperature $\beta$ and chemical potential $\mu$.
The mean-field helicity pairing potentials is defined to be
\begin{equation}
\begin{split}
\tilde{\Delta}^{\ }_{\boldsymbol{k}\lambda}:=&\,
\frac{1}{2}
\sum_{\boldsymbol{p},\lambda'}
J^{\ }_{\boldsymbol{k}-\boldsymbol{p}}
\left[
\lambda\lambda'
\cos
\left(
\varphi^{\ }_{\boldsymbol{p}}
-
\varphi^{\ }_{\boldsymbol{k}}
\right)
-
3
\right]
\tilde{\delta}^{\ }_{\boldsymbol{p}\lambda'}
\\
=&\,
\tilde{\Delta}^{\ }_{-\boldsymbol{k}\lambda}
\end{split}
\label{eq: def sc order and pairing potentials if Heis b}
\end{equation}
\end{subequations}
where $\lambda=\pm$.

Together with the superconducting order parameters%
~(\ref{eq: def sc order and pairing potentials if Heis a}),
they obey the self-consistent conditions
(\ref{eq:selfconsistent conditions}).

The term proportional to $\lambda$ represents the triplet 
component of the pairing potential
while the constant term in the square bracket gives the singlet component.

We have solved self-consistently the gap equation 
with the Dirac dispersion 
(i.e., $\varepsilon_{\boldsymbol{k}}\equiv0$) 
for a pairing interaction that is isotropic in momentum space
$J^{\ }_{\bs{q}}=J^{\ }_{|\bs{q}|}$. 
As with the case of the density-density interaction, 
we have found under the assumption $|\mu\beta|\gg1$ 
that the triplet component never exceeds the singlet component 
of the superconducting pairing potential if the pairing interaction 
$J^{\ }_{|\bs{q}|}$ never changes sign as a function of $|\bs{q}|$.

The Heisenberg interaction is attractive (repulsive) 
in the spin-singlet channel and repulsive (attractive) 
in the spin-triplet channel for 
$J^{\ }_{\bs{q}}>0$ ($J^{\ }_{\bs{q}}<0$). 
In centrosymmetric superconductors, this property leads the way 
toward spin-fluctuation mediated spin-triplet superconductivity for 
$J^{\ }_{\bs{q}}<0$. If inversion symmetry is broken, however, 
spin-singlet and spin-triplet pairing channels are not separated and 
for the case $J^{\ }_{\bs{q}}<0$, the interaction is altogether 
not attractive. Hence, the Heisenberg interaction will not lead to triplet 
(dominated) superconductivity in the same fashion as in centrosymmetric 
superconductors.

\subsection{
Dzyaloshinskii-Moriya interaction
           }
\label{subsec: DM interaction}

Finally, we study a spin-density-spin-density interaction of 
Dzyaloshinskii-Moriya type, 
which requires the breaking of inversion symmetry  to be present. 
Let the coefficient $\boldsymbol{D}^{\ }_{\boldsymbol{q}}$ 
be a three vector with vanishing $z$ component for our case. 
It shares the symmetry of $\boldsymbol{g}^{\ }_{\boldsymbol{q}}$, 
in particular, it is odd under $\boldsymbol{q}\rightarrow-\boldsymbol{q}$. 
The Dzyaloshinskii-Moriya interaction interaction is then
\begin{equation}
\begin{split}
&
H^{\ }_{\mathrm{DM}}:=
\frac{1}{2}
\sum_{\boldsymbol{q}}
\boldsymbol{D}^{\ }_{ \boldsymbol{q}}
\cdot
\left(
\boldsymbol{S}^{\ }_{ \boldsymbol{q}}
\wedge
\boldsymbol{S}^{\ }_{-\boldsymbol{q}}
\right).
\label{eq: def DM interaction}
\end{split}
\end{equation}
The reduced Hamiltonian for the scattering of Cooper pairs
with vanishing center of mass momentum reads
\begin{equation}
\begin{split}
H^{\text{red}}_{\mathrm{DM}}=&\,
\frac{1}{8}
\sum_{\boldsymbol{k},\boldsymbol{p}}
\sum_{s^{\ }_1,s^{\ }_2,s^{\ }_3,s^{\ }_4}
\boldsymbol{D}^{\ }_{\boldsymbol{k}-\boldsymbol{p}}
\cdot\left(
\boldsymbol{\sigma}_{s^{\ }_1,s^{\ }_4}\wedge
\boldsymbol{\sigma}_{s^{\ }_2,s^{\ }_3}\right)\\
&\,\times
c^{\dag}_{ \boldsymbol{k}s^{\ }_1}
c^{\dag}_{-\boldsymbol{k}s^{\ }_2}
c^{\   }_{-\boldsymbol{p}s^{\ }_2}
c^{\   }_{ \boldsymbol{p}s^{\ }_4}.
\end{split}
\label{eq: reduced BCS-DM}
\end{equation}
As is shown in Appendix~\ref{appsec: proofs},
the reduced interaction%
~(\ref{eq: reduced BCS-DM})
has the representation
\begin{equation}
\begin{split}
H^{\text{red}}_{\text{DM}}=&\,
\frac{\text{i}}{8}
\sum_{\boldsymbol{k},\boldsymbol{p}}
\sum_{\lambda,\lambda'=\pm1}
e^{
\text{i}
\left(
\varphi^{\ }_{\boldsymbol{p}}
-
\varphi^{\ }_{\boldsymbol{k}}
\right)
  }
\\
&\,
\times
\boldsymbol{D}^{\ }_{\boldsymbol{k}-\boldsymbol{p}}
\cdot
\left(
\lambda
\frac{
\boldsymbol{g}^{\ }_{\boldsymbol{p}}
     }
     {
|\boldsymbol{g}^{\ }_{\boldsymbol{p}}|
     }
-
\lambda'
\frac{
\boldsymbol{g}^{\ }_{\boldsymbol{k}}
     }
     {
|\boldsymbol{g}^{\ }_{\boldsymbol{k}}|
     }
\right)
a^{\dag}_{ \boldsymbol{k}\lambda}
a^{\dag}_{-\boldsymbol{k}\lambda}
a^{\   }_{-\boldsymbol{p}\lambda'}
a^{\   }_{ \boldsymbol{p}\lambda'}
\\
&\,
+
\cdots.
\label{eq: transformed DM}
\end{split}
\end{equation} 
Once again, the terms $\cdots$ that were omitted involve a pair
of creation or of annihilation operators of opposite
helicities while we keep only pairs made
of time-reversed helicity single-particle
states from Eq.~(\ref{eq: def TR on helicity states}).

The mean-field superconducting order parameters are again defined to be
\begin{subequations}
\label{eq: def sc order and pairing potentials if DM}
\begin{equation}
\tilde{\delta}^{\ }_{\boldsymbol{k}\lambda}:=
\lambda
e^{\text{i}\varphi^{\ }_{-\boldsymbol{k}}}
\left\langle
a^{\ }_{-\boldsymbol{k}\lambda}
a^{\ }_{ \boldsymbol{k}\lambda}
\right\rangle^{\ }_{\beta,\mu}=
+
\tilde{\delta}^{\ }_{-\boldsymbol{k}\lambda}
\label{eq: def sc order and pairing potentials if DM a}
\end{equation}
where $\lambda=\pm$ and the angular bracket 
represents the statistical averaging
at inverse temperature $\beta$ and chemical potential $\mu$.
We also define the mean-field helicity pairing potentials to be
\begin{equation}
\begin{split}
\tilde{\Delta}^{\ }_{\boldsymbol{k}\lambda}:=&\,
\frac{1}{2}
\sum_{\boldsymbol{p},\lambda'}
\boldsymbol{D}^{\ }_{\boldsymbol{k}-\boldsymbol{p}}
\cdot
\left(
\lambda'
\frac{
\boldsymbol{g}^{\ }_{\boldsymbol{p}}
     }
     {
|\boldsymbol{g}_{\boldsymbol{p}}|
     }
-
\lambda
\frac{
\boldsymbol{g}_{\boldsymbol{k}}
     }
     {
|\boldsymbol{g}_{\boldsymbol{k}}|
     }
\right)
\tilde{\delta}^{\ }_{\boldsymbol{p}\lambda'}
\\
=&\,
\tilde{\Delta}^{\ }_{-\boldsymbol{k}\lambda}
\end{split}
\label{eq: def sc order and pairing potentials if DM b}
\end{equation}
\end{subequations}
where $\lambda=\pm$.
Together with the superconducting order parameters%
~(\ref{eq: def sc order and pairing potentials if DM a}),
they obey the self-consistent conditions
(\ref{eq:selfconsistent conditions}).

We have solved self-consistently the gap equation 
with the Dirac-dispersion ($\varepsilon^{\ }_{\boldsymbol{k}}\equiv0$) 
for the model interaction 
$\boldsymbol{D}^{\ }_{\boldsymbol{q}}=
D \boldsymbol{g}^{\ }_{\boldsymbol{q}}
\text{exp}\left(-\boldsymbol{g}^{2}_{\boldsymbol{q}}/a^2\right)
$, where $a$ and $D$ are parameters. 
We have found that, depending on the chemical potential and 
the cutoff parameter $a$, the triplet component can exceed 
the singlet component of the superconducting pairing potential. 
This is in contrast to 
the results from the density-density interaction and the
Heisenberg interaction, where the singlet component is dominant. 
However, as the Dzyaloshinskii-Moriya interaction arises 
in second-order perturbation theory from an exchange interaction, 
it should not be considered on its own.

\subsection{
Superconductivity with in-plane magnetic field
           }
\label{subsec: Sc with inplane field}

A well established result for 2D noncentrosymmetric 
superconductors with the Rashba spin-orbit coupling%
~\eqref{eq: def 2x2 Rashba on lattice c} is that 
the superconducting pair potential acquires 
a real-space modulation in the presence of a 
Zeeman coupling to an in-plane magnetic field.\cite{Edelstein89,Samokhin04a,Samokhin04b,Kaur05}
An in-plane magnetic field shifts the Fermi sea away from the center of 
the Brillouin zone and as a result, the electrons with 
opposite wave vectors are not degenerate in energy anymore.

A similar effect is expected for the 2D Rashba-Dirac model 
subject to this study. We shall demonstrate this for a 
momentum-independent density-density interaction as was discussed 
in Sec.~\ref{subsec: Density-density interaction}.

The noninteracting Hamiltonian~\eqref{eq: def second quantization} 
in the Rashba-Dirac limit, 
i.e., for $\varepsilon^{\ }_{\boldsymbol{k}}\equiv0$,
is altered in the presence of an in-plane magnetic field 
\begin{equation}
\begin{split}
&
\boldsymbol{B}\equiv
B^{\ }_{1}
\boldsymbol{e}^{\ }_{1}
+
B^{\ }_{2}
\boldsymbol{e}^{\ }_{2},
\qquad
\boldsymbol{e}^{\ }_{3}:=
\boldsymbol{e}^{\ }_{1}\wedge\boldsymbol{e}^{\ }_{2},
\\
&
\boldsymbol{g}^{\ }_{\boldsymbol{k}}\equiv
g^{\ }_{\boldsymbol{k}1}
\boldsymbol{e}^{\ }_{1}
+
g^{\ }_{\boldsymbol{k}2}
\boldsymbol{e}^{\ }_{2},
\end{split}
\end{equation} 
according to
\begin{equation}
\begin{split}
&
H^{\boldsymbol{B}}_{0}=
\sum_{\boldsymbol{k}\in\mathrm{BZ}}
\psi^{\dag}_{\boldsymbol{k}} 
\mathcal{H}^{\boldsymbol{B}}_{0;\boldsymbol{k}} 
\psi^{\ }_{\boldsymbol{k}}=
\sum_{\boldsymbol{k}\in\mathrm{BZ}}
\tilde{\psi}^{\dag}_{\boldsymbol{k}}
\tilde{\mathcal{H}}^{\boldsymbol{B}}_{0;\boldsymbol{k}} 
\tilde{\psi}^{\ }_{\boldsymbol{k}},
\\
&
\mathcal{H}^{\boldsymbol{B}}_{0;\boldsymbol{k}}=
-
\mu
\sigma^{\ }_{0}
+ 
\left(
\boldsymbol{g}^{\ }_{\boldsymbol{k}}
-
\boldsymbol{B}\right)
\cdot
\boldsymbol{\sigma},
\\
&
\tilde{\mathcal{H}}^{\boldsymbol{B}}_{0;\boldsymbol{k}}=
\left(\begin{array}{cc}
\xi^{\boldsymbol{B}}_{\boldsymbol{k}+}
&
0
\\
0
&
\xi^{\boldsymbol{B}}_{\boldsymbol{k}-}
\end{array}
\right).
\label{eq: def second quantization with B field}
\end{split}
\end{equation}
The single-particle dispersion is now given by
\begin{equation}
\xi^{\boldsymbol{B}}_{\boldsymbol{k}\pm}=
-
\mu
\pm
|\boldsymbol{g}^{\ }_{\boldsymbol{k}}-\boldsymbol{B}|.
\label{dispersion in mag field}
\end{equation}
Accordingly, 
the phase factor entering the transformation%
~\eqref{eq: def trsf to helicity basis} 
between the laboratory basis
and the helicity basis is changed to
\begin{equation}
e^{
\text{i}\varphi^{\boldsymbol{B}}_{\boldsymbol{k}}
  }
=
\frac{
g^{\ }_{\boldsymbol{k}1}
-
B^{\ }_{1}+
\text{i}
g^{\ }_{\boldsymbol{k}2}
-
\text{i}
B^{\ }_{2}
     }
     {
|\boldsymbol{g}^{\ }_{\boldsymbol{k}}-\boldsymbol{B}|
     }.
\label{eq: def trsf to helicity basis with B-field}
\end{equation}

A pair of electrons on the Fermi surface without
magnetic field with opposite wave vectors 
$\boldsymbol{k}$ and $-\boldsymbol{k}$
is not degenerate in energy anymore in the presence of
$\boldsymbol{B}$, for
\begin{equation}
\xi^{\boldsymbol{B}}_{\boldsymbol{k}\pm}
-
\xi^{\boldsymbol{B}}_{-\boldsymbol{k}\pm}=
\mp
2
\frac{
\boldsymbol{g}^{\ }_{\boldsymbol{k}}
\cdot
\boldsymbol{B}
     }
     {
\left|\boldsymbol{g}^{\ }_{\boldsymbol{k}}\right|
     }
+
\mathcal{O}
\left(
\frac{
|\boldsymbol{B}|^{2}
     }
     {
\left|\boldsymbol{g}^{\ }_{\boldsymbol{k}}\right|^{2}
     }
\right).
\label{eq: expansion dispersion diff if B}
\end{equation}
It might thus be energetically more favorable to pair
electrons with the same energy but with a finite center-of-mass
momentum, than to pair electrons with vanishing center-of-mass
momentum. For simplicity, we also assume that only electrons 
of a single helicity $\lambda$ form Cooper pairs.
{}From here on, we denote the center-of-mass momentum of Cooper pairs by 
$\boldsymbol{q}$ while $\boldsymbol{k}$ and $\boldsymbol{k'}$ 
refer to the relative coordinate of the paired electrons. 
We assume that a single wave vector $\boldsymbol{q}$ 
for the modulation of the pairing potential 
will be energetically favorable, 
rather than a set of degenerate wave vectors. 
With these simplifications, the self-consistent gap equation for 
the pair potential 
$\tilde{\Delta}_{\boldsymbol{k}\lambda}(\boldsymbol{q})$ 
at temperature $T$ close to the superconducting transition temperature
and for $N$ sites is
\begin{equation}
\begin{split}
\tilde{\Delta}^{\ }_{\boldsymbol{k},\boldsymbol{q};\lambda}
=&\,
-
\frac{V}{2}
\frac{T}{N}
\sum_{\boldsymbol{k}',\omega^{\ }_{n}}
\cos
\left(
\frac{
\varphi^{\boldsymbol{B}}_{-\boldsymbol{k}'+\boldsymbol{q}/2}
-
\varphi^{\boldsymbol{B}}_{-\boldsymbol{k} +\boldsymbol{q}/2}
     }
     {
2
     }
\right)
\\
&\,
\times
\cos
\left(
\frac{
\varphi^{\boldsymbol{B}}_{\boldsymbol{k}'+\boldsymbol{q}/2}
-
\varphi^{\boldsymbol{B}}_{\boldsymbol{k} +\boldsymbol{q}/2}
     }
     {
2
     }
\right)
\tilde{\Delta}^{\ }_{\boldsymbol{k}',\boldsymbol{q};\lambda}
\\
&\times
G^{(0)}_{ \boldsymbol{k}'+\boldsymbol{q}/2, \text{i}\omega^{\ }_{n};\lambda}
G^{(0)}_{-\boldsymbol{k}'+\boldsymbol{q}/2,-\text{i}\omega^{\ }_{n};\lambda}
\end{split}
\label{eq: gap eqn with B-field}
\end{equation}
where the single-particle Green's function in the normal state 
for electrons with helicity $\lambda=\pm$ is given by
\begin{equation}
G^{(0)}_{\boldsymbol{k},\text{i}\omega^{\ }_{n};\lambda}=
-
\frac{
1
     }
     {
-\text{i}\omega^{\ }_{n}
+
\xi^{\boldsymbol{B}}_{\boldsymbol{k}\lambda}
     }.
\label{eq:Green funciton B field}
\end{equation}
For $s$-wave pairing, the pairing potential 
$\tilde{\Delta}_{\boldsymbol{k}\lambda}(\boldsymbol{q})$ 
is independent of $\boldsymbol{k}$. 
The gap equation~(\ref{eq:Green funciton B field})
simplifies to, after performing the summation over Matsubara frequencies,
\begin{subequations}
\label{eq: shortened gap eqn with B-field}
\begin{equation}
\begin{split}
1
=&\,
-\frac{V}{2N}
\sum_{\boldsymbol{k}'}
\cos
\left(
\frac{
\varphi^{\boldsymbol{B}}_{-\boldsymbol{k}'+\boldsymbol{q}/2}
-
\varphi^{\boldsymbol{B}}_{-\boldsymbol{k} +\boldsymbol{q}/2}
     }
     {
2
     }
\right)
\\
&\,\times
\cos
\left(
\frac{
\varphi^{\boldsymbol{B}}_{\boldsymbol{k}'+\boldsymbol{q}/2}
-
\varphi^{\boldsymbol{B}}_{\boldsymbol{k} +\boldsymbol{q}/2}
     }
     {
2
     }
\right)
f^{\boldsymbol{B}}_{\boldsymbol{k}',\boldsymbol{q};\lambda}.
\end{split}
\end{equation}
with the function
\begin{equation}
f^{\boldsymbol{B}}_{\boldsymbol{k},\boldsymbol{q};\lambda}:=
\frac{
\tanh
\frac{
\xi^{\boldsymbol{B}}_{
-\boldsymbol{k}
+
(\boldsymbol{q}/2)\lambda
                     }
     }
     {
2T
     }
+
\tanh
\frac{
\xi^{\boldsymbol{B}}_{\boldsymbol{k}+(\boldsymbol{q}/2)\lambda}
     }
     {
2T
     }
     }
     {
2\left(
\xi^{\boldsymbol{B}}_{-\boldsymbol{k}+(\boldsymbol{q}/2)\lambda}
+
\xi^{\boldsymbol{B}}_{\boldsymbol{k}+(\boldsymbol{q}/2)\lambda}
\right)
     }.
\label{thermal factor}
\end{equation}
\end{subequations}
In the Rashba-Dirac continuum limit%
~(\ref{eq: def 2x2 Rashba-Dirac H0}), 
the dispersion%
~\eqref{dispersion in mag field} 
together with the transformation%
~\eqref{eq: def trsf to helicity basis with B-field} 
and the thermal factor%
~\eqref{thermal factor} 
obey the symmetries
\begin{equation}
\begin{split}
&
\xi^{\boldsymbol{B}}_{\boldsymbol{ k}+(\boldsymbol{q}^{\ }_{0}/2)\lambda}=
\xi^{\boldsymbol{B}}_{\boldsymbol{-k}+(\boldsymbol{q}^{\ }_{0}/2)\lambda}=
\xi^{\boldsymbol{B}=0}_{\boldsymbol{ k}\lambda}=
\xi^{\boldsymbol{B}=0}_{\boldsymbol{-k}\lambda},
\\
&
\varphi^{\boldsymbol{B}}_{\boldsymbol{k}+(\boldsymbol{q}^{\ }_{0}/2)}=
\varphi^{\boldsymbol{B}=0}_{\boldsymbol{k}},
\\
&
f^{\boldsymbol{B}  }_{\boldsymbol{k},(\boldsymbol{q}^{\ }_{0}/2);\lambda}=
f^{\boldsymbol{B}=0}_{\boldsymbol{k},\boldsymbol{0};\lambda},
\end{split}
\label{B field dispersion symmetry}
\end{equation}
with 
$\boldsymbol{q}^{\ }_0=
 2\boldsymbol{B}\wedge
 \boldsymbol{e}^{\ }_{3}/(\hbar v^{\ }_{\mathrm{RD}})$
being proportional to the shift of the Fermi surface induced by
$\boldsymbol{B}$.
Hence, the gap equation for the superconducting condensate with the
center-of-mass momentum $\boldsymbol{q}^{\ }_0$ in the presence 
of the in-plane magnetic field $\boldsymbol{B}$ is the same as 
the gap equation in the absence of any in-plane magnetic field 
for a condensate with vanishing center-of-mass momentum. 
A condensate with vanishing center-of-mass momentum has
the largest transition temperature.
Hence, we deduce from the symmetry%
~\eqref{B field dispersion symmetry} 
that a superconducting order parameter with the center of mass momentum 
$\boldsymbol{q}^{\ }_0$ 
nucleates in the presence of an in-plane magnetic field.
The wave vector of the modulated pairing potential is perpendicular 
to the magnetic field in the plane and is independent 
of the chemical potential.
It also follows that the critical temperature of superconductivity 
is not suppressed by the magnetic field
in the Rashba-Dirac continuum limit%
~(\ref{eq: def 2x2 Rashba-Dirac H0}) 
and by the consideration of only one helicity.

In contrast to this simple result, 
the center-of-mass momentum of Cooper pairs in 
2D noncentrosymmetric 
superconductors with the Rashba spin-orbit coupling%
~\eqref{eq: def 2x2 Rashba on lattice c}
is selected by a delicate energetical compromise on how
the two helicity-resolved Fermi surfaces are 
shifted in opposite directions in momentum space.

\section{
Mean-field phase diagram
        }
\label{sec: Mean-field phase diagram}

It is time to explore the mean-field phase diagram that follows from Hamiltonian%
~(\ref{eq: BdG in terms tilde tilde pairings}) in the parameter space spanned 
by the choice made for the normal-state dispersion and for the pair potentials.
To this end, we consider the parameter space spanned by the couplings 
$\mu$, $\Delta^{\ }_{\text{s}}$, and $\Delta^{\ }_{\text{t}}$ 
entering the mean-field Hamiltonian.
A mean-field phase corresponds to a connected region in parameter space characterized 
by a nonvanishing gap. We shall then introduce in Sec.~\ref{sec: Majorana fermions} 
point defects in the mean-field Hamiltonian%
~(\ref{eq: BdG in terms tilde tilde pairings}), i.e., 
superconducting vortices with unit circulation, and compute 
the parity of the number of zero modes they bind to characterize the 
topological nature of the mean-field phases separated in parameter 
space by gap-closing boundaries.

For convenience, we recall that the BdG Hamiltonian is,
in the $\Phi$ representation~(\ref{eq: BdG in terms tilde tilde pairings}),
\begin{subequations}
\label{eq: def BdG Hamiltonian}
\begin{equation}
\mathcal{H}^{\ }_{\boldsymbol{k}}:=
\begin{pmatrix}
\varepsilon^{\ }_{\boldsymbol{k}}-\mu
&
A^{\ }_{\boldsymbol{k}} e^{-\text{i}\varphi^{\ }_{\boldsymbol{k}}}
&
\Delta^{\ }_{\text{s},\boldsymbol{k}}
&
\Delta^{\ }_{\text{t},\boldsymbol{k}}
e^{-\text{i}\varphi^{\ }_{\boldsymbol{k}}}
\\
A^{\ }_{\boldsymbol{k}} e^{\text{i}\varphi^{\ }_{\boldsymbol{k}}}
&
\varepsilon^{\ }_{\boldsymbol{k}}-\mu
&
\Delta^{\ }_{\text{t},\boldsymbol{k}}
e^{+\text{i}\varphi^{\ }_{\boldsymbol{k}}}
&
\Delta^{\ }_{\text{s},\boldsymbol{k}}
\\
\Delta^{\ }_{\text{s},\boldsymbol{k}}
&
\Delta^{\ }_{\text{t},\boldsymbol{k}}
e^{-\text{i}\varphi^{\ }_{\boldsymbol{k}}}
&
-\varepsilon^{\ }_{\boldsymbol{k}}+\mu
&
-A^{\ }_{\boldsymbol{k}} e^{-\text{i}\varphi^{\ }_{\boldsymbol{k}}}
\\
\Delta^{\ }_{\text{t},\boldsymbol{k}}
e^{+\text{i}\varphi^{\ }_{\boldsymbol{k}}}
&
\Delta^{\ }_{\text{s},\boldsymbol{k}}
&
-A^{\ }_{\boldsymbol{k}} e^{+\text{i}\varphi^{\ }_{\boldsymbol{k}}}
&
-\varepsilon^{\ }_{\boldsymbol{k}}+\mu
\end{pmatrix}
\end{equation}
where the normal-state dispersion is specified by
\begin{equation}
\begin{split}
&
\varepsilon^{\ }_{ \boldsymbol{k}}=
\varepsilon^{\ }_{-\boldsymbol{k}}
\in\mathbb{R},
\qquad
\mu\in\mathbb{R},
\\
&
A^{\ }_{ \boldsymbol{k}}\equiv
\left|\boldsymbol{g}^{\ }_{\boldsymbol{k}}\right|=
A^{\ }_{-\boldsymbol{k}}
\geq0,
\qquad
\boldsymbol{g}^{\ }_{ \boldsymbol{k}}=
-
\boldsymbol{g}^{\ }_{-\boldsymbol{k}}
\in\mathbb{R}^{2},
\\
&
\varphi^{\ }_{ \boldsymbol{k}}\equiv
\mathrm{arctan}\frac{g^{\ }_{ \boldsymbol{k};2}}{g^{\ }_{ \boldsymbol{k};1}}
\in[0,2\pi[,
\end{split}
\end{equation}
the singlet-pair potential $\Delta^{\ }_{\text{s},\boldsymbol{k}}$ 
and the triplet-pair potential $\Delta^{\ }_{\text{t},\boldsymbol{k}}$
transform according to any trivial irreducible
representation of the space group consistent with
$
\Delta^{\ }_{\text{t},\boldsymbol{k}}
e^{\pm\text{i}\varphi^{\ }_{\boldsymbol{k}}}
$
being single valued.
In the isotropic continuum limit, we thus assume that
the singlet-pair potential is constant while
the triplet-pair potential 
$\Delta^{\ }_{\text{t}}(\boldsymbol{k})$
can be factorized into a real
number $\Delta^{\ }_{\text{t}}$ times some strictly increasing positive 
function $f$ with at least a first-order zero 
at the origin such that
(i) it saturates to unity for large positive
argument and (ii) is invertible on the positive real axis with the
inverse $f^{-1}$, say, for instance, $f(x):=\tanh x$, i.e.,
\begin{equation}
\Delta^{\ }_{\text{t}}(\boldsymbol{k})=
\Delta^{\ }_{\text{t}}\,
f(|\boldsymbol{k}|/k^{\ }_{\text{t}})
\label{eq: def profile of d}
\end{equation}
\end{subequations}
for some wavelength 
$k^{\ }_{\text{t}}>0$
that defines the size of the core of the vortex
$\exp\big[-\text{i}\varphi(\boldsymbol{k})\big]$ 
at the origin in $\boldsymbol{k}$ space.
The aim of this section is to identify when the quasiparticle
spectral gap vanishes as a function of
the parameters $\Delta^{\ }_{\text{s}},\Delta^{\ }_{\text{t}}$, and
$\mu$ for a given dispersion relation in the isotropic continuum limit,
i.e., we need the eigenvalues of Eq.~(\ref{eq: def BdG Hamiltonian}).

To this end, we first square both sides of 
Eq.~(\ref{eq: def BdG Hamiltonian}), finding the
block diagonal form
\begin{equation}
\begin{split}
&
\mathcal{H}^{2}_{\boldsymbol{k}}=
\begin{pmatrix}
\mathbb{A}^{\ }_{\boldsymbol{k}}
&
0
\\
0
&
\mathbb{A}^{\ }_{\boldsymbol{k}}
\end{pmatrix},
\\
&
\mathbb{A}^{\ }_{\boldsymbol{k}}=
\left[
\left(
\varepsilon^{\ }_{\boldsymbol{k}}
-
\mu
\right)^{2}
+
A^{2}_{\boldsymbol{k}}
+
\Delta^{2}_{\text{s}}
+
\Delta^{2}_{\text{t},\boldsymbol{k}}
\right]
\sigma^{\ }_{0}
\\
&
\hphantom{\mathbb{A}=}
+
2
\left[
\left(
\varepsilon^{\ }_{\boldsymbol{k}}
-
\mu
\right)
A^{\ }_{\boldsymbol{k}}
+
\Delta^{\ }_{\text{s}}
\Delta^{\ }_{\text{t},\boldsymbol{k}}
\right]
\cos\varphi^{\ }_{\boldsymbol{k}}\,
\sigma^{\ }_{1}
\\
&
\hphantom{\mathbb{A}=}
+
2
\left[
\left(
\varepsilon^{\ }_{\boldsymbol{k}}
-
\mu
\right)
A^{\ }_{\boldsymbol{k}}
+
\Delta^{\ }_{\text{s}}
\Delta^{\ }_{\text{t},\boldsymbol{k}}
\right]
\sin\varphi^{\ }_{\boldsymbol{k}}\,
\sigma^{\ }_{2}.
\end{split}
\end{equation}
The four eigenvalues of $\mathcal{H}^{\ }_{\boldsymbol{k}}$ are
\begin{equation}
E^{\ }_{\boldsymbol{k};\lambda,\pm}=
\pm
\sqrt{
\left(
\varepsilon^{\ }_{\boldsymbol{k}}
-
\mu
+
\lambda
A^{\ }_{\boldsymbol{k}}
\right)^{2}
+
\left(
\Delta^{\ }_{\text{s}}
+
\lambda
\Delta^{\ }_{\text{t},\boldsymbol{k}}
\right)^{2}
     },
\label{eq: 4 BdG eigenvalues}
\end{equation}
where $\lambda=\pm$.
All nonvanishing energy eigenvalues come in pairs with opposite
signs. This spectral symmetry 
is a consequence of the particle-hole transformation [$X^{\ }_{\mu\nu}$ was defined in Eq.~\eqref{eq: def of Xmunu}]
\begin{equation}
X^{\ }_{22}\mathcal{H}^{T}_{-\boldsymbol{k}} X^{\ }_{22}=
-
\mathcal{H}^{\ }_{\boldsymbol{k}}.
\label{eq: def PH sym}
\end{equation}
The Hamiltonian $\mathcal{H}^{\ }_{\boldsymbol{k}}$ 
also features a helical symmetry given by
\begin{equation}
\left(
\hat{g}^{\ }_{\boldsymbol{k}1}X^{\ }_{10}
+
\hat{g}^{\ }_{\boldsymbol{k}2}X^{\ }_{20}
\right)
\mathcal{H}^{\ }_{\boldsymbol{k}}
\left(
\hat{g}^{\ }_{\boldsymbol{k}1}X^{\ }_{10}
+
\hat{g}^{\ }_{\boldsymbol{k}2}X^{\ }_{20}
\right)
=
\mathcal{H}^{\ }_{\boldsymbol{k}}.
\label{eq: def helical sym}
\end{equation}
Viewing the Rashba spin-orbit coupling as a fictitious magnetic field along
a $\boldsymbol{k}$-dependent direction, the helical symmetry%
~(\ref{eq: def helical sym})
reflects the
conservation of spin along this direction in momentum space.

For completeness, TRS is nothing but
\begin{equation}
X^{\ }_{20}\mathcal{H}^{*}_{-\boldsymbol{k}} X^{\ }_{20}=
+
\mathcal{H}^{\ }_{\boldsymbol{k}}
\label{eq: def TRS sym}
\end{equation}
in the $\Phi$ basis of Eq.~(\ref{eq: def bispinors}) that we have chosen.

Zero modes are vanishing eigenvalues of
$\mathcal{H}^{\ }_{\boldsymbol{k}}$, 
i.e., they are the solutions to
\begin{equation}
\begin{split}
0=&\,
\mathrm{det}\,\mathcal{H}^{\ }_{\boldsymbol{k}}
\\
=&\,
\left[
\left(
\varepsilon^{\ }_{\boldsymbol{k}}
-
\mu
+
A^{\ }_{\boldsymbol{k}}
\right)^{2}
+
\left(
\Delta^{\ }_{\mathrm{s}}
+
\Delta^{\ }_{\mathrm{t},\boldsymbol{k}}
\right)^{2}
\right]
\\
&\,
\times
\left[
\left(
\varepsilon^{\ }_{\boldsymbol{k}}
-
\mu
-
A^{\ }_{\boldsymbol{k}}
\right)^{2}
+
\left(
\Delta^{\ }_{\mathrm{s}}
-
\Delta^{\ }_{\mathrm{t},\boldsymbol{k}}
\right)^{2}
\right].
\end{split}
\end{equation}
There are thus two possibilities to 
get zero modes. Either
\begin{subequations}
\begin{equation}
\hbox{case ($+$):}\qquad
0=
\varepsilon^{\ }_{\boldsymbol{k}}
-
\mu
+
A^{\ }_{\boldsymbol{k}},
\qquad
0=
\Delta^{\ }_{\mathrm{s}}
+
\Delta^{\ }_{\mathrm{t},\boldsymbol{k}},
\label{eq: condition for + helicity}
\end{equation}
or
\begin{equation}
\hbox{case ($-$):}\qquad
0=
\varepsilon^{\ }_{\boldsymbol{k}}
-
\mu
-
A^{\ }_{\boldsymbol{k}},
\qquad
0=
\Delta^{\ }_{\mathrm{s}}
-
\Delta^{\ }_{\mathrm{t},\boldsymbol{k}}.
\label{eq: condition for - helicity}
\end{equation}
\end{subequations}
Equation~(\ref{eq: condition for + helicity})
requires that the $\lambda=+$ helicity gap vanishes on the
$\lambda=+$ helicity Fermi surface.
Equation~(\ref{eq: condition for - helicity})
requires that the $\lambda=-$ helicity gap vanishes on the
$\lambda=-$ helicity Fermi surface.
The condition 
\begin{subequations}
\begin{equation}
\varepsilon^{\ }_{\boldsymbol{k}}
-
\mu
+
\lambda
A^{\ }_{\boldsymbol{k}}=0
\end{equation}
on the normal-state dispersion determines the Fermi surfaces
\begin{equation}
\text{FS}^{\ }_{\lambda}:=
\left\{
\boldsymbol{k}|
\varepsilon^{\ }_{\boldsymbol{k}}
-
\mu
+ 
\lambda
A^{\ }_{\boldsymbol{k}}=0
\right\}.
\end{equation} 
\end{subequations}
The condition
\begin{subequations}
\begin{equation}
\Delta^{\ }_{\text{s}} 
+
\lambda 
\Delta^{\ }_{\text{t}}\,f^{\ }_{\boldsymbol{k}}=0
\label{eq: condition on pairing}
\end{equation}
on the pairing potentials determines the momenta for which
the superconducting single-particle gap vanishes
\begin{equation}
\text{SC}^{\ }_{\lambda}:=
\left\{
\boldsymbol{k}|
\Delta^{\ }_{\text{s}} 
\pm 
\Delta^{\ }_{\text{t}}\,f^{\ }_{\boldsymbol{k}}=0
\right\}.
\end{equation}
\end{subequations}
Conditions $(+)$ or $(-)$ are satisfied along the points 
\begin{equation}
\text{FS}^{\ }_{\lambda}\cap\,\text{SC}^{\ }_{\lambda} 
\ne 
\emptyset,
\qquad
\lambda=\pm.
\end{equation} 
(in other words, the Fermi surfaces cross the 
superconducting single-particle nodal surfaces).

\begin{figure}
\includegraphics[scale=0.3]{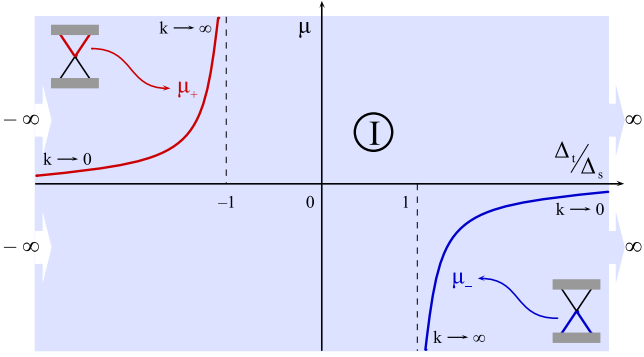}
\caption{
(Color online)
Mean-field phase boundary in the Rashba-Dirac limit~(\ref{eq: def Rashba-Dirac limit}).
        }
\label{fig:phase_dia2d-Dirac}
\end{figure}

\subsection{
Isotropic continuum limit
           }

We work in the continuum limit with the 
upper bound
$\Lambda$ and the 
lower bound $-\Lambda$
to the single-particle mean-field spectrum,
as is appropriate for the surface states of a 3D band insulator. 
We assume that the SRS dispersion
$\varepsilon$,
the Rashba dispersion $A$, 
and profile $f$ of the vortex
$\exp\big[-\text{i}\varphi(\boldsymbol{k})\big]$
at the origin in $\boldsymbol{k}$ space
are smooth functions of $|\boldsymbol{k}|$. 
For the analysis to come, 
it is useful to define the dimensionless quantity
\begin{equation}
\mathsf{k}\equiv|\boldsymbol{k}|/k^{\ }_{\text{t}}.
\end{equation}

We define the 2D parameter space with
$\Delta^{\ }_{\text{t}}/\Delta^{\ }_{\text{s}}$ 
as the horizontal axis and $\mu$ 
as the vertical axis.
For any finite positive singlet pairing potential 
$\Delta^{\ }_{\text{s}}\ne 0$,
we show 
\begin{enumerate}
\item\label{enu: 1}
That there are two nonintersecting curves
(to simplify the notation
$\Lambda\to\infty$)
\begin{equation}
\begin{split}
&
\mu^{\ }_{+}:
(-\infty,-1]
\longrightarrow
\mathbb{R},
\\
&\hphantom{\mu^{\ }_{\lambda}:}
\Delta^{\ }_{\text{t}}/\Delta^{\ }_{\text{s}}
\longmapsto
\mu^{\ }_{+}(\Delta^{\ }_{\text{t}}/\Delta^{\ }_{\text{s}}),
\\
&
\mu^{\ }_{-}:
[1,\infty)
\longrightarrow
\mathbb{R},
\\
&\hphantom{\mu^{\ }_{-}:}
\Delta^{\ }_{\text{t}}/\Delta^{\ }_{\text{s}}
\longmapsto
\mu^{\ }_{-}(\Delta^{\ }_{\text{t}}/\Delta^{\ }_{\text{s}}),
\end{split}
\end{equation}
defined by the condition
(\ref{eq: condition for + helicity})
for $\lambda=+$
and
(\ref{eq: condition for - helicity})
for $\lambda=-$
at which the BdG spectrum%
~(\ref{eq: 4 BdG eigenvalues})
is gapless.
\item\label{enu: 2}
The curves $\mu^{\ }_{\lambda}$
are one-to-one reparameterizations of the
dispersions $\xi^{\ }_{\lambda}(|\boldsymbol{k}|)$
with $\lambda=\pm$.
\item\label{enu: 3}
How the two curves 
$\mu^{\ }_{\lambda}$ 
change upon changing the topology of the Fermi surfaces.
\end{enumerate}

For the superconducting single-particle gap to vanish, we must
choose
\begin{subequations}
\label{eq: main step for nodes}
\begin{equation}
\lambda=
-
\text{sgn}\,
\frac{
\Delta^{\ }_{\text{s}}
     }
     {
\Delta^{\ }_{\text{t}}
     }
\end{equation}
in Eq.~(\ref{eq: condition on pairing})
from which the implicit definition 
\begin{equation}
0\leq 
f(\mathsf{k})=
\left|
\frac{
\Delta^{\ }_{\text{s}}
     }
     {
\Delta^{\ }_{\text{t}}
     }
\right|
\end{equation}
\end{subequations}
of
$\mathsf{k}$ 
follows. Hence, 
$\mathsf{k}$ 
is the function
\begin{equation}
\begin{split}
&
\mathsf{k}:
(1,\infty)
\longrightarrow
\mathbb{R},
\\
&
\hphantom{\mathsf{k}:}
\left|
\frac{
\Delta^{\ }_{\text{t}}
     }
     {
\Delta^{\ }_{\text{s}}
     }
\right|     
\longrightarrow
\mathsf{k}\left(
\left|
\frac{
\Delta^{\ }_{\text{t}}
     }
     {
\Delta^{\ }_{\text{s}}
     }
\right|     
     \right):=
f^{-1}
\left(
\left|
\frac{
\Delta^{\ }_{\text{s}}
     }
     {
\Delta^{\ }_{\text{t}}
     }
\right|
\right),
\end{split}
\end{equation}
which is not defined
whenever the superconducting single-particle gap does not close, 
i.e., when
$|\Delta^{\ }_{\text{t}}|<|\Delta^{\ }_{\text{s}}|$.
Claims \ref{enu: 1} and \ref{enu: 2} follow
with the definition
\begin{equation}
\mu^{\ }_{\lambda}(\mathsf{k}):=
\varepsilon(\mathsf{k})
+
\lambda
A(\mathsf{k})
\end{equation}
where $\mathsf{k}$ and $\lambda$ were 
themselves defined
in Eq.~(\ref{eq: main step for nodes})
and with the momentum core size $k^{\ }_{\text{t}}$ taken to be unity.

To illustrate how the topology of the normal-state dispersion
changes the curves $\mu^{\ }_{\lambda}$ with $\lambda=\pm$,
we make the (electronlike) parabolic approximation
\begin{equation}
\varepsilon(\boldsymbol{k})=
\frac{\hbar^{2}|\boldsymbol{k}|^{2}}{2m},
\qquad
m\geq0,
\qquad
A(\boldsymbol{k})=
\hbar v^{\ }_{\mathrm{RD}}|\boldsymbol{k}|,
\end{equation}
and choose the momentum-vortex profile
\begin{equation}
f(x)=\tanh x 
\end{equation}
with the momentum core size $k^{\ }_{\text{t}}$ taken to be unity.
We consider the Rashba-Dirac limit 
\begin{equation}
m=\infty
\label{eq: def Rashba-Dirac limit}
\end{equation} 
first, as is illustrated in Fig.~\ref{fig:phase_dia2d-Dirac}. 
The two curves
$\mu^{\ }_{\pm}(\Delta^{\ }_{\text{t}}/\Delta^{\ }_{\text{s}})$ 
where the gap vanishes are indicated in 
Fig.~\ref{fig:phase_dia2d-Dirac}.
They are obtained via the
reparameterization of the dispersions $\xi^{\ }_\pm(|\boldsymbol{k}|)$, 
as depicted on the insets on the second and fourth quadrants. There is a
one-to-one correspondence between the thick lines in the insets
and the curves 
$\mu^{\ }_{\pm}(\Delta^{\ }_{\text{t}}/\Delta^{\ }_{\text{s}})$.

In Fig.~\ref{fig:phase_dia2d-Dirac}, 
we look at the mean-field phases that can be
identified given the gap-closing curves 
$\mu^{\ }_{\pm}(\Delta^{\ }_{\text{t}}/\Delta^{\ }_{\text{s}})$. 
Here, one must notice that taking 
$\Delta^{\ }_{\text{t}}/\Delta^{\ }_{\text{s}}\to\infty$, 
for a given chemical potential $\mu$ such that the gap does
not close, is connected to the path originating from 
$\Delta^{\ }_{\text{t}}/\Delta^{\ }_{\text{s}}\to -\infty$. 
For instance, one can send 
$\Delta^{\ }_{\text{s}}\to 0$ 
so that it changes sign while holding 
$\Delta^{\ }_{\text{t}}$ 
constant but with a given $\mu$ such
that the gap does not close. Therefore, the regions depicted in
Fig.~\ref{fig:phase_dia2d-Dirac} are connected upon folding the
horizontal axis into a circle (the plane into a cylinder). Then,
because of the topology of the curves 
$\mu^{\ }_{\pm}(\Delta^{\ }_{\text{t}}/\Delta^{\ }_{\text{s}})$, 
any one region can be connected to any other without crossing these curves, 
and hence there is a single phase for the system, which we denote by ${\rm I}$.

\begin{figure}
\includegraphics[scale=0.3]{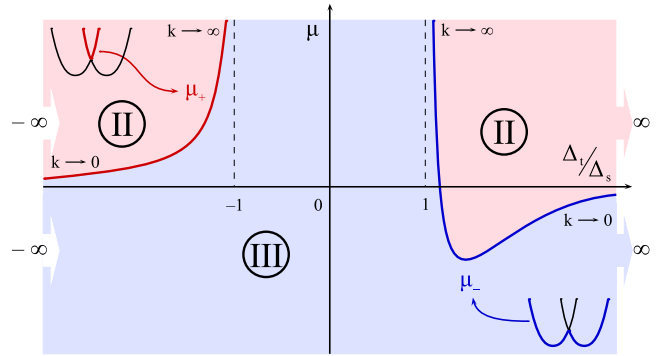}\caption{
(Color online)
Mean-field phase boundary away from the Rashba-Dirac limit,
i.e., when Eq.~(\ref{eq: def non-Dirac}) holds.
        }
\label{fig:phase_dia2d-Fermi}
\end{figure}

For any finite curvature of the dispersion 
$\varepsilon(\boldsymbol{k})$, i.e.,
\begin{equation}
0\leq m<\infty,
\label{eq: def non-Dirac}
\end{equation}
we find the boundaries shown in
Fig.~\ref{fig:phase_dia2d-Fermi}. Again, the two curves
$\mu^{\ }_{\pm}(\Delta^{\ }_{\text{t}}/\Delta^{\ }_{\text{s}})$ 
where the gap vanishes are obtained via the
reparameterization of the dispersions $\xi^{\ }_{\pm}(|\boldsymbol{k}|)$, 
as depicted on the insets on the second and fourth quadrants.
We see that the topology of the curve 
$\mu^{\ }_{+}(|\Delta^{\ }_{\text{t}}/\Delta^{\ }_{\text{s}}|)$
that tracks the normal-state dispersion
$\xi^{\ }_{+}(|\boldsymbol{k}|)$
is insensitive to tuning $m$ from infinity to any finite value.
This is not so for the topology of the curve 
$\mu^{\ }_{-}(|\Delta^{\ }_{\text{t}}/\Delta^{\ }_{\text{s}}|)$
that tracks the normal-state dispersion
$\xi^{\ }_{-}(|\boldsymbol{k}|)$.
This curve is dramatically influenced by the 
nonmonotonous dependence of 
$\xi^{\ }_{-}(|\boldsymbol{k}|)$
on $|\boldsymbol{k}|$
for any finite curvature, i.e., any mass $m<\infty$.
In the Rashba-Dirac limit $m=\infty$,
$\mu^{\ }_{-}(|\Delta^{\ }_{\text{t}}/\Delta^{\ }_{\text{s}}|)$
is strictly negative, and $\mu^{\ }_{-}(1^{\ }_{+})\to-\infty$. 
But when $m$ is finite, 
$\mu^{\ }_{-}(1^{\ }_{+})\to\infty$.

The distinct phases in the $m$ finite case are identified in
Fig.~\ref{fig:phase_dia2d-Fermi}. If the regions with
$\Delta^{\ }_{\text{t}}/\Delta^{\ }_{\text{s}}\to\pm\infty$ 
are identified, then the two regions $\text{II}$ and $\text{III}$ 
are always separated from each other by the two curves 
$\mu^{\ }_{\pm}(|\Delta^{\ }_{\text{t}}/\Delta^{\ }_{\text{s}}|)$. 
There is no path connecting the two regions $\text{II}$ and $\text{III}$
without ever closing the mean-field spectral gap: this is a necessary
(but not sufficient)
condition for these to be two distinct phases.

\subsection{
Anisotropic case
           }

The boundaries in the
$\Delta^{\ }_{\text{t}}$-$\mu$ plane
at which the BdG single-particle spectrum closes
for an anisotropic continuum dispersion or for a
two-dimensional lattice are more difficult to determine.
Indeed, a technical difficulty brought about by the loss of
continuous rotational symmetry is that
it is not possible anymore to characterize the nodes of the
normal-state dispersion or the nodes of the superconducting
gaps with a single wave number. This could result in these
boundaries acquiring a thickness (i.e., a finite area).\cite{Beri09}

For a 2D lattice model,
a qualitative difference with the continuum limit that is
of no relevance to this section is the fermion doubling
and its consequences for the existence and the stability
of Majorana fermions in the core of superconducting vortices.
This is the subject of the ensuing section in which we
search for Majorana modes in the core of defects (vortices) 
of the superconducting order parameter and we probe their
stability under adiabatic changes of the bulk parameters
(i.e., far away from the vortices).

\section{
Majorana fermions
        }
\label{sec: Majorana fermions}

Caroli \emph{et al.} showed in Ref.~\onlinecite{Caroli64}
that isolated vortices 
in a weakly coupled type II $s$-wave superconductor
with TRS and SRS support a discrete set of
finite-energy bound states with a level spacing
of order of the ratio of the squared 
single-particle bulk superconducting gap
to the bandwidth. There is no bound state at the Fermi energy
bound to the core of vortices in this case.

Jackiw and Rossi showed in Ref.~\onlinecite{Jackiw81} that, 
in two space and one time
dimensions quantum electrodynamics (QED$_{2+1}$) coupled to 
one scalar Higgs field, an isolated static defect in the Higgs field,
i.e., a single vortex with vorticity $N$, supports
$N$ bound states that are all pinned to the zero energy.
These $N$ bound states are $N$ Majorana fermions. 
An index theorem for this result was proved by Weinberg.%
\cite{Weinberg81}

Read and Green\cite{Read00} showed that a two-dimensional chiral 
$p^{\ }_{x}\pm\text{i}p^{\ }_{y}$
superconductor supports a Majorana mode bound to the
core of an isolated half vortex.

We are going to show that 
(i)
an isolated vortex with unit vorticity
in the singlet-pair potential binds
a single Majorana mode in region I of 
Fig.~\ref{fig:phase_dia2d-Dirac},
(ii) 
an isolated vortex with unit vorticity
in the triplet-pair potential binds
\textit{two} Majorana modes in region II of 
Fig.~\ref{fig:phase_dia2d-Fermi},
and (iii) isolated vortices in region III of 
Fig.~\ref{fig:phase_dia2d-Fermi}
do not bind Majorana fermions.
We will start by reviewing the derivation
of the Jackiw-Rossi Majorana mode that 
applies to region I of 
Fig.~\ref{fig:phase_dia2d-Dirac}. 
We will then discuss region II and III in
Fig.~\ref{fig:phase_dia2d-Fermi}.

We work in the isotropic continuum limit with the Hamiltonian
in the $\Phi$ representation~(\ref{eq: BdG in terms tilde tilde pairings})
given by
\begin{subequations}
\label{eq: def vortex H}
\begin{widetext}
\begin{equation}
\mathcal{H}^{\ }_{\text{vor}}:=
\begin{pmatrix}
\varepsilon(k,\bar{k})
&
\hbar v^{\ }_{\mathrm{RD}}
\bar{k}
&
\Delta^{\ }_{\text{s}}(z,\bar{z})
&
\frac{1}{2}
\left\{
\Delta^{\ }_{\text{t}}(z,\bar{z}),
f\left(\frac{|k|}{k^{\ }_{\text{t}}}\right)
\frac{\bar{k}}{|k|}
\right\}
\\
\hbar v^{\ }_{\mathrm{RD}}
k
&
\varepsilon(k,\bar{k})
&
\frac{1}{2}
\left\{
\Delta^{\ }_{\text{t}}(z,\bar{z}),
f\left(\frac{|k|}{k^{\ }_{\text{t}}}\right)
\frac{k}{|k|}
\right\}
&
\Delta^{\ }_{\text{s}}(z,\bar{z})
\\
\text{H.c.}
&
\text{H.c.}
&
-\varepsilon(k,\bar{k})
&
-
\hbar v^{\ }_{\mathrm{RD}}
\bar{k}
\\
\text{H.c.}
&
\text{H.c.}
&
-
\hbar v^{\ }_{\mathrm{RD}}
k
&
-\varepsilon(k,\bar{k})
\end{pmatrix}.
\label{eq: def vortex H a}
\end{equation}
\end{widetext}
Here, the SRS normal-state dispersion is parabolic
\begin{equation}
\begin{split}
&
\varepsilon(k,\bar{k}):=
\frac{\hbar^{2}|k|^{2}}{2m}
-
\mu,
\end{split}
\label{eq: def vortex H b}
\end{equation}
\end{subequations}
where the real valued $\mu$ is the chemical potential.
Moreover, the singlet-pair potential 
$\Delta^{\ }_{\text{s}}(z,\bar{z})$
has a unit vortex at the origin of the complex-$z$ plane 
with the characteristic core size 
$\ell^{\ }_{\text{s}}$
and saturates to the bulk value
$\Delta^{\ }_{\text{s}}$ 
for $|z|\gg\ell^{\ }_{\text{s}}$,
as does the triplet-pair potential 
$\Delta^{\ }_{\text{t}}(z,\bar{z})$
with the characteristic core size 
$\ell^{\ }_{\text{t}}$
and the bulk value
$\Delta^{\ }_{\text{t}}$ 
for $|z|\gg\ell^{\ }_{\text{t}}$.
The bulk values $\Delta^{\ }_{\text{s}}$ and
$\Delta^{\ }_{\text{t}}$
share a common phase that can be removed 
by a global gauge transformation up to a relative sign.
The function $f$ that guarantees single valuedness of the Hamiltonian
was defined in Eq.~(\ref{eq: def profile of d}).
The anticommutators in the antidiagonal 
are needed since translation invariance has
been broken. We are using the complex notation
\begin{subequations}
\begin{equation}
\begin{split}
&
k:=k^{\ }_{1}+\text{i}k^{\ }_{2},
\qquad
\bar{k}:=k^{\ }_{1}-\text{i}k^{\ }_{2},
\\
&
z:= z^{\ }_{1}+\text{i}z^{\ }_{2},
\qquad
\bar{z}:= z^{\ }_{1}-\text{i}z^{\ }_{2},
\end{split}
\end{equation}
together with the algebra
\begin{equation}
[z^{\ }_{a},k^{\ }_{b}]=\text{i}\delta^{\ }_{ab},
\qquad
a,b=1,2,
\end{equation}
or, equivalently,
\begin{equation}
[z,\bar{k}]=[\bar{z},k]=2\text{i},
\qquad
[z,k]=[\bar{z},\bar{k}]=0.
\label{eq: complex Heisenberg algebra}
\end{equation}
A representation of the algebra%
~(\ref{eq: complex Heisenberg algebra})
is given by
\begin{equation}
k      =-2\text{i}\partial^{\ }_{\bar{z}},
\qquad
\bar{k}=-2\text{i}\partial^{\ }_{     z }.
\label{eq: real space rep}
\end{equation}
The representation dual to 
Eq.~(\ref{eq: real space rep}) 
is
\begin{equation}
z      =2\text{i}\partial^{\ }_{\bar{k}},
\qquad
\bar{z}=2\text{i}\partial^{\ }_{     k }.
\label{eq: mom space rep}
\end{equation}
\end{subequations}
We shall rely on the polar coordinate
representation
\begin{equation}
k=\kappa e^{+\text{i}\varphi}
\end{equation}
in terms of which
\begin{equation}
\begin{split}
&
z=
2\text{i}\partial^{\ }_{\bar{k}}=
\text{i}e^{+\text{i}\varphi}
\left(
\partial^{\ }_{\kappa}
+
\frac{
\text{i}
     }
     {
\kappa
     }
\partial^{\ }_{\varphi}
\right),
\\
&
\bar{z}=
2\text{i}\partial^{\ }_{k}=
\text{i}e^{-\text{i}\varphi}
\left(
\partial^{\ }_{\kappa}
-
\frac{
\text{i}
     }
     {
\kappa
     }
\partial^{\ }_{\varphi}
\right).
\end{split}
\end{equation}
We choose to represent $z,\bar z$ as differential
operators of functions of $k,\bar k$ instead of the other way around
because this can bring a simplification in the solution of the zero
modes. By solving for the wave functions of the zero modes in momentum
space, we take advantage of the fact that we have a first-order
differential equation instead of a second-order one, which would be
the case had we chosen to solve for the wave functions in position
space. This simplification works because we deform the profile
of the vortex without changing the fact that the solutions are
precisely at energy $E=0$, as we discuss below.

Instead of facing the difficulty to solve analytically 
for the spectrum of the BdG Hamiltonian~(\ref{eq: def vortex H}), 
we are thus going to make approximations
that are motivated by the mean-field phase diagram
of Sec.~\ref{sec: Mean-field phase diagram}.

We are first going to take
the Rashba-Dirac limit at $\mu=0$ (the Rashba-Dirac point)
without triplet-pair potential, $\Delta^{\ }_{\text{t}}=0$. 
This is nothing but
the origin of region I in Fig.~\ref{fig:phase_dia2d-Dirac}.
In this limit, Hamiltonian~(\ref{eq: def vortex H})
is the direct sum of two
$2\times2$ Hamiltonians.

We are then going to take
the Fermi limit $v^{\ }_{\mathrm{RD}}=0$ with $\mu>0$
without singlet-pair potential,
$\Delta^{\ }_{\text{s}}=0$, i.e., 
the vertical half line at infinity in
region I of Fig.~\ref{fig:phase_dia2d-Dirac}.
In this limit, Hamiltonian~(\ref{eq: def vortex H})
is again the direct sum of two
$2\times2$ Hamiltonians.

\begin{figure}[t]
\includegraphics[angle=0,scale=0.3]{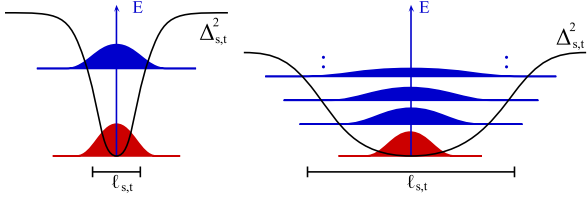}
\caption{
(Color online)
Qualitative comparison of the bound state spectra
of a deep and narrow vortex (left) and a wide and shallow vortex (right).
While the former supports only few finite-energy bound states (blue), 
in the spectrum of the latter more states are allowed.
However, the existence of a zero-energy mode (red) is independent of the 
details of the regime.
         }
\label{fig: spectrum in a vortex}
\end{figure}

Even after these simplifications, the spectrum with
a vortex at the origin is difficult to compute. 
One must solve a system of two coupled partial differential 
equations that depends on the non universal details
encoded by the profile of the vortices 
in the superconducting pair potentials
and by the profile $f$ for the vortex in the 
$\boldsymbol{d}$ vector. 
This microscopic information does influence the
discrete spectrum that represents the 
bound states with nonvanishing energies
attached to the core of the vortex. For example,
if the profile of the vortex is deep 
due to a large bulk gap and narrow due to a small 
characteristic size, there should be very few 
bound states with nonvanishing energies 
(see Fig.~\ref{fig: spectrum in a vortex}). 
On the other hand, in the opposite
limit of a shallow and smooth profile for the vortex,
many bound states with nonvanishing energies
extending far away from the
vortex core are to be expected
(see Fig.~\ref{fig: spectrum in a vortex}). 

However, we are not after the full spectrum of states
bound to the core of a vortex. We are only seeking
the conditions under which Majorana bound states,
i.e., bound states pinned at the normal-state chemical potential, 
are present. The very existence of a Majorana state does
not depend on the profiles of the vortices in real and 
momentum space as long as the nonvanishing energy spectrum
of bound states remains discrete and separated from the zero energy. 
This suggests choosing the vortex profile
\begin{subequations}
\label{eq: linear vortex profile}
\begin{equation}
\Delta^{\ }_{\text{s},\text{t}}(z,\bar{z})=
\Delta^{\ }_{\text{s},\text{t}}
\frac{z}{\ell^{\ }_{\text{s},\text{t}}}
\label{eq: vortex linearization}
\end{equation}
for the singlet (s) or triplet (t) component of the 
pair potential in real space, respectively, and the vortex profile
\begin{equation}
f\left(\frac{|k|}{k^{\ }_{\text{t}}}\right)=
\frac{|k|}{k^{\ }_{\text{t}}}
\end{equation}
\end{subequations}
for the $\boldsymbol{d}$ vector in momentum space.
This approximation has the merit of linearizing the
spectral eigenvalue problem.

\subsection{
The Rashba-Dirac limit
           }
\label{sec: Majorana fermions a}

The Rashba-Dirac limit is defined by the condition 
\begin{subequations}
\label{eq: def Rashba-Dirac limit + Rashba-Dirac point}
\begin{equation}
m=\infty.
\end{equation}
The Rashba-Dirac-point limit is defined by the additional condition
\begin{equation}
\mu=0.
\end{equation}
\end{subequations}
In this limit and when the singlet-pair potential vanishes,
the bulk gap closes so that vortices in the triplet-pair
potential are ill defined.

In the limit~(\ref{eq: def Rashba-Dirac limit + Rashba-Dirac point})
and when the triplet-pair potential vanishes, 
after setting $\hbar=v^{\ }_{\mathrm{RD}}=1$,
\begin{subequations}
\label{eq: Rashba-Dirac limit+Dirac point+ only singlet}
\begin{equation}
\mathcal{H}^{\ }_{\text{vor}}:=
\begin{pmatrix}
0
&
\bar{k}
&
\frac{
\Delta^{\ }_{\text{s}}
     }
     {
\ell^{\ }_{\text{s}}
     }
 z
&
0
\\
k
&
0
&
0
&
\frac{
\Delta^{\ }_{\text{s}}
     }
     {
\ell^{\ }_{\text{s}}
     }
z
\\
\text{H.c.}
&
0
&
0
&
-
\bar{k}
\\
0
&
\text{H.c.}
&
-
k
&
0
\end{pmatrix}
\end{equation}
decomposes into the direct sum of the $2\times2$
Hermitian operators
\begin{equation}
\mathcal{H}^{(1)}_{\text{vor}}:=
\begin{pmatrix}
\bar{k}
&
\frac{
\Delta^{\ }_{\text{s}}
     }
     {
\ell^{\ }_{\text{s}}
     }
 z
\\
\frac{
\bar{\Delta}^{\ }_{\text{s}}
     }
     {
\ell^{\ }_{\text{s}}
     }
\bar{z}
&
-
k
\end{pmatrix}
\end{equation}
and
\begin{equation}
\mathcal{H}^{(2)}_{\text{vor}}:=
\begin{pmatrix}
k
&
\frac{
\Delta^{\ }_{\text{s}}
     }
     {
\ell^{\ }_{\text{s}}
     }
z
\\
\frac{
\bar{\Delta}^{\ }_{\text{s}}
     }
     {
\ell^{\ }_{\text{s}}
     }
\bar{z}
&
-
\bar{k}
\end{pmatrix}.
\end{equation}
\end{subequations}

Jackiw and Rossi showed that Hamiltonian~(\ref{eq: def vortex H a})
with a vortex in $\Delta^{\ }_{\text{s}}(z,\bar{z})$ satisfying 
$\ell^{\ }_{\text{s}}<\infty$ and $|\Delta^{\ }_{\text{s}}(z,\bar{z})|_{|z|\rightarrow \infty}<\infty$, supports one and only one bound state
and that this bound state is pinned to the chemical potential,
i.e., to the midgap of the BdG Hamiltonian. We will show explicitly that
Hamiltonian~(\ref{eq: Rashba-Dirac limit+Dirac point+ only singlet}),
with an unbounded vortex profile, also has a singly degenerate solution,
and thus there is one and only one Majorana fermion
in the Rashba-Dirac limit at the Rashba-Dirac point if the pair potential
is pure singlet. The stability of this Majorana fermion  
away from the Rashba-Dirac point or in the presence of a triplet-pair potential is guaranteed by the fact that the particle-hole symmetry
of the eigenvalue spectrum precludes the migration of the zero mode
as long as the gap does not close. For large values of the
chemical potential, it is natural to anticipate that
many more bound states will have peeled off from the continuum
with a level spacing \`a la Caroli-de-Gennes. This expectation
is consistent with the computation from Ref.~\onlinecite{Khaymovich09}
of the states bound to a vortex 
as a function of the chemical potential
for pure-singlet superconducting graphene.

\begin{proof}
Let
\begin{equation}
c:=\frac{
\Delta^{\ }_{\text{s}}
     }
     {
\ell^{\ }_{\text{s}}
     }.
\end{equation}
We seek the solutions to
\begin{subequations}
\label{eq: Rashba-Dirac limit, point zero mode step 1}
\begin{equation}
\begin{split}
&
0=
\bar{k}
u^{(1)}
+
c
 z
v^{(1)},
\\
&
0=
\bar{c}
\bar{z}
u^{(1)}
-
k
v^{(1)},
\end{split}
\label{eq: Rashba-Dirac limit, point zero mode step 1 a}
\end{equation}
and
\begin{equation}
\begin{split}
&
0=
k
u^{(2)}
+
c
 z
v^{(2)},
\\
&
0=
\bar{c}
\bar{z}
u^{(2)}
-
\bar{k}
v^{(2)},
\end{split}
\label{eq: Rashba-Dirac limit, point zero mode step 1 b}
\end{equation}
\end{subequations}
respectively.
If we take the complex conjugate of the second condition
in Eqs.~(\ref{eq: Rashba-Dirac limit, point zero mode step 1 a})
and (\ref{eq: Rashba-Dirac limit, point zero mode step 1 b}),
respectively, we get
\begin{subequations}
\begin{equation}
\begin{split}
&
0=
\bar{k}
u^{(1)}
+
c
 z
v^{(1)},
\\
&
0=
-
\bar{k}
\bar{v}^{(1)}
+
c
z
\bar{u}^{(1)},
\end{split}
\end{equation}
and
\begin{equation}
\begin{split}
&
0=
k
u^{(2)}
+
c
 z
v^{(2)},
\\
&
0=
-
k
\bar{v}^{(2)}
+
c
z
\bar{u}^{(2)},
\end{split}
\end{equation}
\end{subequations}
respectively. For any $j=1,2$, if
\begin{equation}
\begin{pmatrix}
u^{(j)}(k,\bar{k})
\\
v^{(j)}(k,\bar{k})
\end{pmatrix}
\end{equation}
is a zero mode, so is
\begin{equation}
\pm
\begin{pmatrix}
\bar{v}^{(j)}(-k,-\bar{k})
\\
\bar{u}^{(j)}(-k,-\bar{k})
\end{pmatrix}.
\end{equation}
Hence, we try the ansatz
\begin{subequations}
\begin{equation}
\Phi^{(j)}_{\pm}(k,\bar{k}):=
\begin{pmatrix}
u^{(j)}(k,\bar{k})
\\
\pm \bar{u}^{(j)}(-k,-\bar{k})
\end{pmatrix}
\end{equation}
where
\begin{equation}
\begin{split}
0=&\,
\bar{k}
u^{(1)}(k,\bar{k})
+
(\pm)
c
 z
\bar{u}^{(1)}(-k,-\bar{k})
\\
=&\,
\kappa e^{-\text{i}\varphi}
u^{(1)}(\kappa,\varphi)
\\
&\,
+
(\pm)
\text{i}
c
e^{+\text{i}\varphi}
\left(
\partial^{\ }_{\kappa}
+
\frac{
\text{i}
     }
     {
\kappa
     }
\partial^{\ }_{\varphi}
\right)
\bar{u}^{(1)}(\kappa,\varphi+\pi)
\end{split}
\end{equation}
and
\begin{equation}
\begin{split}
0=&\,
k
u^{(2)}(k,\bar{k})
+
(\pm)
c
 z
\bar{u}^{(2)}(-k,-\bar{k})
\
\\
=&\,
\kappa e^{\text{i}\varphi}
u^{(2)}(\kappa,\varphi)
\\
&\,
+
(\pm)
\text{i}
c
e^{+\text{i}\varphi}
\left(
\partial^{\ }_{\kappa}
+
\frac{
\text{i}
     }
     {
\kappa
     }
\partial^{\ }_{\varphi}
\right)
\bar{u}^{(2)}(\kappa,\varphi+\pi).
\end{split}
\end{equation}
\end{subequations}
We choose a gauge in which
\begin{equation}
\tilde{c}:=
\text{i}c
\end{equation}
is real and make the ansatz
\begin{subequations}
\begin{equation}
\begin{split}
&
u^{(1)}_{\pm}(\kappa,\varphi)=
e^{\text{i}\varphi}
w^{(1)}_{\pm}(\kappa),
\\
&
u^{(2)}_{\pm}(\kappa,\varphi)=
w^{(2)}_{\pm}(\kappa),
\end{split}
\label{eq: dif eq for Rashba-Dirac a}
\end{equation}
where the real-valued $w^{(j)}_{\pm}(\kappa)$ satisfy
\begin{equation}
\begin{split}
0=&\,
\left[
\kappa
+
(\mp)
\tilde{c}
\left(
\partial^{\ }_{\kappa}
+
\frac{1}{\kappa}
\right)
\right]
w^{(1)}_{\pm}(\kappa)
\end{split}
\label{eq: dif eq for Rashba-Dirac b}
\end{equation}
and
\begin{equation}
\begin{split}
0=&\,
\left[
\kappa
+
(\pm)
\tilde{c}
\partial^{\ }_{\kappa}
\right]
w^{(2)}_{\pm}(\kappa),
\end{split}
\label{eq: dif eq for Rashba-Dirac c}
\end{equation}
\end{subequations}
respectively.
The formal solutions to
Eqs.~(\ref{eq: dif eq for Rashba-Dirac b})
and (\ref{eq: dif eq for Rashba-Dirac c})
are
\begin{subequations}
\begin{equation}
w^{(1)}_{\pm}(\kappa)=
w^{(1)}_{\pm}(\kappa^{\ }_{0})
\exp
\left\{
\int\limits_{\kappa^{\ }_{0}}^{\kappa}
d\kappa'
\left[
(\pm)
\frac{\kappa'}{\tilde{c}}
-
\frac{1}{\kappa'}
\right]
\right\}
\end{equation}
and
\begin{equation}
w^{(2)}_{\pm}(\kappa)=
w^{(2)}_{\pm}(\kappa^{\ }_{0})
\exp
\left[
-
\int\limits_{\kappa^{\ }_{0}}^{\kappa}
d\kappa'
(\pm)
\frac{\kappa'}{\tilde{c}}
\right],
\end{equation}
\end{subequations}
respectively. Only
\begin{subequations}
\begin{equation}
w^{(2)}_{\text{sgn}\,\tilde{c}}(\kappa)=
w^{(2)}_{\text{sgn}\,\tilde{c}}(\kappa^{\ }_{0})
\exp
\left(
-
\frac{\kappa^{2}-\kappa^{2}_{0}}{2|\tilde{c}|}
\right)
\end{equation}
is normalizable. We conclude that 
\begin{equation}
\Phi^{\ }_{\text{sgn}\,\tilde{c}}(\kappa,\varphi):=
\begin{pmatrix}
1\\
0\\
0\\
\text{sgn}\,\tilde{c}
\end{pmatrix}
w^{(2)}_{\text{sgn}\,\tilde{c}}(\kappa)
\end{equation}
\end{subequations}
is a Majorana state with the eigenvalue
$\text{sgn}\,\tilde{c}$ under the particle-hole
transformation~(\ref{eq: def PH sym}).
The uniqueness of this Majorana state, up to a
normalization factor,
can be proved along the same lines as is done in
Appendix~\ref{appsec: zero modes}.
\end{proof}

\subsection{
The Fermi limit
           }
\label{sec: Majorana fermions b}

The Fermi limit is defined by the condition
\begin{equation}
v^{\ }_{\mathrm{RD}}=0.
\label{eq: def pure parabolic limit}
\end{equation}
In this limit and when the triplet-pair potential vanishes,
isolated vortices support finite-energy Caroli-de-Gennes-Matricon 
bound states in the weak-coupling limit 
$\Delta^{\ }_{\text{s}}/\mu\ll1$. 
No Majorana fermions are to be found
tied to the core of an isolated vortex.

In the limit~(\ref{eq: def pure parabolic limit})
with a vanishing singlet-pair potential,
\begin{equation}
\mathcal{H}^{\ }_{\text{vor}}=
\begin{pmatrix}
\varepsilon(k,\bar{k})
&
0
&
0
&
\frac{
\Delta^{\ }_{\text{t}}
     }
     {
2
\ell^{\ }_{\text{t}}
k^{\ }_{\text{t}}
     }\{z,\bar{k}\}
\\
0
&
\varepsilon(k,\bar{k})
&
\frac{
\Delta^{\ }_{\text{t}}
     }
     {
2
\ell^{\ }_{\text{t}}
k^{\ }_{\text{t}}
     }
\{z,k\}
&
0
\\
0
&
\text{H.c.}
&
-\varepsilon(k,\bar{k})
&
0
\\
\text{H.c.}
&
0
&
0
&
-\varepsilon(k,\bar{k})
\end{pmatrix}
\end{equation}
decomposes into the direct sum of the $2\times2$
Hermitian operators
\begin{subequations}
\label{eq: triplet done right}
\begin{equation}
\mathcal{H}^{(1)}_{\text{vor}}:=
\begin{pmatrix}
\varepsilon(k,\bar{k})
&
\frac{
\Delta^{\ }_{\text{t}}
     }
     {
2\ell^{\ }_{\text{t}}
k^{\ }_{\text{t}}
     }
\{z,\bar{k}\}
\\
\frac{
\bar{\Delta}^{\ }_{\text{t}}
     }
     {
2
\ell^{\ }_{\text{t}}
k^{\ }_{\text{t}}
     }
\{z,\bar{k}\}^{\dag}
&
-\varepsilon(k,\bar{k})
\end{pmatrix}
\label{eq: triplet done right a}
\end{equation}
and
\begin{equation}
\mathcal{H}^{(2)}_{\text{vor}}:=
\begin{pmatrix}
\varepsilon(k,\bar{k})
&
\frac{
\Delta^{\ }_{\text{t}}
     }
     {
2
\ell^{\ }_{\text{t}}
k^{\ }_{\text{t}}
     }
\{z,k\}
\\
\frac{
\bar{\Delta}^{\ }_{\text{t}}
     }
     {
2
\ell^{\ }_{\text{t}}
k^{\ }_{\text{t}}
     }
\{z,k\}^{\dag}
&
-\varepsilon(k,\bar{k})
\end{pmatrix}.
\label{eq: triplet done right b}
\end{equation}
\end{subequations}
We claim that Hamiltonian $\mathcal{H}^{\ }_{\text{vor}}$
supports two normalized zero modes
if and only if (iff) the chemical potential $\mu>0$.

\begin{proof}
Define the short-hand notation
\begin{equation}
c:=
\frac{
\Delta^{\ }_{\text{t}}
     }
     {
\ell^{\ }_{\text{t}}
k^{\ }_{\text{t}}
     }
\in\mathbb{C}.
\end{equation}

We need
\begin{equation}
\begin{split}
\{z,\bar{k}\}=&\,
z\bar{k}
+
\bar{k}z
\\
=&\,
2\bar{k}z
+
[z,\bar{k}]
\\
=&\,
2\bar{k}z
+
2\text{i}
\end{split}
\end{equation}
and
\begin{equation}
\begin{split}
\{z,\bar{k}\}^{\dag}=&\,
k\bar{z}
+
\bar{z}k
\\
=&\,
2k\bar{z}
+
[\bar{z},k]
\\
=&\,
2k\bar{z}
+
2\text{i}.
\end{split}
\end{equation}
Equation (\ref{eq: triplet done right}) becomes
\begin{subequations}
\label{eq: triplet done right bis}
\begin{equation}
\mathcal{H}^{(1)}_{\text{vor}}=
\begin{pmatrix}
\varepsilon(k,\bar{k})
&
c
\left(
\bar{k}z
+
\text{i}
\right)
\\
\bar{c}
\left(
k\bar{z}
+
\text{i}
\right)
&
-\varepsilon(k,\bar{k})
\end{pmatrix}
\label{eq: triplet done right bis a}
\end{equation}
and
\begin{equation}
\mathcal{H}^{(2)}_{\text{vor}}=
\begin{pmatrix}
\varepsilon(k,\bar{k})
&
ckz
\\
\bar{c}\bar{k}\bar{z}
&
-\varepsilon(k,\bar{k})
\end{pmatrix}.
\label{eq: triplet done right bis b}
\end{equation}
\end{subequations}
We are going to show that
operator~(\ref{eq: triplet done right bis a})
has one and only one zero mode iff $\mu>0$.
We will then show that the same is true for
operator~(\ref{eq: triplet done right bis b}).

We seek a solution to
\begin{equation}
0=
\mathcal{H}^{(1)}_{\text{vor}}
\begin{pmatrix}
u^{(1)}
\\
v^{(1)}
\end{pmatrix}.
\label{eq: triplet zero mode is defined}
\end{equation}
We must solve
\begin{subequations}
\label{eq: zero mode up cond}
\begin{eqnarray}
&&
0=
\varepsilon(k,\bar{k}) 
u^{(1)}
+
c
\left(
\bar{k}2\text{i}\partial^{\ }_{\bar{k}}
+
\text{i}
\right)
v^{(1)},
\label{eq: zero mode up cond a}
\\
&&
0=
\bar{c}
\left(
k2\text{i}\partial^{\ }_{k}
+
\text{i}
\right)
u^{(1)}
-
\varepsilon(k,\bar{k})
v^{(1)}.
\label{eq: zero mode up cond b}
\end{eqnarray}
\end{subequations}
If we take the complex conjugate of
Eq.~(\ref{eq: zero mode up cond b}),
we get 
\begin{subequations}
\begin{eqnarray}
&&
0=
\varepsilon(k,\bar{k}) 
u^{(1)}
+
c
\left(
\bar{k}2\text{i}\partial^{\ }_{\bar{k}}
+
\text{i}
\right)
v^{(1)},
\\
&&
0=
-
c
\left(
\bar{k}2\text{i}\partial^{\ }_{\bar{k}}
+
\text{i}
\right)
\bar{u}^{(1)}
-
\varepsilon(k,\bar{k})
\bar{v}^{(1)}.
\label{eq: zero mode up cond compl conj}
\end{eqnarray}
\end{subequations}
We thus infer that a solution to 
Eq.~(\ref{eq: triplet zero mode is defined}),
if it exists, is given by
\begin{subequations}
\begin{equation}
\Phi^{(1)}_{\pm}=
\begin{pmatrix}
u^{(1)}_{\pm}
\\
\pm
\bar{u}^{(1)}_{\pm}
\end{pmatrix}
\end{equation}
where $u^{(1)}_{\pm}$ is the solution to
\begin{equation}
0=
\varepsilon(k,\bar{k}) 
u^{(1)}_{\pm}
+
(\pm)
c
\left(
\bar{k}2\text{i}\partial^{\ }_{\bar{k}}
+
\text{i}
\right)
\bar{u}^{(1)}_{\pm}.
\label{zero modes block1}
\end{equation}
\end{subequations}

Zero modes, if they exist, can be labeled by
their angular momentum. We seek a zero mode with vanishing
angular momentum, i.e., independent of $\varphi$. We must then
solve
\begin{subequations}
\begin{equation}
\Phi^{(1)}_{\pm}(\kappa)=
\begin{pmatrix}
u^{(1)}_{\pm}(\kappa)
\\
\pm
\bar{u}^{(1)}_{\pm}(\kappa)
\end{pmatrix},
\qquad
0\leq\kappa<\infty,
\end{equation}
where $u^{(1)}_{\pm}$ is the solution to
\begin{equation}
0=
\varepsilon(\kappa) u^{(1)}_{\pm}
+
(\pm)
(\text{i}c)
\left(
\kappa\partial^{\ }_{\kappa}
+
1
\right)
\bar{u}^{(1)}_{\pm}.
\end{equation}
\end{subequations}
With the help of a global gauge transformation,
we can always choose $c$ so that 
\begin{equation}
\tilde{c}:=\text{i}c
\end{equation}
is real valued.
Hence,
\begin{equation}
0=
\varepsilon(\kappa) 
u^{(1)}_{\pm}(\kappa)
+
(\pm)
\tilde{c}
\left(
\kappa\partial^{\ }_{\kappa}
+
1
\right)
\bar{u}^{(1)}_{\pm}(\kappa)
\label{eq: intermediary step}
\end{equation}
with $0\leq\kappa<\infty$
admits a purely real or a purely imaginary solution
since all coefficients of this first-order
differential equation are real valued.
We choose the real-valued solution.
We divide Eq.~(\ref{eq: intermediary step})
by $(\pm)\tilde{c}\kappa$ to obtain the condition
\begin{subequations}
\begin{equation}
0=
\left\{
\partial^{\ }_{\kappa}
+
\left[
1
+
(\pm)
\frac{
\varepsilon(\kappa)
     }
     {
\tilde{c}
     }
\right]
\frac{1}{\kappa}
\right\}
u^{(1)}_{\pm}
\end{equation}
whose formal solution is given by
\begin{equation}
u^{(1)}_{\pm}(\kappa)=
u^{(1)}_{\pm}(\kappa^{\ }_{0})
\times
\exp
\left\{
-
\int\limits^{\kappa}_{\kappa^{\ }_{0}} 
\frac{d\kappa'}{\kappa'}
\left[
1
+
(\pm)
\frac{
\varepsilon(\kappa')
     }
     {
\tilde{c}
     }
\right]
\right\}.
\label{eq: the formal solution}
\end{equation}
\end{subequations}

The formal solution~(\ref{eq: the formal solution})
is admissible  iff it is normalizable, i.e.,
if
\begin{equation}
\int\limits_{0}^{\infty}
d\kappa\, \kappa\,
\left|u^{(1)}_{\pm}(\kappa)\right|^{2}
<\infty.
\end{equation}
Define
\begin{equation}
\begin{split}
F^{\ }_{\pm}(\kappa):=&\,
\int\limits^{\kappa}_{\kappa^{\ }_{0}} 
\frac{d\kappa'}{\kappa'}
\left[
1
+
(\pm)
\frac{
\varepsilon(\kappa')
     }
     {
\tilde{c}
     }
\right]
\\
=&\,
\left[
1
-
(\pm)
\frac{\mu}{\tilde{c}}
\right]
\ln\frac{\kappa}{\kappa^{\ }_{0}}
+
(\pm)
\frac{
\kappa^{2}
-
\kappa^{2}_{0}
     }
     {
4m\tilde{c}
     }.
\end{split}
\end{equation}
For large $\kappa$,
\begin{equation}
F^{\ }_{\pm}(\kappa)\sim
(\pm)
\frac{
\kappa^{2}
     }
     {
4m\tilde{c}
     }
\end{equation}
so that normalizability
imposes the choice
\begin{equation}
\pm=
\text{sgn}\,\tilde{c}
\end{equation}
and the formal solution~(\ref{eq: the formal solution})
becomes
\begin{equation}
u^{(1)}_{\text{sgn}\,\tilde{c}}(\kappa)=
u^{(1)}_{\text{sgn}\,\tilde{c}}(\kappa^{\ }_{0})
\times
\left(
\frac{\kappa}{\kappa^{\ }_{0}}
\right)^{\frac{\mu}{|\tilde{c}|}-1}
\times
e^{
-
\frac{
\kappa^{2}-\kappa^{2}_{0}
     }
     {
4m|\tilde{c}|
     }
  }.
\label{eq: the formal solution bis}
\end{equation}
For small $\kappa$,
\begin{equation}
F^{\ }_{\text{sgn}\,\tilde{c}}(\kappa)\sim
\left(
1
-
\frac{
\mu
     }
     {
|\tilde{c}|
     }
\right)
\ln\frac{\kappa}{\kappa^{\ }_{0}}
\end{equation}
so that normalizability demands convergence of
\begin{equation}
\int\limits_{0}^{\kappa^{\ }_{0}}
d\kappa\,
\kappa^{2(\mu/|\tilde{c}|)-1}
\end{equation}
i.e., 
\begin{equation}
\mu>0.
\end{equation}
We conclude that for any choice of gauge such that
$\tilde{c}\equiv\text{i}c$ is real valued
\begin{subequations}
\label{eq: final block 1 zero mode}
\begin{equation}
\Phi^{(1)}_{\text{sgn}\,\tilde{c}}(\kappa)=
\begin{pmatrix}
u^{(1)}_{\text{sgn}\,\tilde{c}}(\kappa)
\\
{}
\\
(\text{sgn}\,\tilde{c})\,
\bar{u}^{(1)}_{\text{sgn}\,\tilde{c}}(\kappa)
\end{pmatrix},
\qquad
0\leq\kappa<\infty,
\label{eq: final block 1 zero mode a}
\end{equation}
where 
\begin{equation}
u^{(1)}_{\text{sgn}\,\tilde{c}}(\kappa)=
u^{(1)}_{\text{sgn}\,\tilde{c}}(\kappa^{\ }_{0})
\times
\left(
\frac{\kappa}{\kappa^{\ }_{0}}
\right)^{\frac{\mu}{|\tilde{c}|}-1}
\times
e^{
-
\frac{
\kappa^{2}-\kappa^{2}_{0}
     }
     {
4m|\tilde{c}|
     }
  }
\label{eq: final block 1 zero mode b}
\end{equation}
\end{subequations}
is a normalizable Majorana mode iff $\mu>0$.
Observe that $\Phi^{(1)}_{\text{sgn}\,\tilde{c}}(\boldsymbol{k})$ is 
an eigenstate with the eigenvalue $-\text{sgn}\,\tilde{c}$ 
of the particle-hole transformation
defined in Eq.~\eqref{eq: def PH sym}, i.e.,
\begin{equation}
X^{\ }_{22}\Phi^{(1)*}_{\text{sgn}\,\tilde{c}}(-\boldsymbol{k})
=
-\text{sgn}\,\tilde{c}\,
\Phi^{(1)}_{\text{sgn}\,\tilde{c}}(\boldsymbol{k}).
\end{equation}

To show that solution~(\ref{eq: final block 1 zero mode})
is unique, up to a normalization,
one expands 
Eq.~(\ref{eq: triplet zero mode is defined})
in polar harmonics labeled
by the angular quantum number $n\in\mathbb{Z}$
(see Appendix~\ref{appsec: zero modes}).
Modes with angular quantum number $\pm n$ are
pairwise coupled. A formal zero mode of the
form Eq.~(\ref{eq: final block 1 zero mode})
whereby the function $u$ is substituted by a doublet, i.e.,
\begin{subequations}
\begin{equation}
\mathbb{U}^{(1)}_{\pm,n}(\kappa)=
\exp
\Big[
-
\mathbb{F}^{(1)}_{\pm,n}(\kappa)
\Big]
\mathbb{U}^{(1)}_{\pm,n}(\kappa^{\ }_{0})
\label{eq: the formal solution if m neq 0}
\end{equation}
with
\begin{equation}
\mathbb{F}^{(1)}_{\pm,n}(\kappa)=
\int\limits^{\kappa}_{\kappa^{\ }_{0}} 
\frac{d\kappa'}{\kappa'}
\mathbb{G}^{(1)}_{\pm,n}(\kappa')
\end{equation}
\end{subequations}
and $\mathbb{G}^{(1)}_{\pm,n}(\kappa')$ both $2\times2$ matrices,
follows. However, it is not normalizable.

It is time to seek a solution to
\begin{equation}
0=
\mathcal{H}^{(2)}_{\text{vor}}
\begin{pmatrix}
u^{(2)}
\\
v^{(2)}
\end{pmatrix}.
\label{eq: triplet zero mode is defined block 2}
\end{equation}
We must solve
\begin{subequations}
\label{eq: zero mode block 2}
\begin{eqnarray}
&&
0=
\varepsilon(k,\bar{k}) 
u^{(2)}
+
c
k2\text{i}\partial^{\ }_{\bar{k}}
v^{(2)},
\label{eq: zero mode block 2 a}
\\
&&
0=
\bar{c}
\bar{k}2\text{i}\partial^{\ }_{k}
u^{(2)}
-
\varepsilon(k,\bar{k})
v^{(2)}.
\label{eq: zero mode block 2 b}
\end{eqnarray}
\end{subequations}
If we take the complex conjugate of
Eq.~(\ref{eq: zero mode block 2 b}),
we get 
\begin{subequations}
\begin{eqnarray}
&&
0=
\varepsilon(k,\bar{k}) 
u^{(2)}
+
c
k2\text{i}\partial^{\ }_{\bar{k}}
v^{(2)},
\\
&&
0=
-
c
k2\text{i}\partial^{\ }_{\bar{k}}
\bar{u}^{(2)}
-
\varepsilon(k,\bar{k})
\bar{v}^{(2)}.
\end{eqnarray}
\end{subequations}
We thus infer that a solution to 
Eq.~(\ref{eq: triplet zero mode is defined block 2}),
if it exists, is given by
\begin{subequations}
\begin{equation}
\Phi^{(2)}_{\pm}=
\begin{pmatrix}
u^{(2)}_{\pm}
\\
\pm
\bar{u}^{(2)}_{\pm}
\end{pmatrix}
\label{spinor for 2nd block}
\end{equation}
where $u^{(2)}_{\pm}$ is the solution to
\begin{equation}
\begin{split}
0=&\,
\varepsilon(\kappa) 
u^{(2)}_{\pm}
+
(\pm)
\tilde{c}
\kappa e^{2\text{i}\varphi}
\left(
\partial^{\ }_{\kappa}
+
\frac{\text{i}}{\kappa}
\partial^{\ }_{\varphi}
\right)
\bar{u}^{(2)}_{\pm}.
\end{split}
\label{DE block 2}
\end{equation}
\end{subequations}
When expanding the solution in angular momentum modes 
$\exp(\text{i}n\varphi)$, $n\in \mathbb{Z}$,
the mode $n=1$ turns out to be the only mode that 
does not couple to other modes 
via Eq.~\eqref{DE block 2}.
The ansatz
\begin{subequations}
\label{eq: final block 2 zero mode} 
\begin{equation}
u^{(2)}_{\text{sgn}\,\tilde{c}}(\kappa,\varphi)=
e^{\text{i}\varphi}
w^{(2)}_{\text{sgn}\,\tilde{c}}(\kappa)
\label{eq: final block 2 zero mode a} 
\end{equation}
casts Eq.~\eqref{DE block 2} 
in the same form as Eq.~\eqref{eq: intermediary step} 
so that
\begin{equation}
w^{(2)}_{\text{sgn}\,\tilde{c}}(\kappa)=
w^{(2)}_{\text{sgn}\,\tilde{c}}(\kappa^{\ }_{0})
\times
\left(
\frac{\kappa}{\kappa^{\ }_{0}}
\right)^{\frac{\mu}{|\tilde{c}|}-1}
\times
e^{
-
\frac{
\kappa^{2}-\kappa^{2}_{0}
     }
     {
4m|\tilde{c}|
     }
  }.
\label{eq: final block 2 zero mode b} 
\end{equation}
\end{subequations}
Observe that $\Phi^{(2)}_{\text{sgn}\,\tilde{c}}(\boldsymbol{k})$ 
with $u^{(2)}_{\pm}(\boldsymbol{k})$ given
in Eq.~\eqref{eq: final block 2 zero mode} is 
an eigenstate with the eigenvalue $-\text{sgn}\,\tilde{c}$ 
of the particle-hole transformation
defined in Eq.~\eqref{eq: def PH sym}, i.e.,
\begin{equation}
X^{\ }_{22}\Phi^{(2)*}_{\text{sgn}\,\tilde{c}}(-\boldsymbol{k})
=
-
\text{sgn}\,\tilde{c}\,
\Phi^{(2)}_{\text{sgn}\,\tilde{c}}(\boldsymbol{k}).
\end{equation}

It remains to verify that the spinor~\eqref{eq: final block 2 zero mode}
is single valued in real space, i.e., that the Fourier transform of the
function%
~(\ref{eq: final block 2 zero mode a})
vanishes at the origin of the complex-$z=r\exp(\text{i}\theta)$ plane.
Hence, we need the small $r$ expansion of the Fourier transform
\begin{equation}
\begin{split}
u^{(2)}_{\text{sgn}\,\tilde{c}}(r,\theta)\propto&\,
\int\limits_{0}^{\infty} d\kappa\kappa
\int\limits_{0}^{2\pi}d\varphi
e^{\text{i}r\kappa\cos(\varphi-\theta)}
\times
e^{\text{i}\varphi}
w^{(2)}_{\text{sgn}\,\tilde{c}}(\kappa)
\\
=&\,
e^{\text{i}\theta}
\int\limits_{0}^{\infty} d\kappa\kappa
\int\limits_{0}^{2\pi}d\phi\,
e^{\text{i}r\kappa\cos\phi}
\times
e^{\text{i}\phi}
w^{(2)}_{\text{sgn}\,\tilde{c}}(\kappa).
\end{split}
\end{equation}
The $\kappa$ integral is well behaved for large $\kappa$
because of the Gaussian factor. Moreover, an upper
cutoff to this integral can be used up to Gaussian accuracy. 
If so, a Taylor expansion of 
$\exp(\text{i}r\kappa\cos\phi)$ in the integrand
can be performed to capture the leading dependence on $r$. 
The integral over $\phi$ 
eliminates the term independent of $r$ so that
\begin{equation}
u^{\ }_{\text{sgn}\,\tilde{c}}(r,\theta)\sim
re^{\text{i}\theta}
+
\mathcal{O}(r^{2})
\end{equation} 
is single valued at the origin $r=0$ and thus an admissible solution.

We conclude that there are two
Majorana states 
\begin{equation}
\Phi^{\ }_{\text{sgn}\,\tilde{c}}(\kappa,\varphi)=
\left[
A
\begin{pmatrix}
1
\\
0
\\
0
\\
\text{sgn}\,\tilde{c}
\end{pmatrix}
+
B
\begin{pmatrix}
0
\\
e^{\text{i}\varphi}
\\
\text{sgn}\,\tilde{c}\,
e^{-\text{i}\varphi}
\\
0
\end{pmatrix}
\right]
u^{\ }_{\text{sgn}\,\tilde{c}}(\kappa)
\label{eq: main result for pair Majorana}
\end{equation}
($A$ and $B$ are real numbers)
bound to the core of 
an isolated vortex satisfying the linear profile%
~(\ref{eq: linear vortex profile})
in the triplet-pair potential.
The Majorana state weighted by the coefficient
$A$ is 
related to the  Majorana state weighted by the
coefficient $B$ through the 
helical symmetry defined in Eq.~(\ref{eq: def helical sym}) and not by the
operation of time reversal defined
in Eq.~(\ref{eq: def TRS sym}). 
This is expected 
since TRS is broken by the vortex.
\end{proof}

Lu and Yip in Ref.~\onlinecite{Lu08} (see also
Sato and Fujimoto in Ref.~\onlinecite{Sato09})
also found two Majorana fermions bound to the core of a vortex with unit 
vorticity in a weakly coupled 
(i.e., a large chemical potential compared to
the pairing potentials) 
2D TRS noncentrosymmetric superconductor
with dominant triplet-pair potential. 
Their first zero mode is the real-space counterpart to
the mode~(\ref{eq: final block 1 zero mode}).
Their second zero mode carries angular momentum $n=1$
and is the real-space counterpart to the mode%
~(\ref{eq: final block 2 zero mode}).

In Ref.~\onlinecite{Qi09a}, 
Qi et al.\ have studied zero modes bound to the core of vortices in TRS 
$p^{\ }_{x}\pm\text{i}p^{\ }_{y}$ superconductors as well.
Viewing the system as a combination of a 
$p^{\ }_{x}+\text{i}p^{\ }_{y}$ superfluid (which corresponds to 
$\mathcal{H}^{(1)}_{\text{vor}}$)
and its time-reversed partner, a $p^{\ }_{x}-\text{i}p^{\ }_{y}$
superfluid (which corresponds to $\mathcal{H}^{(2)}_{\text{vor}}$), 
they simultaneously introduced a vortex in the former and an antivortex 
in the latter. In contrast to our study of a TRS-breaking vortex, 
this configuration of a pair of vortex and antivortex
is TRS and the two Majorana modes obtained by
Qi et al.\ are connected by the operation of time reversal. 
Whereas the Majorana fermions~(\ref{eq: main result for pair Majorana})
are not robust to a generic perturbation that breaks translation invariance,
the TRS-protected pair of Majorana fermions
obtained by Qi et al.\ is robust to any
weak perturbation that preserves TRS.

\subsection{
Away from the Rashba-Dirac and Fermi limits
           }
\label{sec: Majorana fermions c}

Majorana fermions tied to vortices
are robust to continuous changes in the 
BdG Hamiltonian as long as the spectral gap does not close,
for all nonvanishing energy eigenvalues occur pairwise with
the energies $\pm E$ owing to the particle-hole
symmetry~(\ref{eq: def PH sym}).
There will be one (two) Majorana fermion(s) tied
to the core of an isolated vortex carrying vorticity one
in regions I (II) of
Fig.~\ref{fig:phase_dia2d-Dirac}
(Fig.~\ref{fig:phase_dia2d-Fermi}).
By the same reasoning, region III
of Fig.~\ref{fig:phase_dia2d-Fermi}
does not admit Majorana fermions bound to the core of 
unit vortices.\cite{footnote:acknowledgement-mit}
Regions II and III in Fig.~\ref{fig:phase_dia2d-Fermi}
differ by the even number of Majorana fermions
that TRS-breaking vortices can accommodate.
This distinction is not robust to any generic perturbation that
breaks translation symmetry, for it is not protected by TRS.

\section{
Discussion
        }
\label{sec: Discussion} 

In this paper, we studied the possible superconducting phases of the
surface states of 3D TRS topological insulators and 2D TRS
noncentrosymmetric metals. Both systems share remarkable
magneto-electric effects, however their bulk superconducting phases
differ in important ways. The difference stems from the topology of
the bands. Surface states of 3D TRS topological insulators are
topologically equivalent to a single species of Rashba-Dirac fermions 
while noncentrosymmetric metals are Fermi like with two
Fermi surfaces for large chemical potentials. As a result, 
we find that there is a unique superconducting phase 
in the case of the Rashba-Dirac limit while there are
two phases in the Fermi limit.

We studied the phase diagram as a function of the strengths of the
mean-field pair potentials $\Delta^{\ }_{\mathrm{s}}$ (singlet) and
$\Delta^{\ }_{\mathrm{t}}$ (triplet), as well as $\mu$ (chemical
potential). In the Rashba-Dirac limit, a 
single Majorana fermion is bound to the
core of an isolated and TRS-breaking vortex with unit winding number
in the superconducting order parameter everywhere in
the phase diagram in the 
\mbox{$\Delta^{\ }_{\mathrm{t}}/\Delta^{\ }_{\mathrm{s}}$--$\mu$} plane, 
with the exceptions where the gap
closes. The gap-closing lines do not separate distinct phases in the
Rashba-Dirac limit, because one can always connect two sides of a gap-closing
line by, instead of crossing the line directly, going through the
point at infinity ($\Delta^{\ }_{\mathrm{s}}=0$) without closing the
gap. Evidently, gap closing is a necessary but not sufficient
condition to have two distinct phases. 

In the Fermi limit, we find instead that there are two 
superconducting phases. 
These phases correspond to singlet or triplet dominated
physics. In the singlet-dominated phase, we find 
that an isolated TRS-breaking vortex with unit winding number
(a full vortex) does not bind Majorana fermions.
In the triplet-dominated phase, we find 
a pair of Majorana fermions bound to an isolated full vortex. 
Hence, these Majorana states have a distinct
origin from those found for half vortices in the 
$p^{\ }_{x}\pm\text{i}p^{\ }_{y}$
superconductors. The physical reason for this difference
is that TRS forces the spin-resolved pairing
amplitudes $\Delta^{\ }_{\uparrow\uparrow}$ and
$\Delta^{\ }_{\downarrow\downarrow}$ to be related,
and thus one cannot introduce
vorticity in one but not the other, as can be done with
half vortices in the $p^{\ }_{x}\pm\text{i}p^{\ }_{y}$
superconductors. 

\section*{
Acknowledgments
         }

We thank S. Ryu and A. Furusaki for many useful discussions. 
This work is supported in part by the DOE under Grant No. DE-FG02-06ER46316 
(C.C.).
C.M. thanks the Condensed Matter Theory Visitor's Program
at Boston University for support.
T.N. acknowledges the German National Academic Foundation 
for financial support.

\appendix

\begin{widetext}

\section{
Proof of Eqs.~(\ref{eq: transformed V}), 
(\ref{eq: transformed Heisenberg}), 
and~(\ref{eq: transformed DM})
        }
\label{appsec: proofs}

All reduced interaction Hamiltonians 
\eqref{eq: reduced BCS}, 
\eqref{eq: reduced BCS_Heisenberg}, 
and \eqref{eq: reduced BCS-DM}  
have summands which can be represented in terms of the
helicity basis using the transformation%
~\eqref{eq:Spinor Trafo}.
To do this explicitly,
let $\mu$ and $\nu$ run from 0 to 3 with 
$\sigma^{\ }_0$ the $2\times2$ unit matrix
and write
\begin{equation}
\begin{split}
\left(
\psi^{\dag}_{\boldsymbol{k}}
\sigma^{(\mu)}
\psi^{\ }_{\boldsymbol{p}}
\right)
\left(
\psi^{\dag}_{-\boldsymbol{k}}
\sigma^{(\nu)} 
\psi^{\ }_{-\boldsymbol{p}}
\right)=&\,
\sum_{s^{\ }_1,s^{\ }_2,s^{\ }_3,s^{\ }_4}
\sigma^{(\mu)}_{s^{\ }_1,s^{\ }_4}
\sigma^{(\nu)}_{s^{\ }_2,s^{\ }_3}
c^{\dag}_{ \boldsymbol{k}s^{\ }_1}
c^{\dag}_{-\boldsymbol{k}s^{\ }_2}
c^{\   }_{-\boldsymbol{p}s^{\ }_3}
c^{\   }_{ \boldsymbol{p}s^{\ }_4}
+
\cdots
\\
=&\,
\left(
\tilde{\psi}^{\dag}_{\boldsymbol{k}}
\Pi^{\dag}_{\boldsymbol{k}} 
\sigma^{(\mu)}
\Pi^{\ }_{\boldsymbol{p}} 
\tilde{\psi}^{\ }_{\boldsymbol{p}}
\right)
\left(
\tilde{\psi}^{\dag}_{-\boldsymbol{k}}
\Pi^{\dag}_{-\boldsymbol{k}}
\sigma^{(\nu)} 
\Pi^{\ }_{-\boldsymbol{p}} 
\tilde{\psi}^{\ }_{-\boldsymbol{p}}
\right).
\end{split}
\label{eq:Products Helicity Basis}
\end{equation}
The $\cdots$ stands for two-fermion contributions that result 
from anticommuting the operators.

We first evaluate the matrix products
\begin{equation}
\begin{split}
\Pi^{\dag}_{\pm\boldsymbol{k}} 
\sigma^{(0)}
\Pi^{\ }_{\pm\boldsymbol{p}}=&\,
\hphantom{\pm}\
\exp
\left(
\text{i}
\frac{\varphi^{\ }_{\boldsymbol{p}}-\varphi^{\ }_{\boldsymbol{k}}}{2}
\right)
\left[
\cos{\frac{\varphi^{\ }_{\boldsymbol{p}}-\varphi^{\ }_{\boldsymbol{k}}}{2}}
\sigma^{(0)}
-
\text{i}
\sin{\frac{\varphi^{\ }_{\boldsymbol{p}}-\varphi^{\ }_{\boldsymbol{k}}}{2}}
\sigma^{(1)}
\right],
\\
\Pi^{\dag}_{\pm\boldsymbol{k}} 
\sigma^{(1)}
\Pi^{\ }_{\pm\boldsymbol{p}}=&\,
\pm 
\exp
\left(
\text{i}
\frac{\varphi^{\ }_{\boldsymbol{p}}-\varphi^{\ }_{\boldsymbol{k}}}{2}
\right)
\left[
\cos{\frac{\varphi^{\ }_{\boldsymbol{p}}+\varphi^{\ }_{\boldsymbol{k}}}{2}}
\sigma^{(3)}
+
\hphantom{\text{i}}
\sin{\frac{\varphi^{\ }_{\boldsymbol{p}}+\varphi^{\ }_{\boldsymbol{k}}}{2}}
\sigma^{(2)}
\right],
\\
\Pi^{\dag}_{\pm\boldsymbol{k}} 
\sigma^{(2)}
\Pi^{\ }_{\pm\boldsymbol{p}}=&\,
\mp
\exp
\left(
\text{i}
\frac{\varphi^{\ }_{\boldsymbol{p}}-\varphi^{\ }_{\boldsymbol{k}}}{2}
\right)
\left[
\cos{\frac{\varphi^{\ }_{\boldsymbol{p}}+\varphi^{\ }_{\boldsymbol{k}}}{2}}
\sigma^{(2)}
-
\hphantom{\text{i}}
\sin{\frac{\varphi^{\ }_{\boldsymbol{p}}+\varphi^{\ }_{\boldsymbol{k}}}{2}}
\sigma^{(3)}
\right],
\\
\Pi^{\dag}_{\pm\boldsymbol{k}} 
\sigma^{(3)}
\Pi^{\ }_{\pm\boldsymbol{p}}=&\,
\hphantom{\pm}\
\exp
\left(
\text{i}
\frac{\varphi^{\ }_{\boldsymbol{p}}-\varphi^{\ }_{\boldsymbol{k}}}{2}
\right)
\left[
\cos{\frac{\varphi^{\ }_{\boldsymbol{p}}-\varphi^{\ }_{\boldsymbol{k}}}{2}}
\sigma^{(1)}
-
\text{i}
\sin{\frac{\varphi^{\ }_{\boldsymbol{p}}-\varphi^{\ }_{\boldsymbol{k}}}{2}}
\sigma^{(0)}
\right].
\end{split}
\label{eq: lab to helicity 1}
\end{equation}
Here, we observe that any of the right-hand sides 
in Eq.~(\ref{eq: lab to helicity 1})
involves one diagonal and one off-diagonal Pauli matrix.
At this point we can partly settle our constraint to have 
only Cooper pairs made of electrons of the same helicity. 
For this case, only products of two diagonal or 
two off-diagonal terms contribute in the product%
~\eqref{eq:Products Helicity Basis}.

Second, we introduce the notation
\begin{equation}
\eta^{\ }_{\boldsymbol{k}\lambda|\boldsymbol{p}\lambda'}:=
a^{\dag}_{ \boldsymbol{k}\lambda }
a^{\dag}_{-\boldsymbol{k}\lambda }
a^{\   }_{-\boldsymbol{p}\lambda'}
a^{\   }_{ \boldsymbol{p}\lambda'},
\qquad
\lambda,\lambda=\pm,
\end{equation} 
in terms of which we find
\begin{subequations}
\begin{eqnarray}
\tilde{\psi}^{\dag}_{\boldsymbol{k}}
\sigma^{(0)}
\tilde{\psi}^{\ }_{\boldsymbol{p}}
\tilde{\psi}^{\dag}_{-\boldsymbol{k}}
\sigma^{(0)} 
\tilde{\psi}^{\ }_{-\boldsymbol{p}} 
&=&
+
\tilde{\psi}^{\dag}_{\boldsymbol{k}}
\sigma^{(3)}
\tilde{\psi}^{\ }_{\boldsymbol{p}}
\tilde{\psi}^{\dag}_{-\boldsymbol{k}}
\sigma^{(3)} 
\tilde{\psi}^{\ }_{-\boldsymbol{p}} 
=
\eta^{\ }_{\boldsymbol{k}+|\boldsymbol{p}+}
+
\eta^{\ }_{\boldsymbol{k}-|\boldsymbol{p}-}
+\cdots,
\label{SigmaSpinorProduct0033}
\\
\tilde{\psi}^{\dag}_{\boldsymbol{k}}
\sigma^{(1)}
\tilde{\psi}^{\ }_{\boldsymbol{p}}
\tilde{\psi}^{\dag}_{-\boldsymbol{k}}
\sigma^{(1)} 
\tilde{\psi}^{\ }_{-\boldsymbol{p}} 
&=&-
\tilde{\psi}^{\dag}_{\boldsymbol{k}}
\sigma^{(2)}
\tilde{\psi}^{\ }_{\boldsymbol{p}}
\tilde{\psi}^{\dag}_{-\boldsymbol{k}}
\sigma^{(2)} 
\tilde{\psi}^{\ }_{-\boldsymbol{p}} 
=
\eta^{\ }_{\boldsymbol{k}+|\boldsymbol{p}-}
+
\eta^{\ }_{\boldsymbol{k}-|\boldsymbol{p}+}
+\cdots,
\label{SigmaSpinorProduct1122}
\\
\tilde{\psi}^{\dag}_{\boldsymbol{k}}
\sigma^{(0)}
\tilde{\psi}^{\ }_{\boldsymbol{p}}
\tilde{\psi}^{\dag}_{-\boldsymbol{k}}
\sigma^{(3)} 
\tilde{\psi}^{\ }_{-\boldsymbol{p}} 
&=&
+
\tilde{\psi}^{\dag}_{\boldsymbol{k}}
\sigma^{(3)}
\tilde{\psi}^{\ }_{\boldsymbol{p}}
\tilde{\psi}^{\dag}_{-\boldsymbol{k}}
\sigma^{(0)} 
\tilde{\psi}^{\ }_{-\boldsymbol{p}} 
=
\eta^{\ }_{\boldsymbol{k}+|\boldsymbol{p}+}
-
\eta^{\ }_{\boldsymbol{k}-|\boldsymbol{p}-}
+\cdots,
\label{SigmaSpinorProduct0330}
\\
\tilde{\psi}^{\dag}_{\boldsymbol{k}}
\sigma^{(1)}
\tilde{\psi}^{\ }_{\boldsymbol{p}}
\tilde{\psi}^{\dag}_{-\boldsymbol{k}}
\sigma^{(2)} 
\tilde{\psi}^{\ }_{-\boldsymbol{p}} 
&=&
+
\tilde{\psi}^{\dag}_{\boldsymbol{k}}
\sigma^{(2)}
\tilde{\psi}^{\ }_{\boldsymbol{p}}
\tilde{\psi}^{\dag}_{-\boldsymbol{k}}
\sigma^{(1)} 
\tilde{\psi}^{\ }_{-\boldsymbol{p}} 
=
-
\text{i}
\left(
\eta^{\ }_{\boldsymbol{k}+|\boldsymbol{p}-}
-
\eta^{\ }_{\boldsymbol{k}-|\boldsymbol{p}+},
\right)
+\cdots.
\label{SigmaSpinorProduct1221}
\end{eqnarray}\end{subequations}
Here, $\cdots$ stands for contributions which would lead to Cooper 
pairs made up of two electrons of different helicity.
We are left with the task of collecting 
the phase and trigonometric multiplicative factors
from Eq.~(\ref{eq: lab to helicity 1}). 

For the density-density interaction~\eqref{eq: reduced BCS}, 
we have to compute Eq.~\eqref{eq:Products Helicity Basis}
with $\mu=\nu=0$.
According to 
Eq.~(\ref{eq: lab to helicity 1})
this involves collecting 
the phase and trigonometric multiplicative factors
for Eqs.~\eqref{SigmaSpinorProduct0033} 
and \eqref{SigmaSpinorProduct1122}. 
We find
\begin{equation}
\begin{split}
H^{\text{red}}_{V}=&\,
\frac{1}{2}
\sum_{\boldsymbol{k},\boldsymbol{p}}
V^{\ }_{\boldsymbol{k}-\boldsymbol{p}}
\left(
\tilde{\psi}^{\dag}_{\boldsymbol{k}}
\Pi^{\dag}_{\boldsymbol{k}} 
\Pi^{\ }_{\boldsymbol{p}} 
\tilde{\psi}^{\ }_{\boldsymbol{p}}
\right)
\left(
\tilde{\psi}^{\dag}_{-\boldsymbol{k}}
\Pi^{\dag}_{-\boldsymbol{k}}
\Pi^{\ }_{-\boldsymbol{p}} 
\tilde{\psi}^{\ }_{-\boldsymbol{p}}
\right)
\\
=&\,
2\sum_{\boldsymbol{k}\boldsymbol{p}}
V^{\ }_{\boldsymbol{k}-\boldsymbol{p}}
e^{\text{i}(\varphi^{\ }_{\boldsymbol{p}}-\varphi^{\ }_{\boldsymbol{k}})}
\left[
\cos^2
\frac{
\varphi^{\ }_{\boldsymbol{p}}-\varphi^{\ }_{\boldsymbol{k}}
     }
     {
2
     }
\left(
\eta_{\boldsymbol{k}+|\boldsymbol{p}+}
+
\eta_{\boldsymbol{k}-|\boldsymbol{p}-}
\right)
-
\sin^2
\frac{
\varphi^{\ }_{\boldsymbol{p}}
-
\varphi^{\ }_{\boldsymbol{k}}
     }
     {
2
     }
\left(
\eta^{\ }_{\boldsymbol{k}+|\boldsymbol{p}-}
+
\eta^{\ }_{\boldsymbol{k}-|\boldsymbol{p}+}
\right)
\right]
\label{eq: V Psi expansion}
\end{split}
\end{equation}
from which Eq.~\eqref{eq: transformed V} follows.

For the Heisenberg interaction~\eqref{eq: reduced BCS_Heisenberg}, 
we have to compute Eq.~\eqref{eq:Products Helicity Basis}
with $\mu=\nu=1,2,3$.
According to 
Eq.~(\ref{eq: lab to helicity 1}) 
this involves collecting 
the phase and trigonometric multiplicative factors
for Eqs.~\eqref{SigmaSpinorProduct0033} 
and \eqref{SigmaSpinorProduct1122}. 
We find
\begin{equation}
\begin{split}
H^{\text{red}}_{\text{H}}=&\,
\frac{1}{8}
\sum_{\boldsymbol{k},\boldsymbol{p}}
\sum_{j=1}^{3}
J^{\ }_{\boldsymbol{k}-\boldsymbol{p}}
\left(
\tilde{\psi}^{\dag}_{\boldsymbol{k}}
\Pi^{\dag}_{\boldsymbol{k}} 
\sigma^{(j)}
\Pi^{\ }_{\boldsymbol{p}} 
\tilde{\psi}^{\ }_{\boldsymbol{p}}
\right)
\left(
\tilde{\psi}^{\dag}_{-\boldsymbol{k}}
\Pi^{\dag}_{-\boldsymbol{k}}
\sigma^{(j)}
\Pi^{\ }_{-\boldsymbol{p}} 
\tilde{\psi}^{\ }_{-\boldsymbol{p}}
\right)
\\
=&\,
\frac{1}{8}
\sum_{\boldsymbol{k},\boldsymbol{p}}
J^{\ }_{\boldsymbol{k}-\boldsymbol{p}}
e^{\text{i}(\varphi^{\ }_{\boldsymbol{p}}-\varphi^{\ }_{\boldsymbol{k}})}
\left[
\eta^{\ }_{\boldsymbol{k}+|\boldsymbol{p}-}
+
\eta^{\ }_{\boldsymbol{k}-|\boldsymbol{p}+}
-
\eta^{\ }_{\boldsymbol{k}+|\boldsymbol{p}+}
-
\eta^{\ }_{\boldsymbol{k}-|\boldsymbol{p}-}
+
\cos^2
\frac{
\varphi^{\ }_{\boldsymbol{p}}-\varphi^{\ }_{\boldsymbol{k}}
     }
     {
2
     }
\left(
\eta^{\ }_{\boldsymbol{k}+|\boldsymbol{p}-}
+
\eta^{\ }_{\boldsymbol{k}-|\boldsymbol{p}+}
\right)
\right.
\\
&\,
\left.
-
\sin^2
\frac{
\varphi^{\ }_{\boldsymbol{p}}
-
\varphi^{\ }_{\boldsymbol{k}}
     }
     {
2
     }
\left(
\eta^{\ }_{\boldsymbol{k}+|\boldsymbol{p}+}
+
\eta^{\ }_{\boldsymbol{k}-|\boldsymbol{p}-}
\right)
\right]
\end{split}
\label{eq: Heisenberg Psi expansion}
\end{equation}
from which Eq.~\eqref{eq: transformed Heisenberg} follows. 

Finally, the Dzyaloshinskii-Moriya interaction%
~\eqref{eq: reduced BCS-DM} involves terms of the type
~\eqref{eq:Products Helicity Basis}
with $(\mu,\nu)=(2,3)$, $(3,2)$, $(1,3)$, and $(3,1)$ 
which in turn lead to Eqs.~\eqref{SigmaSpinorProduct0330} 
and~\eqref{SigmaSpinorProduct1221}. The calculations yields
\begin{eqnarray}
H^{\text{red}}_{\mathrm{DM}}&=&
\frac{1}{8}
\sum_{\boldsymbol{k},\boldsymbol{p}}
\sum_{j,l=1\ldots3}^{m=1,2}
\epsilon_{jlm} D^{(m)}_{\boldsymbol{k}-\boldsymbol{p}}
\left(
\tilde{\psi}^{\dag}_{\boldsymbol{k}}
\Pi^{\dag}_{\boldsymbol{k}} 
\sigma^{(j)}
\Pi^{\ }_{\boldsymbol{p}} 
\tilde{\psi}^{\ }_{\boldsymbol{p}}
\right)
\left(
\tilde{\psi}^{\dag}_{-\boldsymbol{k}}
\Pi^{\dag}_{-\boldsymbol{k}}
\sigma^{(l)}
\Pi^{\ }_{-\boldsymbol{p}} 
\tilde{\psi}^{\ }_{-\boldsymbol{p}}
\right)
\nonumber\\
&=&\frac{\text{i}}{4}
\sum_{\boldsymbol{k},\boldsymbol{p}}
e^{\text{i}(\varphi^{\ }_{\boldsymbol{p}}-\varphi^{\ }_{\boldsymbol{k}})}
\biggl[
\left(
\eta_{\boldsymbol{k}+|\boldsymbol{p}-}
-
\eta_{\boldsymbol{k}-|\boldsymbol{p}+}
\right)
\left(
D^{(1)}_{\boldsymbol{k}-\boldsymbol{p}}
\cos\frac{\varphi^{\ }_{\boldsymbol{p}}+\varphi^{\ }_{\boldsymbol{k}}}{2}
\cos\frac{\varphi^{\ }_{\boldsymbol{p}}-\varphi^{\ }_{\boldsymbol{k}}}{2}
+
D^{(2)}_{\boldsymbol{k}-\boldsymbol{p}}
\sin\frac{\varphi^{\ }_{\boldsymbol{p}}+\varphi^{\ }_{\boldsymbol{k}}}{2}
\cos\frac{\varphi^{\ }_{\boldsymbol{p}}-\varphi^{\ }_{\boldsymbol{k}}}{2}
\right)
\nonumber\\
&&
-
\left(
\eta_{\boldsymbol{k}+|\boldsymbol{p}+}
-
\eta_{\boldsymbol{k}-|\boldsymbol{p}-}
\right)
\left(
D^{(1)}_{\boldsymbol{k}-\boldsymbol{p}}
\sin\frac{\varphi^{\ }_{\boldsymbol{p}}+\varphi^{\ }_{\boldsymbol{k}}}{2}
\sin\frac{\varphi^{\ }_{\boldsymbol{p}}-\varphi^{\ }_{\boldsymbol{k}}}{2}
-
D^{(2)}_{\boldsymbol{k}-\boldsymbol{p}}
\cos\frac{\varphi^{\ }_{\boldsymbol{p}}+\varphi^{\ }_{\boldsymbol{k}}}{2}
\sin\frac{\varphi^{\ }_{\boldsymbol{p}}-\varphi^{\ }_{\boldsymbol{k}}}{2}
\right)
\biggr]
\nonumber\\
&&
\label{eq: DM Psi expansion}
\end{eqnarray}
from which Eq.~\eqref{eq: transformed DM} follows. 

\section{
Unnormalizability of higher angular momentum zero modes
        }
\label{appsec: zero modes}

We are going to show that the solutions to 
Eq.~\eqref{zero modes block1} with nonzero angular momentum
and the solutions to Eq.~\eqref{DE block 2} 
with angular momentum other than $+1$ are not normalizable.
We expand
\begin{equation}
\begin{split}
&
u^{(1)}_{\pm}(\kappa,\varphi)=
\sum_{n\geq0}
\left[
e^{ \text{i}n\varphi}f^{(1)}_{\pm,n}(\kappa)
+
e^{-\text{i}n\varphi}g^{(1)}_{\pm,n}(\kappa)
\right],
\qquad
u^{(2)}_{\pm}(\kappa,\varphi)=
\sum_{n\geq-1}
\left[
e^{ \text{i}(n+2)\varphi}f^{(2)}_{\pm,n}(\kappa)
+
e^{-\text{i} n   \varphi}g^{(2)}_{\pm,n}(\kappa)
\right],
\end{split}
\end{equation}
for Eqs.~\eqref{zero modes block1} and~\eqref{DE block 2}, respectively.
The differential equations mutually couples 
two and only two angular momentum modes.
As all coefficients of the differential equation are purely real, 
the expansion parameters 
$f^{(j)}_{\pm,n}(\kappa)$ and $g^{(j)}_{\pm,n}(\kappa)$ 
can be chosen to be either purely real or purely imaginary numbers. 
Without loss of generality, we make the former choice. 
In terms of the doublet 
$\mathbb{U}^{(j)}_{\pm,n}=\left[f^{(j)}_{\pm,n},g^{(j)}_{\pm,n}\right]^T$
that represents the two coupled modes labeled by $n$, we find
\begin{subequations}
\begin{equation}
\partial^{\ }_{\kappa}\mathbb{U}^{(j)}_{\pm,n}(\kappa)=
-\frac{1}{\kappa}
\mathbb{G}^{(j)}_{\pm,n}(\kappa)
\mathbb{U}^{(j)}_{\pm,n}(\kappa).
\label{eq:matrix equn both blocks}
\end{equation}
The matrices are given by
\begin{equation}
\mathbb{G}^{(j)}_{\pm,n}(\kappa)=
\left(
\begin{array}{cc}
	2-(j+n)&\pm\frac{\varepsilon(\kappa)}{\tilde{c}}\\
	\pm\frac{\varepsilon(\kappa)}{\tilde{c}}&j+n
\end{array}
\right).
\end{equation}
\end{subequations}
Hence, the doublet solution 
can be written as
\begin{subequations}
\begin{equation}
\mathbb{U}^{(j)}_{\pm,n}(\kappa)=
\exp
\Big[
-
\mathbb{F}^{(j)}_{\pm,n}(\kappa)
\Big]
\mathbb{U}^{(j)}_{\pm,n}(\kappa^{\ }_{0})
\label{eq: the formal solution doublet}
\end{equation}
with
\begin{equation}
\mathbb{F}^{(j)}_{\pm,n}(\kappa)=
\int\limits^{\kappa}_{\kappa^{\ }_{0}} 
\frac{d\kappa'}{\kappa'}
\mathbb{G}^{(j)}_{\pm,n}(\kappa).
\end{equation}
\end{subequations}

The demand of normalizability reads
\begin{equation}
\begin{split}
\int\limits_{0}^\infty
d\kappa
\kappa
\left[\mathbb{U}^{(j)}_{\pm,n}(\kappa)\right]^T
\mathbb{U}^{(j)}_{\pm,n}(\kappa)
=
\int\limits_{0}^\infty
d\kappa
\kappa
\left[\mathbb{U}^{(j)}_{\pm,n}(\kappa^{\ }_{0})\right]^T
\exp
\Big(
-2
\mathbb{F}^{(j)}_{\pm,n}(\kappa)
\Big)
\mathbb{U}^{(j)}_{\pm,n}(\kappa^{\ }_{0})
<
\infty.
\end{split}
\label{eq:condition for both blocks}
\end{equation}
In the limit of large $\kappa$, both matrices 
$\mathbb{G}^{(j)}_{\pm,n}(\kappa)$ ($j=1,2$) 
obey the same behavior. 
The matrix in the exponent becomes for both cases $j=1,2$,
\begin{equation}
\mathbb{F}^{(j)}_{\pm,n}(\kappa)\rightarrow
\pm\frac{\kappa^2}{2m\tilde{c}}\sigma^{\ }_{1}.
\end{equation}
Upon exponentiation it, 
the condition~\eqref{eq:condition for both blocks} then reads
\begin{equation}
\begin{split}
\int\limits_{0}^\infty
d\kappa
\kappa
\left\{
\cosh{\left(\frac{\kappa^2}{m\tilde{c}}\right)}
\left[\left(f^{(j)}_{\pm,n}(\kappa^{\ }_{0})\right)^2+\left(g^{(j)}_{\pm,n}(\kappa^{\ }_{0})\right)^2\right]
\right.
\,\left.
\mp2\sinh{\left(\frac{\kappa^2}{m\tilde{c}}\right)}
f^{(j)}_{\pm,n}(\kappa^{\ }_{0})g^{(j)}_{\pm,n}(\kappa^{\ }_{0})
\right\}
<\infty,
\end{split}
\label{eq:condition for both blocks big kappa}
\end{equation}
which gives a condition for the initial values
\begin{equation}
f^{(j)}_{\pm,n}(\kappa^{\ }_{0})=
\pm\text{sgn}\,\tilde{c} g^{(j)}_{\pm,n}(\kappa^{\ }_{0}).
\label{eq:cond1 both blocks}
\end{equation}
In the opposite limit of small $\kappa$, 
we find for the matrix that has to be exponentiated
\begin{equation}
\mathbb{F}^{(j)}_{\pm,n}(\kappa)\rightarrow 
\mathbb{G}^{(j)}_{\pm,n}(0)\ln{\frac{\kappa}{\kappa^{\ }_{0}}}.
\end{equation}
Upon exponentiating it, 
the condition~\eqref{eq:condition for both blocks} reads
\begin{equation}
\begin{split}
\infty>&\,
\int\limits_{0}^\infty
d\kappa
\kappa
\frac{
\kappa^{-2\left(1+\sqrt{(n+j-1)^2+\frac{\mu^2}{\tilde{c}^2}}\right)}
     }
     {
\sqrt{(n+j-1)^2+\frac{\mu^2}{\tilde{c}^2}}
     }
\left\{
\left\{
\left[f^{(j)}_{\pm,n}(\kappa^{\ }_{0})\right]^2
+
\left[g^{(j)}_{\pm,n}(\kappa^{\ }_{0})\right]^2
\right\}
\sqrt{
(n+j-1)^2+\frac{\mu^2}{\tilde{c}^2}
     }
\right.
\\
&\,
\left.
+
\left[
\left(f^{(j)}_{\pm,n}(\kappa^{\ }_{0})\right)^2
-
\left(g^{(j)}_{\pm,n}(\kappa^{\ }_{0})\right)^2
\right]
(n+j-1)
\mp
2\frac{\mu}{\tilde{c}} 
f^{(j)}_{\pm,n}(\kappa^{\ }_{0})g^{(j)}_{\pm,n}(\kappa^{\ }_{0}) 
+ 
\mathcal{O}
\left(
\kappa^{
4\sqrt{
(n+j-1)^2
+
\frac{\mu^2}{\tilde{c}^2}
      }
       }
\right)
\right\}.
\end{split}
\label{eq:condition for both blocks small kappa}
\end{equation}
All terms that are given explicitly in the curly bracket 
have to vanish in order to achieve normalizability. 
This amounts to
\begin{equation}
f^{(j)}_{\pm,n}(\kappa^{\ }_{0})=
\pm\frac{\tilde{c}}{\mu}\left[\sqrt{(n+j-1)^2+\frac{\mu^2}{\tilde{c}^2}}
-(n+j-1)\right] g^{(j)}_{\pm,n}(\kappa^{\ }_{0}).
\label{eq:cond2 both blocks}
\end{equation}
Both conditions~\eqref{eq:cond1 both blocks} and~\eqref{eq:cond2 both blocks} 
are only satisfied simultaneously if $n=1-j$. 
For this mode not to be vanishing, the sign in
Eq.~\eqref{eq:cond1 both blocks} must be chosen 
$\pm=\text{sgn}\,\tilde{c}$, 
which is only compatible with Eq.~\eqref{eq:cond2 both blocks} for $\mu>0$.
This corresponds to the solutions discussed in 
Sec.~\ref{sec: Majorana fermions} 
and we conclude that these are the only normalizable 
zero modes for each of the blocks $j=1,2$.
\end{widetext}


\begin{thebibliography}{99}
\bibitem{footnote: alpha to-alpha}
Changing our convention $\alpha>0$ to $\alpha<0$
amounts to exchanging the helicities through
$\boldsymbol{g}^{\ }_{\boldsymbol{k}}\to
-\boldsymbol{g}^{\ }_{\boldsymbol{k}}$.
\end{thebibliography}
\end{document}